\documentclass[prd,aps,twocolumn,showpacs,preprintnumbers,
superscriptaddress,floatfix,amssymb,eqsecnum]{revtex4} 
\usepackage{epsfig}
\usepackage{graphicx}
\newcommand{\be}{\begin{equation}}
\newcommand{\ee}{\end{equation}}

\begin{document}

\preprint{
\vbox{
\hbox{ADP-08-09/T669}
\hbox{Edinburgh 2008/50}
}}

\title{Precision electromagnetic structure of decuplet baryons in the chiral regime}

\author{S.~Boinepalli}\affiliation{Department of Physics and
  Mathematical Physics and \\
  Special Research Centre for the Subatomic Structure of Matter, \\
  University of Adelaide, 5005, Australia}
\author{D.B.~Leinweber}\affiliation{Department of Physics and
  Mathematical Physics and \\
  Special Research Centre for the Subatomic Structure of Matter, \\
  University of Adelaide, 5005, Australia}
\author{P.J.~Moran}\affiliation{Department of Physics and
  Mathematical Physics and \\
  Special Research Centre for the Subatomic Structure of Matter, \\
  University of Adelaide, 5005, Australia}
\author{A.G.~Williams}\affiliation{Department of Physics and
  Mathematical Physics and \\
  Special Research Centre for the Subatomic Structure of Matter, \\
  University of Adelaide, 5005, Australia}

\author{J.M.~Zanotti}\affiliation{School of Physics \& Astronomy,
  University of Edinburgh, Edinburgh EH9 3JZ, UK}

\author{J.B.~Zhang}\affiliation{Department of Physics, Zhejiang
  University, Hangzhou, Zhejiang 310027, P.R.~China}

\begin{abstract}
  The electromagnetic properties of the baryon decuplet are calculated
  in quenched QCD on a $20^3 \times 40$ lattice with a lattice spacing
  of 0.128~fm using the fat-link irrelevant clover (FLIC) fermion
  action with quark masses providing a pion mass as low as 300~MeV.
  Magnetic moments and charge radii are extracted from the electric
  and magnetic form factors for each individual quark sector.  From
  these, the corresponding baryon properties are constructed.
  We present results for the higher order moments of the spin-3/2
  baryons, including the electric quadrupole moment $E2$ and the magnetic
  octupole moment $M3$. The world's first determination of a non-zero
  $M3$ form factor for the $\Delta$ baryon is presented. With these
  results we provide a conclusive analysis which shows that decuplet
  baryons are deformed.
  We compare the decuplet baryon results from a similar lattice
  calculation of the octet baryons. We establish that the environment
  sensitivity is far less pronounced in the case of the decuplet
  baryons compared to that in the octet baryons.
  A surprising result is that the charge radii of the decuplet baryons
  are generally smaller than that of the octet baryons.
  The magnetic moment of the $\Delta^+$ reveals a turn over in the low
  quark mass region, making it smaller than the proton magnetic
  moment. These results are consistent with the expectations of
  quenched chiral perturbation theory.
  A similar turn over is also noticed in the magnetic moment of the
  $\Sigma^{*0}$, but not for $\Xi^*$ where only kaon loops can appear
  in quenched QCD.
 The electric quadrupole moment of the $\Omega^-$ baryon is positive
 when the negative charge factor is included, and is equal to $0.86
 \pm 0.12 \times 10^{-2}~{\rm fm^2 }$, indicating an oblate shape.
\end{abstract}


\maketitle


\section{INTRODUCTION}
\label{sec:intro}

The study of the electromagnetic properties of baryons provides
valuable insight into the non-perturbative structure of QCD (see
Refs.~\cite{Gao:2003ag,Hyde-Wright:2004gh,Arrington:2006zm,Perdrisat:2006hj,deJager:2006nt}
for recent reviews).
Baryon charge radii and magnetic moments provide an excellent
opportunity to observe the non-analytic behavior predicted by chiral
effective field theory ($\chi$EFT).
Since these are inherently non-perturbative properties of hadrons,
first-principles calculations on the lattice are essential for our
understanding of hadronic structure, and indeed there has been much
progress in this direction, mainly for the nucleon and pseudoscalar
states (see~\cite{Zanotti_PoS007} for a review).
For decuplet baryons, however, there has been very little progress
since Ref.~\cite{Leinweber:1992hy} which appeared almost 15 years ago.
However, renewed interest appeared
recently~\cite{Alexandrou:2007we,Alexandrou:2008bn}.

The Adelaide group has been investigating the electromagnetic
structure of hadrons for several years now. 
In Refs.~\cite{Leinweber:2004tc,Leinweber:2006ug}, we presented a
novel method for determining the strange quark contribution to the
nucleon's electromagnetic form factors, the results of which were
later confirmed by an improved analysis of old experimental data
\cite{Young:2006jc} and new data from parity violating experiments at
JLab \cite{Acha:2006my}.
This was followed by an in-depth study in quenched QCD of the
electromagnetic properties of the octet baryons
\cite{Boinepalli:2006xd}.
Of particular interest was an observed environmental isospin
dependence of the strange quark distributions in $\Lambda^0$ and
$\Sigma^0$.  
More recently, we performed an investigation into the pseudoscalar and
vector meson electromagnetic form factors \cite{Hedditch:2007ex}.
Here we determined that the $\rho^{+}$ meson has a negative quadrupole
moment, indicating that the $\rho$ meson is oblate.

In this paper we continue our study of the electromagnetic structure
of hadrons and present a quenched lattice QCD calculation of the
electromagnetic form factors of $SU(3)_{\rm flavor}$ decuplet baryons. 
From these form factors we determine magnetic moments, charge and
magnetic radii and present results for the electric quadrupole and
magnetic octupole moments.

On the lattice, decuplet baryons are stable as a result of the
unphysical large quark masses that are used in present calculations
and the finite volume of the lattice.
Decay to a pion and an octet baryon is forbidden by energy
conservation.
However, stability of decuplet baryons is common to
most hadronic models.  
In this sense, lattice results provide a useful forum in which the
strengths and weaknesses of various models may be identified.
The lattice results also provide access to observables not readily
available with present experiments such as the higher-order multipole
moments of the $\Omega^-$ which is stable to strong interactions.

An examination of decuplet baryon structure in lattice QCD enables
one to study new aspects of non-perturbative quark-gluon dynamics.  
In analyzing the results we make comparisons within the baryon
decuplet and with the octet results~\cite{Boinepalli:2006xd} which
provide insights into the spin dependence of quark interactions.

The $E2$ and $M3$ moments accessible in spin-3/2 systems provide
insights into the shape of the decuplet baryon ground state.  
These higher-order moments also have the potential to discriminate
between various model descriptions of hadronic phenomena.
 
To put our results into perspective, we compare our calculations with
experimental measurements where available, and with the predictions of
Quenched Chiral Perturbation Theory (Q$\chi$PT).

The decuplet baryon interpolating fields used in the correlation
functions are discussed in Sec.~\ref{subsec:IntFields}.  
The extraction of baryon mass and electromagnetic form factors
proceeds through a calculation of two and three-point correlation
functions. 
These are discussed in Sec.~\ref{subsec:CF}.
The two and three point functions for decuplet baryons are discussed
in Secs.~\ref{subsec:2ptFn} and \ref{subsec:3ptFn}.
Throughout this analysis we employ the lattice techniques introduced
in \cite{Leinweber:1992hy}, and these are summarized in
Sec.~\ref{sec:LatTech}.
In Sec.~\ref{sec:corrFun} we outline the methods used in our
analysis of the lattice two and three point functions.
Our results are presented and discussed in
Sec.~\ref{sec:Results}, and summarized in
Sec.~\ref{sec:Summary}.


\section{Theoretical Formalism}
\label{sec:theory}
\subsection{Interpolating Fields}
\label{subsec:IntFields}

The commonly used interpolating field for exciting the $\Delta^{++}$
resonance from the QCD vacuum takes the long established
\cite{Ioffe:1981kw,Chung:1981cc} form of
\begin{equation}
\chi_\mu^{\Delta^{++}}(x) =
\epsilon^{abc} \left ( u^{Ta}(x) C \gamma_\mu
                                u^b(x) \right ) u^c(x).  
\end{equation}
Unless otherwise noted, we follow the notation of Sakurai
\cite{SAKURAI}.
The Dirac gamma matrices are Hermitian and satisfy $\left \{
  \gamma_\mu , \gamma_\nu \right \} = 2 \, \delta_{\mu \nu}$, with
$\sigma_{\mu \nu} = {1 \over 2i} \left [ \gamma_\mu , \gamma_\nu
\right ] $.  $C = \gamma_4 \gamma_2$ is the charge conjugation matrix,
$a,\ b,\ c$ are color indices, $u(x)$ is a $u$-quark field, and the
superscript $T$ denotes transpose.  
The generalization of this interpolating field for the $\Delta^+$
composed of two $u$ quarks and one $d$ quark has the form
\FL
\begin{eqnarray}
\chi_\mu^{\Delta^{+}}(x) =
{1 \over \sqrt{3} } \; \epsilon^{abc} \Bigl [
\!\!\!&2&\!\!\! \left ( u^{Ta}(x) C \gamma_\mu d^b(x) \right )
u^c(x) \nonumber \\
\!\!\!&+&\!\!\! \left ( u^{Ta}(x) C \gamma_\mu u^b(x) \right )
d^c(x)\ \Bigr ] \, .\,\,\,\,\,\,
\label{deltapif}
\end{eqnarray}
Other decuplet baryon interpolating fields are obtained with the
appropriate substitutions of $u(x),\ d(x)\ \to\ u(x),\ d(x)$ or
$s(x)$.  
The interpolating field for $\Sigma^{*0}$ is given by the symmetric
generalization
\FL
\begin{eqnarray}
\chi_\mu^{\Sigma^{*0}}(x) =
\sqrt{2 \over 3} \; \epsilon^{abc} \Bigl [
\!\!\!& &\!\!\! \left ( u^{Ta}(x) C \gamma_\mu d^b(x) \right )
s^c(x) \nonumber \\
\!\!\!&+&\!\!\! \left ( d^{Ta}(x) C \gamma_\mu s^b(x) \right )
u^c(x)  \nonumber\\
\!\!\!&+&\!\!\! \left ( s^{Ta}(x) C \gamma_\mu u^b(x) \right )
d^c(x)\ \Bigr ] \, .\,\,\,\,\,\,
\end{eqnarray}
The $SU(2)$-isospin symmetry relationship for $\Sigma^*$ form factors
\begin{equation}
\Sigma^{*0} = {\Sigma^{*+} + \Sigma^{*-} \over 2} \, ,
\end{equation}
may be easily seen in the $\Sigma^{*0}$ interpolating field by noting
\begin{eqnarray}
\lefteqn{\epsilon^{abc} \left ( s^{Ta}(x) C \gamma_\mu u^b(x) \right )
d^c(x) =} \qquad \qquad \qquad \nonumber \\
&& \epsilon^{abc}
\left ( u^{Ta}(x) C \gamma_\mu s^b(x) \right ) d^c(x) .
\end{eqnarray}

\subsection{Correlation functions}
\label{subsec:CF}

Two-point correlation functions at the quark level are obtained
through the standard procedure of contracting pairs of quark fields.
Considering the $\Delta^+$ correlation function at the quark level and
performing all possible quark field contractions gives the two-point
function as
\begin{widetext}
\begin{eqnarray}
\left < T \left ( \chi_\mu^{\Delta^+}(x)
\overline \chi_\nu^{\Delta^+}(0) \right ) \right > 
{1 \over 3} \; \epsilon^{abc} \epsilon^{a'b'c'}  \Bigl \{ 
4 S_u^{a a'} \,
                       \gamma_\nu \, C S_u^{T b b'} C \, \gamma_\mu \,
S_d^{c c'} 
\!\!&+&\!\!  4 S_u^{a a'}\, \gamma_\nu \, C S_d^{T b b'} C \, \gamma_\mu \,
S_u^{c c'} \nonumber \\
+  4 S_d^{a a'}\, \gamma_\nu \, C S_u^{T b b'} C \, \gamma_\mu \,
S_u^{c c'} 
\!\!&+&\!\!  2 S_u^{a a'} \, {\rm tr} \left [ \gamma_\nu \,
  C S_u^{T b b'} C \, \gamma_\mu \, S_d^{c c'} \right ] \label{delta2pf} \\
+  2 S_u^{a a'} \, {\rm tr} \left [ \gamma_\nu \,
  C S_d^{T b b'} C \, \gamma_\mu \, S_u^{c c'} \right ]
\!\!&+&\!\!  2 S_d^{a a'} \, {\rm tr} \left [ \gamma_\nu \,
  C S_u^{T b b'} C \, \gamma_\mu \, S_u^{c c'} \right ] \Bigr \} \,
\nonumber
\end{eqnarray}
\end{widetext}
where the quark-propagator $S_u^{a a'} = T \left ( u^a(x) \overline
u^{a'}(0) \right )$ and similarly for other quark flavors.
$SU(3)_{\rm flavor}$ symmetry is clearly displayed in this equation.
\begin{figure}[tbp]
\vspace*{0.2cm}
{\includegraphics[height=3.3cm,angle=90]{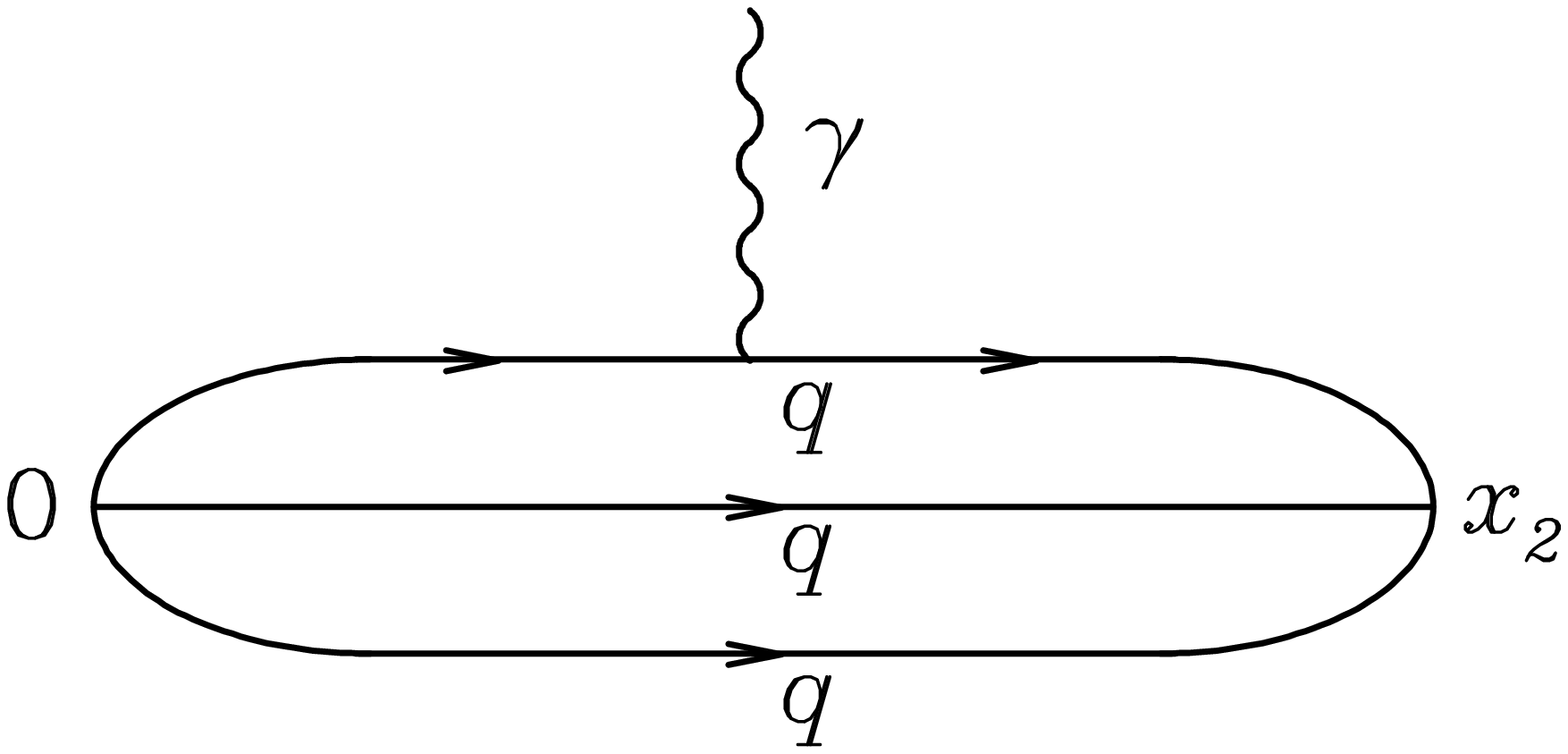}\hspace{0.8cm}
 \includegraphics[height=3.3cm,angle=90]{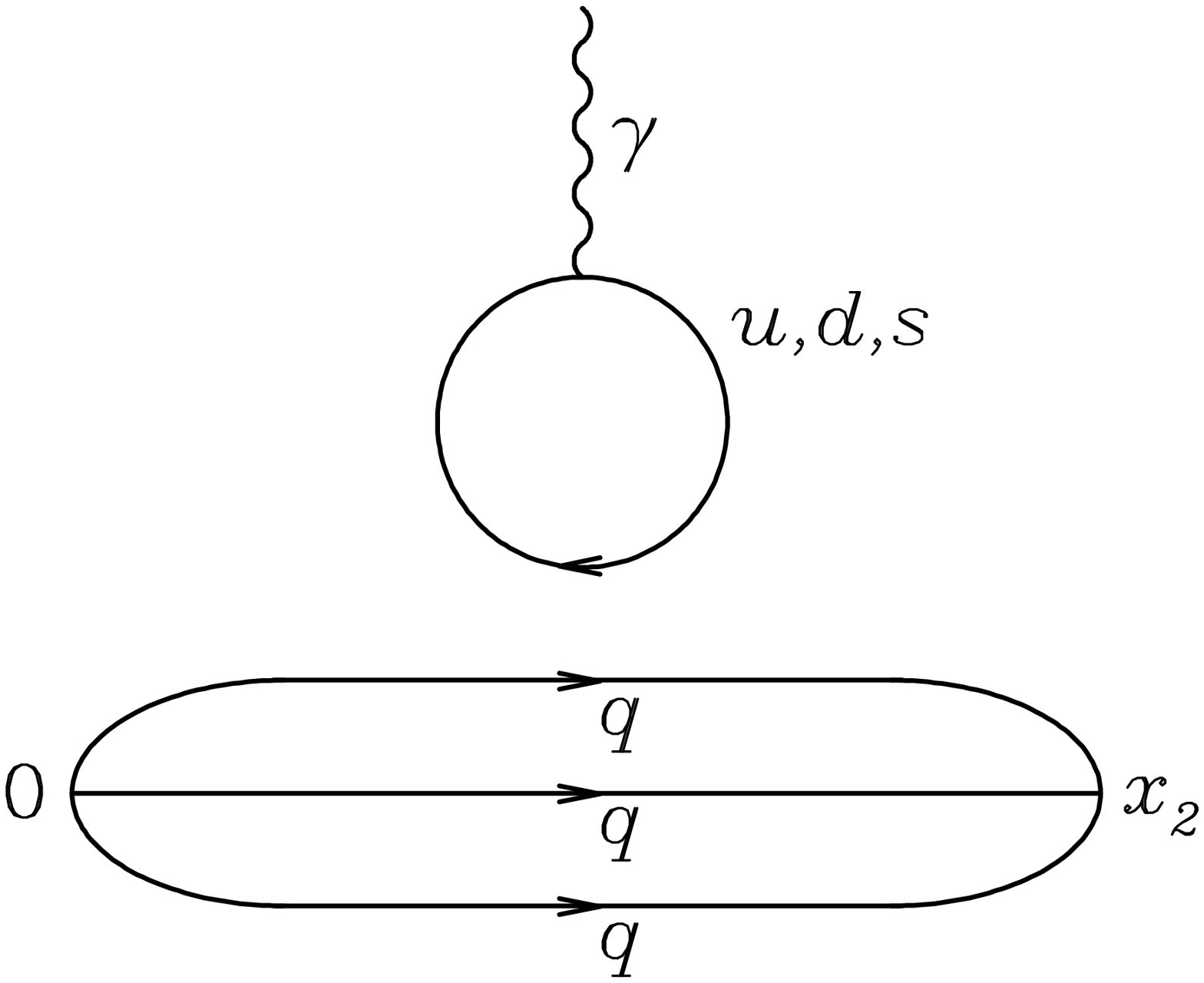}}
\vspace{0.7cm}
\caption{Diagrams illustrating the two topologically different
insertions of the current within the framework of lattice QCD.  
}
\label{topology}
\end{figure}

In determining the three point function, one encounters two
topologically different ways of performing the current insertion.
Figure \ref{topology} displays skeleton diagrams for these two
insertions.
These diagrams may be dressed with an arbitrary number of gluons.
Diagram (a) illustrates the connected insertion of the current to one
of the valence quarks of the baryon.
%
%
Diagram (b) accounts for the alternative time ordering where the
current first produces a disconnected $q \, \overline q$ pair which in
turn interacts with the valence quarks of the baryon via gluons.

The number of terms in the three-point function is four times that in
Eq.~(\ref{delta2pf}).
The correlation function relevant for a $\Delta^+$ three-point
function is
\begin{widetext}
\begin{eqnarray}
\lefteqn{
T \left ( \chi_\mu^{\Delta^+}(x_2) \, j^\alpha(x_1) \,
\overline \chi_\nu^{\Delta^+}(0) \right )  =
{1 \over 3} \; \epsilon^{abc} \epsilon^{a'b'c'}  \biggl \{ }
\nonumber \\
&& \;\;\: 4 \widehat S_u^{a a'} \, \gamma_\nu \, C S_u^{T b b'} C \,
\gamma_\mu \, S_d^{c c'}
+ 4 \widehat S_u^{a a'}\, \gamma_\nu \, C S_d^{T b b'} C \,
\gamma_\mu \, S_u^{c c'}
+ 4 \widehat S_d^{a a'}\, \gamma_\nu \, C S_u^{T b b'} C \,
\gamma_\mu \, S_u^{c c'} \nonumber \\
&& + 4 S_u^{a a'} \, \gamma_\nu \, C \widehat S_u^{T b b'} C \,
\gamma_\mu \, S_d^{c c'}
   + 4 S_d^{a a'}\, \gamma_\nu \, C \widehat S_u^{T b b'} C \,
\gamma_\mu \, S_u^{c c'}
   + 4 S_u^{a a'}\, \gamma_\nu \, C \widehat S_d^{T b b'} C \,
\gamma_\mu \, S_u^{c c'} \nonumber \\
&& + 4 S_u^{a a'}\, \gamma_\nu \, C S_d^{T b b'} C \, \gamma_\mu \,
\widehat S_u^{c c'}
   + 4 S_d^{a a'}\, \gamma_\nu \, C S_u^{T b b'} C \, \gamma_\mu \,
\widehat S_u^{c c'}
   + 4 S_u^{a a'} \, \gamma_\nu \, C S_u^{T b b'} C \,
\gamma_\mu \, \widehat S_d^{c c'} \nonumber \\
&& + 2 \widehat S_u^{a a'} \, tr \left [ \gamma_\nu \,
  C S_u^{T b b'} C \, \gamma_\mu \, S_d^{c c'} \right ]
   + 2 \widehat S_u^{a a'} \, tr \left [ \gamma_\nu \,
  C S_d^{T b b'} C \, \gamma_\mu \, S_u^{c c'} \right ]
   + 2 \widehat S_d^{a a'} \, tr \left [ \gamma_\nu \,
  C S_u^{T b b'} C \, \gamma_\mu \, S_u^{c c'} \right ]
\nonumber \\
&& + 2 S_u^{a a'} \, tr \left [ \gamma_\nu \,
  C \widehat S_u^{T b b'} C \, \gamma_\mu \, S_d^{c c'} \right ]
   + 2 S_d^{a a'} \, tr \left [ \gamma_\nu \,
  C \widehat S_u^{T b b'} C \, \gamma_\mu \, S_u^{c c'} \right ]
   + 2 S_u^{a a'} \, tr \left [ \gamma_\nu \,
  C \widehat S_d^{T b b'} C \, \gamma_\mu \, S_u^{c c'} \right ]
\nonumber \\
&& + 2 S_u^{a a'} \, tr \left [ \gamma_\nu \,
  C S_d^{T b b'} C \, \gamma_\mu \, \widehat S_u^{c c'} \right ]
   + 2 S_d^{a a'} \, tr \left [ \gamma_\nu \,
  C S_u^{T b b'} C \, \gamma_\mu \, \widehat S_u^{c c'} \right ]
   +  2 S_u^{a a'} \, tr \left [ \gamma_\nu \,
  C S_u^{T b b'} C \, \gamma_\mu \, \widehat S_d^{c c'} \right ]
\biggr \}
\nonumber \\
&& + \sum_{q = u,\, d,\, s} e_q\, \sum_i
tr \left [ S_q^{ii}(x_1,x_1) \, \gamma_\mu \right ]
{1 \over 3} \; \epsilon^{abc} \epsilon^{a'b'c'}  \biggl \{
\nonumber \\
&& \;\;\: 4 S_u^{a a'} \,
                       \gamma_\nu \, C S_u^{T b b'} C \, \gamma_\mu \,
S_d^{c c'}
  + 4 S_u^{a a'}\, \gamma_\nu \, C S_d^{T b b'} C \, \gamma_\mu \,
S_u^{c c'}
  + 4 S_d^{a a'}\, \gamma_\nu \, C S_u^{T b b'} C \, \gamma_\mu \,
vS_u^{c c'} \nonumber \\
&& + 2 S_u^{a a'} \, tr \left [ \gamma_\nu \,
  C S_u^{T b b'} C \, \gamma_\mu \, S_d^{c c'} \right ]
   + 2 S_u^{a a'} \, tr \left [ \gamma_\nu \,
  C S_d^{T b b'} C \, \gamma_\mu \, S_u^{c c'} \right ]
   +  2 S_d^{a a'} \, tr \left [ \gamma_\nu \,
  C S_u^{T b b'} C \, \gamma_\mu \, S_u^{c c'} \right ]
\biggr \}
\nonumber \\
\label{Delta3pt}
\end{eqnarray}
where
\begin{equation}
\widehat S_q^{aa'}(x_2,x_1,0) = e_q\, \sum_{i} S_q^{ai}(x_2,x_1) \,
\gamma_\alpha \, S_q^{i a'}(x_1,0) \, ,
\end{equation}
\end{widetext}
denotes the connected insertion of the probing current to a quark of
charge $e_q$.  
Here we have explicitly selected the electromagnetic current.
However, the present discussion may be generalized to any
quark-field-based current operator bilinear in the quark fields.

The latter term of Eq.~(\ref{Delta3pt}) accounts for the disconnected
quark loop contribution depicted in Fig.~\ref{topology}b.
The sum over the quarks running around the loop has been restricted to
the flavors relevant to the ground state baryon octet and decuplet.
In the $SU(3)_{\rm flavor}$ limit the sum vanishes for the
electromagnetic current.  
However, the heavier strange quark mass allows for a nontrivial
result.
Due to the technical difficulties of numerically estimating $M^{-1}$
for the squared lattice volume of diagonal spatial indices, these
contributions have been omitted from lattice QCD calculations of
electromagnetic structure in the spirit of Q$\chi$PT, and
we will also do so here.
For other observables such as the scalar density or forward matrix
elements of the axial vector current relevant to the spin of the
baryon, the ``charges'' running around the loop do not sum to zero.
In this case the second term of Eq.~(\ref{Delta3pt}) can be just as
significant as the connected term
\cite{Kuramashi:1993ka,Dong:1993pk}.

An examination of Eq.~(\ref{Delta3pt}) reveals complete symmetry among
the quark flavors in the correlation function.  
For example, wherever a $d$ quark appears in the correlator, a $u$
quark also appears in the same position in another term.  
An interesting consequence of this is that the connected insertion of
the electromagnetic current for $\Delta^0$ vanishes.  
All electromagnetic properties of the $\Delta^0$ have their origin
strictly in the disconnected loop contribution.  
Physically, what this means is that the valence wave function for each
of the quarks in the $\Delta$ resonances are identical under charge
symmetry.

\subsection{Two-point Green functions}
\label{subsec:2ptFn}

In this and the following subsection discussing correlation functions
at the hadronic level, the Dirac representation of the
$\gamma$-matrices is used to facilitate calculations of the
$\gamma$-matrix algebra.
It is then a simple task to account for the differences in
$\gamma$-matrix and metric definitions in reporting the final results
using Sakurai's notation.

The extraction of baryon mass and electromagnetic form factors
proceeds through the calculation of the ensemble average (denoted
$\bigm < \cdots \bigm >$) of two and three-point Green functions. 
The two-point function for decuplet baryons is defined as
\begin{eqnarray}
\lefteqn{\bigm < G^{BB}_{\sigma \tau} (t;\vec{p}; \Gamma) \bigm > = }
\qquad \nonumber \\
\!\!\!&&\!\!\! \sum_{\vec{x}} e^{-i
\vec{p} \cdot \vec{x} } \Gamma^{\beta \alpha} \bigm < \Omega \bigm |
T \left ( \chi^\alpha_{\sigma}(x)
\overline \chi^\beta_{\tau}(0) \right )
\bigm | \Omega \bigm > .\,\,
\end{eqnarray}
Here $\Omega$ represents the QCD vacuum, $\Gamma$ is a $4 \times 4$
matrix in Dirac space and $\alpha, \, \beta$ are Dirac indices.  
The subscripts $\sigma, \, \tau$ are the Lorentz indices of the
spin-3/2 interpolating fields.  
At the hadronic level one proceeds by inserting a complete set of
states\break $\bigm | B, p, s \bigm >$ and defining
\begin{equation}
\bigm < \Omega \bigm | \chi_{\sigma} (0) \bigm | B, p, s \bigm > \, =
   Z_B(p) \sqrt{M \over E_p} \, u_{\sigma}(p,s) ,
\label{interp}
\end{equation}
where $Z_B$ represents the coupling strength of $\chi_\sigma(0)$ to
the baryon $B$.  Our use of smeared interpolators makes this momentum
dependent. Momentum is denoted by $p$, spin by $s$, and
$u_\alpha(p,s)$ is a spin-vector in the Rarita-Schwinger formalism.
$E_p = \sqrt{\vec{p}^2 + M^2}$ and Dirac indices have been suppressed.
Using the Rarita-Schwinger spin sum,
\begin{widetext}
\begin{eqnarray}
\sum_s u_\sigma(p,s) \overline u_\tau(p,s) &=&
- { \gamma \cdot p + M \over 2 M } \left \{ g_{\sigma \tau} - { 1
\over 3} \gamma_\sigma \gamma_\tau - { 2 p_\sigma p_\tau \over 3 M^2 }
+ { p_\sigma \gamma_\tau - p_\tau \gamma_\sigma \over 3 M} \right \} ,
\label{rsss} \\
&\equiv& \Lambda_{\sigma \tau} , \nonumber
\end{eqnarray}
our usual definitions for $\Gamma$,
\begin{equation}
\Gamma_j = {1 \over 2} \left ( \begin{array}{cc} \sigma_j &\quad 0 \\
                                                 0        &\quad 0 \\
                               \end{array} \right ) \quad ;
\quad \Gamma_4 = {1 \over 2} \left ( \begin{array}{cc} I &\quad 0 \\
                                                       0 &\quad 0 \\
                                     \end{array} \right ) ,
\label{gammadef}
\end{equation}
and $\vec p = (p,0,0)$, the large Euclidean time limit of the two
point function takes the form
\begin{equation}
   \bigm < G^{BB}_{\sigma \tau} (t;\vec{p}, \Gamma_4) \bigm > \,
   = Z_B(p)\overline{Z}_B(p) {M \over E_p}
   e^{- E_p t} \, {\rm tr} \left [\; \Gamma_4 \; \Lambda_{\sigma \tau}
     \;
   \right ] ,
\end{equation}
where
\begin{mathletters}
\begin{eqnarray}
   \bigm < G^{BB}_{00} (t;\vec{p}, \Gamma_4) \bigm >
   &=& Z_B(p)\overline{Z_B}(p)  \,
   {2 \over 3} \, { |\vec p|^2 \over M_B^2 }
   \left ( {E_p + M_B \over 2 E_p} \right ) e^{- E_p t} ,
   \label{gbb00} \\
   \bigm < G^{BB}_{11} (t;\vec{p}, \Gamma_4) \bigm >
   &=&  Z_B(p)\overline{Z_B}(p) \,
   {2 \over 3} \, { E_p^2 \over M_B^2 }
   \left ( {E_p + M_B \over 2 E_p} \right ) e^{- E_p t} ,
   \label{gbb11} \\
   \bigm < G^{BB}_{22} (t;\vec{p}, \Gamma_4) \bigm >
   &=&  Z_B(p)\overline{Z_B}(p) \,
   {2 \over 3}
   \left ( {E_p + M_B \over 2 E_p} \right ) e^{- E_p t} ,
   \label{gbb22} \\
   \bigm < G^{BB}_{33} (t;\vec{p}, \Gamma_4) \bigm >
   &=&  Z_B(p)\overline{Z_B}(p) \,
   {2 \over 3}
   \left ( {E_p + M_B \over 2 E_p} \right ) e^{- E_p t} .
   \label{gbb33}
\end{eqnarray}
\end{mathletters}
\end{widetext}
Here $\overline{Z_B}(p)$ denotes the overlap associated with our
smeared source. $Z_B(p)$ is associated with the sink which need not
have the same smearing.

Equations (\ref{gbb00}) through (\ref{gbb33}) provide four correlation
functions from which a baryon mass may be extracted.
All baryon masses extracted from the different selections of Lorentz
indices agree within statistical uncertainties.  
The combination providing the smallest statistical fluctuations is
$\bigm < G^{BB}_{22} (t;\vec{p}, \Gamma_4) + G^{BB}_{33} (t;\vec{p},
\Gamma_4) \bigm >$ and these results are presented in
section~\ref{sec:Results}.

It should be noted that the spin-3/2 interpolating field also has
overlap with spin-1/2 baryons.
For the $\Delta$ baryons and $\Omega^-$ this poses no problem as these
baryons are the lowest lying baryons in the mass spectrum having the
appropriate isospin and strangeness quantum numbers.  
However, $\Sigma^*$ and $\Xi^*$ correlation functions may have lower
lying octet spin-1/2 components allowed by flavor-symmetry breaking,
$m_u = m_d \ne m_s$. Therefore it is desirable to use
the spin-3/2 projection operator \cite{Benmerrouche:1989uc}
\begin{equation}
P^{3/2}_{\mu \nu}(p) = g_{\mu \nu} -
                      {1 \over 3} \gamma_\mu \gamma_\nu -
{1 \over 3 p^2} \left ( \gamma \cdot p \, \gamma_\mu \, p_\nu + p_\mu
\, \gamma_\nu \, \gamma \cdot p \right ) .
\label{proj}
\end{equation}
However, our precision results for baryon two-point functions give no
indication of a low-lying spin-1/2 component being excited by the
spin-3/2 interpolating fields, and conclude such excitations are
negligible.

\subsection{Three-point functions and multipole form factors}
\label{subsec:3ptFn}

Here we begin with a brief overview of the results of Ref.\ 
\cite{Nozawa:1990gt}, where the multipole form factors are defined in
terms of the covariant vertex functions and in terms of the current
matrix elements.
The electromagnetic current matrix element for spin-3/2 particles may
be written as
\begin{equation}
\bigm < p',s' \bigm | j^{\mu}(0) \bigm | p,s \bigm > =
\sqrt{ M_B^2 \over E_p E_{p'} } \,
\overline{u}_{\alpha}(p',s') {\cal O}^{\alpha \mu \beta}
u_{\beta}(p,s) .
\label{cvf}
\end{equation}
Here $p$ and $p'$ ($s$ and $s'$) denote the momentum (spin) of the
initial and final states, respectively, and $u_\alpha(p,s)$ is a
Rarita-Schwinger spin-vector.  
The following Lorentz covariant form for the tensor
\begin{eqnarray}
 {\cal O}^{\alpha \mu \beta} =
&-& g^{\alpha \beta} \left \{ a_{1} \gamma^{\mu}
+ {a_{2} \over 2 M_B} P^{\mu} \right \} \nonumber \\
&-& \frac{ q^{\alpha} q^{\beta} }{(2M_{B})^{2}}
 \left\{ c_{1} \gamma^{\mu} + \frac{c_{2}}{2M_{B}} P^{\mu} \right\} ,
\label{tensor}
\end{eqnarray}
where $P = p'+p$, $q=p'-p$ and $M_{B}$ is the mass of the baryon,
satisfies the standard requirements of invariance under time reversal,
parity, G-parity and gauge transformations.
The parameters $a_{1}$, $a_{2}$, $c_{1}$ and $c_{2}$ are independent
covariant vertex functions.

\begin{widetext}

The multipole form factors are defined in terms of the covariant
vertex functions
through the following Lorentz invariant expressions
\cite{Nozawa:1990gt},
\begin{mathletters}
\begin{eqnarray}
{\cal G}_{E0}(q^{2}) & = & ( 1 + \frac{2}{3} \tau ) \left\{
                       a_{1} + (1+\tau) a_{2} \right\}
 - \frac{1}{3} \tau (1+\tau) \left\{ c_{1} +
                    (1+\tau)c_{2} \right\} \, , \\
{\cal G}_{E2}(q^{2}) & = & \left\{ a_{1} + (1+\tau) a_{2} \right\}
 - \frac{1}{2} (1+\tau) \left\{ c_{1} +
               (1+\tau)c_{2}  \right\} \, , \\
{\cal G}_{M1}(q^{2}) & = & ( 1 + \frac{4}{5} \tau ) a_{1}
                       - \frac{2}{5} \tau (1+\tau) c_{1} \, , \\
{\cal G}_{M3}(q^{2}) & = &  a_{1} - \frac{1}{2} (1+\tau) c_{1} \, ,
\end{eqnarray}
\end{mathletters}
with $\tau = -q^2/(2 M_B)^2\ (\ge 0)$.
The multipole form factors ${\cal G}_{E0}$, ${\cal G}_{E2}$, ${\cal
  G}_{M1}$ and ${\cal G}_{M3}$ are referred to as charge ($E0$),
electric-quadrupole ($E2$), magnetic-dipole ($M1$) and
magnetic-octupole ($M3$) multipole form factors, respectively.

In a manner similar to that for the two-point function, the
three-point Green function for the electromagnetic current is defined
as
\begin{equation}
\bigm < G^{B j^\mu B}_{\sigma \tau}
               (t_2, t_1; \vec{p}\,', \vec{p}; \Gamma) \bigm > \,
   =
\sum_{\vec{x_2}, \vec{x_1}}
   e^{-i \vec{p'} \cdot \vec{x_2} } e^{+ i \left (
   \vec{p'} - \vec{p} \right ) \cdot \vec{x_1} }
   \Gamma^{\beta \alpha} \bigm < \Omega \bigm | T
   \left ( \chi^\alpha_\sigma(x_2) j^\mu(x_1)
   \overline \chi^\beta_\tau(0) \right )
   \bigm | \Omega \bigm > .
\end{equation}
Once again, the subscripts $\sigma, \, \tau$ are the Lorentz indices
of the spin-3/2 interpolating fields.
For large Euclidean time separations $t_2 - t_1 >\!> 1$ and $t_1 >\!>
1$ the three-point function at the hadronic level takes the limit
\begin{equation}
\bigm < G^{B j^\mu B}_{\sigma \tau} (t_2, t_1;\vec{p}\,', \vec{p};
                                     \Gamma) \bigm > =
\sum_{s, s^\prime} e^{-E_{p'} (t_2-t_1)} e^{-E_p t_1}
 \Gamma^{\beta \alpha} \bigm < \Omega \bigm |
    \chi^\alpha_\sigma \bigm | p', s^\prime \bigm >
    \bigm < p', s^\prime \bigm |
    j^\mu \bigm | p, s \bigm > \bigm < p, s \bigm |
    \overline \chi^\beta_\tau  \bigm | \Omega \bigm > ,
\end{equation}
where the matrix element of the electromagnetic current is defined in
(\ref{cvf}), and the matrix elements of the interpolating fields are
defined by Eq.~(\ref{interp}).

The time dependence of
the three-point function may be eliminated through the use of the
two-point functions.
Maintaining the lattice Ward identity, which guarantees the lattice
electric form factor reproduces the total charge of the baryon at
$q^2=0$, provides an indispensable guide to the optimum ratio of Green
functions.
The preferred ratio of two- and three-point Green functions
is~\cite{Leinweber:1992hy}
\begin{eqnarray}
   R_\sigma{}^\mu{}_\tau (t_2,t_1; \vec{p}\,', \vec{p};
                          \Gamma ) &=&
    \left ( {
\bigm < G^{B j^\mu B}_{\sigma \tau}
      (t_2,t_1; \vec{p}\,', \vec{p}; \Gamma) \bigm >
\bigm < G^{B j^\mu B}_{\sigma \tau}
      (t_2,t_1; -\vec{p}, -\vec{p}\,'; \Gamma) \bigm >
\over
\bigm < G^{BB}_{\sigma \tau} (t_2; \vec{p}\,'; \Gamma_4 ) \bigm >
\bigm < G^{BB}_{\sigma \tau} (t_2; -\vec{p}; \Gamma_4 ) \bigm > }
\right )^{1/2} , \label{ratio} \\
&\simeq&
   \left ( { E_p + M \over 2 E_p } \right )^{1/2}
   \left ( { E_{p'} + M \over 2 E_{p'} } \right )^{1/2}
   \overline R_\sigma{}^\mu{}_\tau (\vec{p}\,', \vec{p}; \Gamma )
   \, , \label{redratio}
\end{eqnarray}
where we have defined the reduced ratio $\overline
R_\sigma{}^\mu{}_\tau (\vec{p}\,', \vec{p}; \Gamma )$.  
Note that there is no implied sum over $\sigma$ and $\tau$ in
Eq.~(\ref{ratio}). Also, the square root in Eq.~(\ref{ratio}) spoils
the covariant/contraviant nature of $R_\sigma{}^\mu{}_\tau$ and no meaning
should be attached to the location of the indices. We still prefer
this notation due to the close connection with $G^{B j^\mu B}_{\sigma
  \tau}$.

Using our standard definitions for $\Gamma$ given in
Eq.~(\ref{gammadef}) and the Rarita-Schwinger spin sum of
Eq.~(\ref{rsss}), the multipole form factors may be isolated and
extracted from the following combinations of $\overline
R_\sigma{}^\mu{}_\tau (\vec{p}\,', \vec{p}; \Gamma )$
\begin{mathletters}
\begin{eqnarray}
{\cal G}_{E0}(q^2) &=& {1 \over 3} \left (
                      \overline R_1{}^4{}_1(\vec q_1,0; \Gamma_4)
                 +    \overline R_2{}^4{}_2(\vec q_1,0; \Gamma_4)
                 +    \overline R_3{}^4{}_3(\vec q_1,0; \Gamma_4)
                      \right ) , \label{ge0} \\
{\cal G}_{E2}(q^2) &=& 2{M (E+M) \over |\vec q_1|^2} \left (
                      \overline R_1{}^4{}_1(\vec q_1,0; \Gamma_4)
                 +    \overline R_2{}^4{}_2(\vec q_1,0; \Gamma_4)
               - 2 \, \overline R_3{}^4{}_3(\vec q_1,0; \Gamma_4)
                      \right ) , \label{ge2} \\
{\cal G}_{M1}(q^2) &=& - \, {3 \over 5}{E+M \over |\vec q_1|}
                       \left (
                      \overline R_1{}^3{}_1(\vec q_1,0; \Gamma_2)
                 +    \overline R_2{}^3{}_2(\vec q_1,0; \Gamma_2)
                 +    \overline R_3{}^3{}_3(\vec q_1,0; \Gamma_2)
                      \right ) , \label{gm1} \\
{\cal G}_{M3}(q^2) &=& - \, 4 {M (E+M)^2 \over |\vec q_1|^3}
                      \left (
                      \overline R_1{}^3{}_1(\vec q_1,0; \Gamma_2)
                 +    \overline R_2{}^3{}_2(\vec q_1,0; \Gamma_2)
   -    {3 \over 2}\, \overline R_3{}^3{}_3(\vec q_1,0; \Gamma_2)
                      \right ) ,  \label{gm3}
\end{eqnarray}
\label{formfactors}
\end{mathletters}
where $\vec q_1 = (q,0,0)$. 
We note that smaller statistical uncertainties may be obtained for
${\cal G}_{E2}$ by using the symmetry
\begin{equation}
\overline R_2{}^4{}_2(\vec q_1,0; \Gamma_4) = \overline
R_3{}^4{}_3(\vec q_1,0; \Gamma_4)\, ,
\label{sym_R}
\end{equation}
in Eq.~(\ref{ge2}).
Hence, we define an average $\overline{R}^4_{\rm avg}$ as
\begin{equation}
 \overline{R}^4_{\rm avg}(\vec q_1,0;
 \Gamma_4)=\frac{1}{2}\left[\overline R_2{}^4{}_2(\vec q_1,0;
 \Gamma_4)+\overline R_3{}^4{}_3(\vec q_1,0; \Gamma_4)\right]\, .
\end{equation}
With this definition the expression for ${\cal G}_{E2}(q^2)$ used in
our simulations is
\begin{equation}
{\cal G}_{E2}(q^2) = 2{M (E+M) \over |\vec q_1|^2} \left (
                      \overline R_1{}^4{}_1(\vec q_1,0; \Gamma_4)
                 -   \overline R_{\rm avg}^4(\vec q_1,0;
                 \Gamma_4)\right )\,. \label{ge2_mod} 
\end{equation}
\end{widetext}
%


\section{Lattice Techniques}
\label{sec:LatTech}

The three-point functions discussed in section~\ref{sec:theory} are
constructed using the sequential source technique outlined in
Refs.~\cite{Leinweber:1990dv,Leinweber:1992hy,Wilcox:1991cq}.
Our quenched gauge fields are generated with the ${\mathcal O}(a^2)$
mean-field improved L\"uscher-Weisz plaquette plus rectangle gauge
action \cite{Luscher:1984xn} using the plaquette measure for the mean
link.
The simulations are performed on a $20^3 \times 40$ lattice with a
lattice spacing of 0.128 fm as determined by the Sommer scale
\cite{Sommer:1993ce} $r_0=0.50$ fm.
This large volume lattice ensures a good density of low-lying momenta
which are key to giving rise to chiral non-analytic behavior in the
observables simulated on the lattice
\cite{Leinweber:2004tc,Leinweber:2006ug}.

We perform a high-statistics analysis using a large sample of 400
configurations for our lightest eight quark masses.  We also consider
a subset of 200 configurations for our three heaviest quark masses to
explore the approach to the heavy-quark regime.  
A small sub-ensemble bias correction is applied multiplicatively to
the heavy quark results, by matching the central values of the 200
configuration sub-ensemble and 400 configuration ensemble averages at
$\kappa = 0.12780$.
All tables display the raw, unbiased data for the first four kappa
values. The first row of the $\kappa = 0.12780$ results gives the
results from the 200 configuration sub-ensembles, the second gives the
400 configuration ensemble results. The scaled results from the 200
configuration sub-ensembles are shown in the figures.

We use the fat-link irrelevant clover (FLIC) Dirac operator
\cite{Zanotti:2001yb} which provides
${\mathcal O}(a)$ improvement \cite{Zanotti:2004dr}.  
The improved chiral properties of FLIC fermions allow efficient access
to the light quark-mass regime \cite{Boinepalli:2004fz}, making them
ideal for dynamical fermion simulations now underway
\cite{Kamleh:2004xk}.
For the vector current, we an ${\cal O}(a)$-improved FLIC
conserved vector current \cite{Boinepalli:2006xd}. 
We use a smeared source at $t_{0} = 8$, and a current insertion
centered at $t_1 = 14$. Complete details are described
in Ref.~\cite{Boinepalli:2006xd}.

Table~\ref{tab:baryonmasses} provides the kappa values used in our
simulations, together with the calculated $\pi$ and decuplet baryon
masses.  
While we refer to $m_\pi^2$ to infer the quark masses, we note that
the critical value where the pion mass vanishes is $\kappa_{\rm cr} =
0.13135$.

We select $\kappa = 0.12885$ to represent the strange quark in this
simulation.  At this $\kappa$ the $s\bar s$ pseudo scalar mass is
0.697~GeV, which compares well with the experimental value of $2\, m_{\rm
K}^2 - m_\pi^2 = ( 0.693\ {\rm GeV} )^2$, motivated by leading order
chiral perturbation theory.

The error analysis of the correlation function ratios is performed via
a third-order, single-elimination jackknife, with the $\chi^2$ per
degree of freedom $(\chi^{2}_{\rm dof})$ obtained via covariance
matrix fits.
We perform a series of fits through the ratios after the current
insertion at $t_1 = 14$.
By examining the $\chi^2_{\rm dof}$ we are able to establish a valid
window through which we may fit in order to extract our observables.
In all cases, we required a value of $\chi^2_{\rm dof}$ no larger than
1.5.
The values of the static quantities quoted in this paper on a per
quark-sector basis correspond to values for single quarks of unit
charge.

\begin{table*}[tbp]
\caption{Hadron masses in units of GeV for various values
  of the hopping parameter, $\kappa$. Pion masses are in $\rm GeV^2$
  while the baryon masses are in $\rm GeV$.}
\label{tab:baryonmasses}
\begin{ruledtabular}
\begin{tabular}{ccccc}
\noalign{\smallskip}
$\kappa$ & $\mathit{m_\pi}^2$ & $\Delta $ & $\Sigma^*$  & $\Xi^*$  \\
\hline
\noalign{\smallskip}
$0.12630$ & $0.9960(56)$ & $1.999(9) $  & $1.908(10)$ & $1.815(11)$  \\
$0.12680$ & $0.8936(56)$ & $1.945(10)$  & $1.871(11)$ & $1.797(12)$ \\
$0.12730$ & $0.7920(55)$ & $1.890(10)$  & $1.834(11)$ & $1.779(13)$ \\
$0.12780$ & $0.6920(54)$ & $1.836(11)$  & $1.798(12)$ & $1.761(13)$ \\
$0.12780$ & $0.6910(35)$ & $1.845(10)$  & $1.807(11)$ & $1.770(11)$ \\
$0.12830$ & $0.5925(33)$ & $1.791(11)$  & $1.771(11)$ & $1.752(12)$ \\
$0.12885$ & $0.4854(31)$ & $1.732(12)$  & $1.732(12)$ & $1.732(12)$ \\
$0.12940$ & $0.3795(31)$ & $1.673(14)$  & $1.693(13)$ & $1.712(13)$ \\
$0.12990$ & $0.2839(33)$ & $1.622(16)$  & $1.659(15)$ & $1.695(13)$ \\
$0.13205$ & $0.2153(35)$ & $1.592(17)$  & $1.638(15)$ & $1.685(13)$ \\
$0.13060$ & $0.1384(43)$ & $1.565(18)$  & $1.620(16)$ & $1.676(14)$ \\
$0.13080$ & $0.0939(44)$ & $1.549(19)$  & $1.609(16)$ & $1.670(14)$ \\
experiment & $0.0196$ & $1.232$ & $1.382$ & $1.531$ \\
\end{tabular}
\end{ruledtabular}
\end{table*}

When extracting form factors from the lattice correlation functions
via the ratios defined in Eqs.~(\ref{ge0}) to (\ref{gm3}) in
Sec.~\ref{subsec:3ptFn}, we employ the advanced analysis techniques
outlined in detail in Ref.~\cite{Boinepalli:2006xd}.

The following calculations are performed in the lab frame $\vec
p=0,\,\,\vec{p}\,'= \vec q = | \vec q | \, \widehat x$ at $|\vec q|a =
2 \pi / L_x$ with $L_x = 20$, the minimum nonzero momentum available
on our lattice.
While $q^2$ is dependent on the mass of the baryon, we find this mass
dependence to be small.
Indeed all form factors may be regarded as being calculated at $Q^2 =
-q^2 = 0.230 \pm 0.001\ {\rm GeV}^2$ where the error is dominated by
the mass dependence of the target baryon.
Where a spatial direction of the electromagnetic current is required,
it is chosen to be the $z$-direction.

%
\section{Correlation Function Analysis}
\label{sec:corrFun}

\subsection{Baryon Masses}
\begin{figure}[tbp!]
\begin{center}
  {\includegraphics[height=\hsize,angle=90]{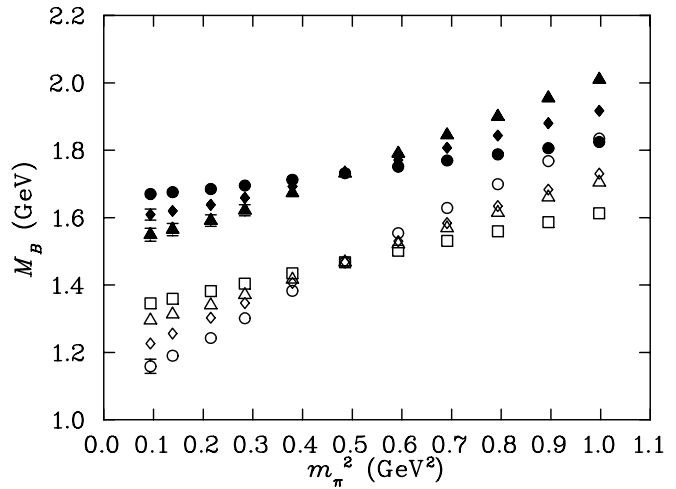}}
\end{center}
\caption{Masses of the decuplet and the octet
  baryons~\cite{Boinepalli:2006xd} at different quark masses.  The
  octet baryon masses are the open symbols while the symbols of the
  decuplet baryons are filled.  At the lowest quark mass the lowest
  point is the Nucleon, followed by $\Lambda$, $\Sigma$ and the $\Xi$.
  The decuplet baryon with lowest mass is the $\Delta$ followed by
  $\Sigma^*$ and $\Xi^*$.}
\label{fig:baryonmasses}
\end{figure}

Figure \ref{fig:baryonmasses} is a plot of the decuplet baryon masses
along with the masses of the octet baryon from our previous
calculation~\cite{Boinepalli:2006xd}.
We observe the $SU(3)_{\rm flavor}$ limit at our sixth quark mass when
$m_\pi^2 = 0.485(3)~\rm GeV^2$.  The mass of the $\Omega^-$ is the
mass of the $\Delta$ at the $SU(3)_{\rm flavor}$ limit, i.e., $1.732
\pm 0.012~\mathrm{GeV}$ which differs from the experimentally
measured value of $1.67~\mathrm{GeV}$ 
by only about $3.6\%$.  The higher value from the quenched simulation
is in accord with the expectations of quenched
$\chi$EFT~\cite{Young:2002cj,Labrenz:1996jy}.
The mass of the $\Delta$ baryon shows an upward chiral curvature as
the $m_\pi^2$ becomes smaller. 
This behavior has already been discussed in
Refs.~\cite{Boinepalli:2004fz,Young:2002cj,Leinweber:2002bw}.

\subsection{Form Factor Correlators}
\label{ffCorr}

The baryon form factors are calculated on a quark-sector by
quark-sector basis with each sector normalized to the contribution of
a single quark with unit charge.
Hence to calculate the corresponding baryon property, each quark
sector contribution should be multiplied by the appropriate charge and
quark number.
Under such a scheme for a generic form factor $f$, the $\Delta^+$ form
factor, $f_{\Delta^+}$, is obtained from the $u$- and $d$-quark
sectors normalized for a single quark of unit charge via
\begin{equation}
\mathit{f_{\Delta^+}} = 2 \times \frac{2}{3} \times \mathit{f_u} + 1
\times\left(-\frac{1}{3}\right) \times \mathit{f_d}\,.
\label{eq:normsinglequark}
\end{equation}

\begin{figure}[tbp!]
\begin{center}
  {\includegraphics[height=\hsize,angle=90]{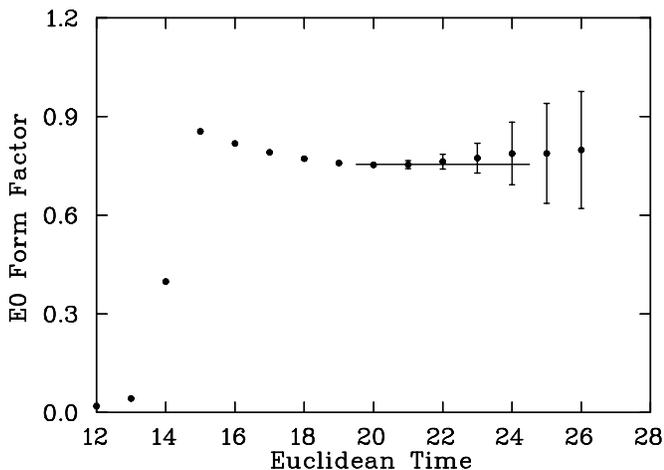}}
\end{center}
\caption{$E0$ electric form factor of the $u$ quark in the $\Delta
  $ at $Q^2=0.230(1)~\mathrm{GeV}^2$ as a function of Euclidean
  time~$(t_2)$ for $\mathit{m}_{\pi}^{2}=0.4854~\mathrm{GeV}^2$, the 
  $SU(3)_{\rm flavor}$ limit. The line indicates the fitting window and the
  best fit value.}
\label{cfe0K3}
\end{figure}

Figure~\ref{cfe0K3} depicts the electric form factor $E0$ of the $u$
quark in the $\Delta$ as a function of time at the $SU(3)_{\rm
  flavor}$ limit.
The $u$ and $d$ quarks in the $\Delta$ states are identical as
discussed in regard to Eq.~(\ref{Delta3pt}).
The straight lines indicate the fits which were selected using
the $\chi^2_{\rm dof}$ considerations outlined in
Ref.~\cite{Boinepalli:2006xd}.
\begin{figure}[tbp!]
\begin{center}
  {\includegraphics[height=\hsize,angle=90]{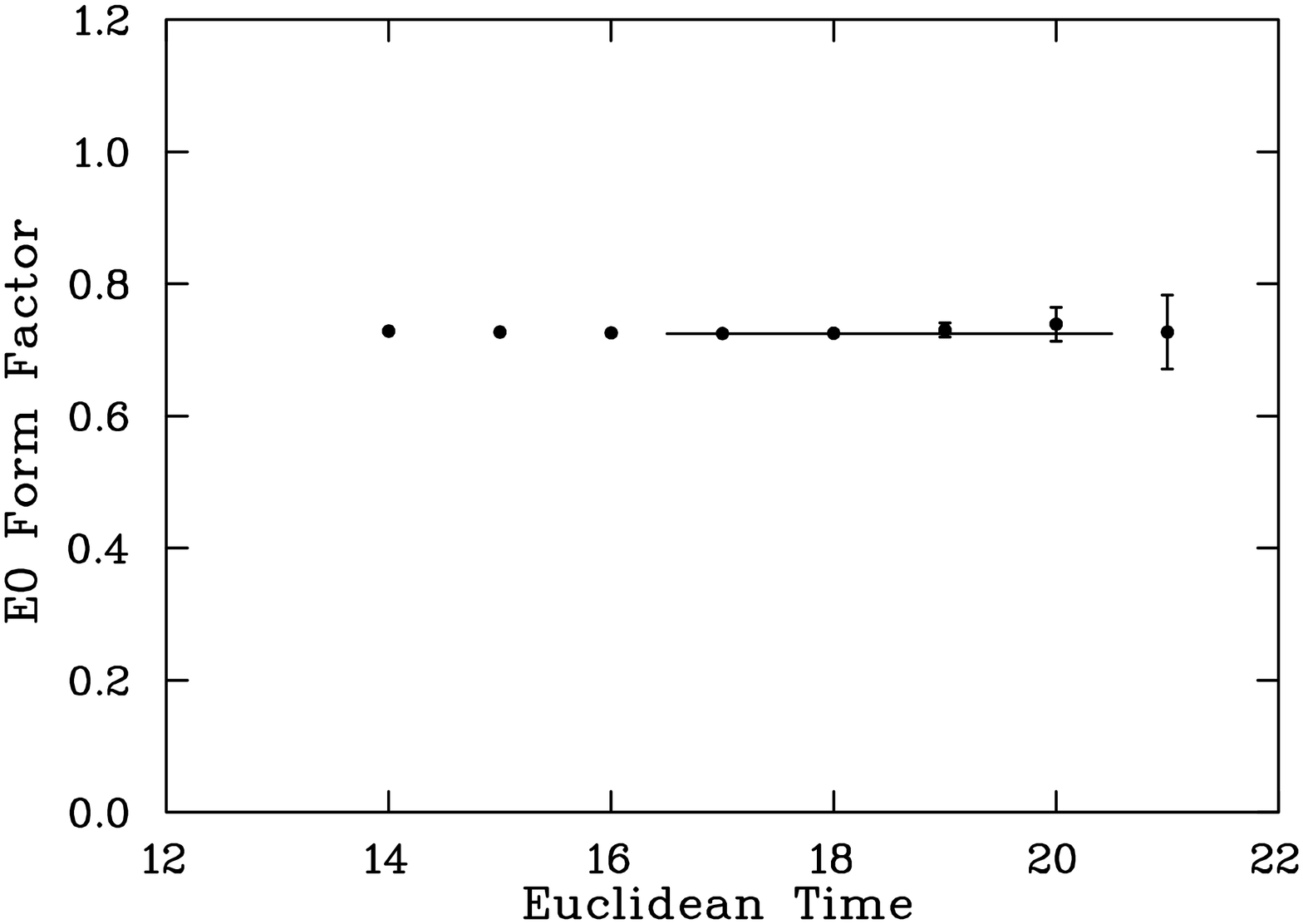}}
\end{center}
\caption{$E0$ electric form factor of the  $u$ quark in the $\Delta$
  at $Q^2=0.230(1)$ $\mathrm{GeV}^2$ as a function of Euclidean
  time $(t_2)$ at the ninth quark mass where $m_\pi^2 = 0.215(4)\ 
  {\rm GeV}^2$.  The correlator is obtained from the splitting between
  the ninth and eighth quark mass results.  The line indicates the
  fitting window and the best fit value.}
\label{cfe0K6DSp}
\end{figure}
For light quark masses smaller than the strange quark mass, we fit the
change in the form factor correlation functions from one quark
mass to the next and add this to the previous result at the heavier
quark mass. This provides significant cancellation of correlated
systematic errors and makes the selection of the fit regime
transparent.

Figure~\ref{cfe0K6DSp} shows the fitting of the electric form factor
splitting for the $\Delta^+$ between the eighth and ninth quark
masses, the latter having $m_\pi^2 = 0.215(4)\ \mathrm{GeV}^2$.
The improvement of the plateau is apparent in Fig.~\ref{cfe0K6DSp}.
Still substantial Euclidean time evolution is required to obtain an
acceptable $\chi^2_{\rm dof}$. 

Tables~\ref{tab:E0D} to \ref{tab:E0XS} list the electric form factors
for all the decuplet baryons at the quark level for the eleven quark
masses considered.
In the tables, the selected time frame, the fit value and the
associated $\chi^2_{\rm dof}$ are indicated.
Table~\ref{tab:barE0} provides collected results for the various decuplet
baryons.

\begin{table*}[tbp]
\caption{Quark sector contributions to the electric form factor $E0$
  of the $\Delta$ at $Q^2 = 0.230(1)\ {\rm GeV}^2$. Sector contributions
  are for a single quark of unit charge. The fit windows are
  selected using the criteria outlined in Ref.~\cite{Boinepalli:2006xd}. The quark
  contribution at the SU(3) limit when $m_\pi^2=0.485(3)\,{\rm GeV^2}$
  provides the $s$ quark contribution in $\Omega^-$.}
\label{tab:E0D}
\begin{ruledtabular}
\begin{tabular}{ccccccc}
\noalign{\smallskip}
$\mathit m_\pi^2$  &\multicolumn{3}{c}{$\mathit u_{\Delta}$}  &
\multicolumn{3}{c}{ $\mathit d_{\Delta}$ }\\
\noalign{\smallskip}
$({\rm GeV}^2)$ & fit value & fit window & $\chi^2_{\rm dof}$ &fit
value & fit window & $\chi^2_{\rm dof}$   \\
\hline
\noalign{\smallskip}
$0.9960(56)$ & $0.800(5)$ & $20-24$  & $1.47$  &  $0.800(5)$   & $20-24$  & $1.47$ \\
$0.8936(56)$ & $0.794(6)$ & $20-24$  & $0.95$  &  $0.794(6)$   & $20-24$  & $0.95$ \\
$0.7920(55)$ & $0.788(7)$ & $20-24$  & $0.79$  &  $0.788(7)$   & $20-24$  & $0.79$ \\
$0.6920(54)$ & $0.783(8)$ & $20-24$  & $0.59$  &  $0.783(8)$   & $20-24$  & $0.59$ \\
$0.6910(35)$ & $0.773(6)$ & $20-24$  & $1.46$  &  $0.773(6)$   & $20-24$  & $1.46$ \\
$0.5925(33)$ & $0.764(7)$ & $20-24$  & $1.07$  &  $0.764(7)$   & $20-24$  & $1.07$ \\
$0.4854(31)$ & $0.755(10)$ & $20-24$ & $0.53$  &  $0.755(10)$  & $20-24$  & $0.53$ \\
$0.3795(31)$ & $0.744(11)$ & $17-20$ & $0.73$  &  $0.744(11)$  & $17-20$  & $0.73$ \\
$0.2839(33)$ & $0.733(13)$ & $17-20$ & $0.79$  &  $0.733(13)$  & $17-20$  & $0.79$ \\
$0.2153(35)$ & $0.725(16)$ & $17-19$ & $0.46$  &  $0.725(16)$  & $17-19$  & $0.46$ \\
$0.1384(43)$ & $0.717(22)$ & $17-19$ & $0.23$  &  $0.717(22)$  & $17-19$  & $0.23$ \\
$0.0939(44)$ & $0.693(33)$ & $17-19$ & $0.34$  &  $0.693(33)$  & $17-19$  & $0.34$ \\
\end{tabular}
\end{ruledtabular}
\vspace{-3pt}
\end{table*}

\begin{table*}[tbp]
\caption{Quark sector contributions to the electric form factor $E0$ of
  $\Sigma^*$ at $Q^2 = 0.230(1)\ {\rm GeV}^2$.  Sector contributions
  are for a single quark having unit charge. The fit windows are
  selected using the criteria outlined in Ref.~\cite{Boinepalli:2006xd}.}
\label{tab:E0SS}
\begin{ruledtabular}
\begin{tabular}{ccccccc}
\noalign{\smallskip}
$\mathit m_\pi^2$  &\multicolumn{3}{c}{$\mathit u_{\Sigma^*}$ or $\mathit d_{\Sigma^*}$}   & \multicolumn{3}{c}{ $\mathit s_{\Sigma^*}$ }\\
\noalign{\smallskip}
$({\rm GeV}^2)$ & fit value & fit window & $\chi^2_{\rm dof}$ &fit value & fit window & $\chi^2_{\rm dof}$   \\
\hline
\noalign{\smallskip}
$0.9960(56)$ & $0.804(6)$ & $20-24$  & $0.85$  &  $0.759(9)$   & $20-24$  & $0.54$ \\
$0.8936(56)$ & $0.798(7)$ & $20-24$  & $0.65$  &  $0.761(9)$   & $20-24$  & $0.50$ \\
$0.7920(55)$ & $0.792(8)$ & $20-24$  & $0.63$  &  $0.763(10)$   & $20-24$  & $0.49$ \\
$0.6920(54)$ & $0.786(9)$ & $20-24$  & $0.56$  &  $0.766(10)$   & $20-24$  & $0.48$ \\
$0.6910(35)$ & $0.774(6)$ & $20-24$  & $1.15$  &  $0.752(8)$   & $20-24$  & $0.97$ \\
 $0.5925(33)$ & $0.764(8)$ & $20-24$  & $0.91$  &  $0.753(8)$    & $20-24$  &  $0.79$ \\
$0.4854(31)$ & $0.755(10)$ & $20-24$  & $0.53$  &  $0.755(10)$  & $20-24$  & $0.53$  \\
$0.3795(31)$ & $0.744(11)$ & $17-20$ & $0.89$  &  $0.754(10)$  & $17-20$  & $0.22$ \\
$0.2839(33)$ & $0.733(12)$ & $17-20$ & $0.81$  &  $0.754(11)$  & $17-20$  & $0.37$ \\
$0.2153(35)$ & $0.727(14)$ & $17-19$ & $0.15$  &  $0.753(11)$  & $17-19$  & $0.21$ \\
$0.1384(43)$ & $0.719(18)$ & $17-19$ & $0.07$  &  $0.753(12)$  & $17-19$  & $0.05$ \\
$0.0939(44)$ & $0.710(23)$ & $17-19$ & $0.22$  &  $0.746(14)$  & $17-19$  & $0.63$ \\
\end{tabular}
\end{ruledtabular}
\vspace{-3pt}
\end{table*}

\begin{table*}[tbp]
\caption{Quark sector contributions to the electric form factor $E0$ of
  $\Xi^*$ at $Q^2 = 0.230(1)\ {\rm GeV}^2$.  Sector contributions
  are for a single quark having unit charge. The fit windows are
  selected using the criteria outlined in Ref.~\cite{Boinepalli:2006xd}.}
\label{tab:E0XS}
\begin{ruledtabular}
\begin{tabular}{ccccccc}
\noalign{\smallskip}
$\mathit m_\pi^2$  &\multicolumn{3}{c}{$\mathit s_{\Xi^*}$}  & \multicolumn{3}{c}{ $\mathit u_{\Xi^*}$ }\\
\noalign{\smallskip}
 $({\rm GeV}^2)$ & fit value & fit window & $\chi^2_{\rm dof}$ &fit value & fit window & $\chi^2_{\rm dof}$   \\
\hline
\noalign{\smallskip}
$0.9960(56)$ & $0.765(10)$ & $20-24$  & $0.47$  &  $0.809(8)$   & $20-24$  & $0.66$ \\
$0.8936(56)$ & $0.766(11)$ & $20-24$  & $0.47$  &  $0.802(8)$   & $20-24$  & $0.58$ \\
$0.7920(55)$ & $0.767(11)$ & $20-24$  & $0.49$  &  $0.795(9)$   & $20-24$  & $0.61$ \\
$0.6920(54)$ & $0.769(12)$ & $20-24$  & $0.50$  &  $0.788(10)$   & $20-24$  & $0.58$ \\
$0.6910(35)$ & $0.753(9)$ & $20-24$  & $0.73$  &  $0.775(7)$   & $20-24$  & $0.89$ \\
$0.5925(33)$ & $0.754(9)$ & $20-24$  & $0.66$  &  $0.765(8)$    & $20-24$  &  $0.75$ \\
$0.4854(31)$ &  $0.755(10)$ & $20-24$  & $0.53$  &  $0.755(10)$  & $20-24$  & $0.53$  \\
$0.3795(31)$ &  $0.754(10)$ & $17-20$ & $0.19$  &  $0.744(10)$  & $17-20$  & $1.19$ \\
$0.2839(33)$ & $0.754(10)$ & $17-20$ & $0.22$  &  $0.734(11)$  & $17-20$  & $0.77$ \\
$0.2153(35)$ & $0.754(10)$ & $17-19$ & $0.03$  &  $0.727(13)$  & $17-19$  & $0.09$ \\
$0.1384(43)$ & $0.755(11)$ & $17-19$ & $0.11$  &  $0.720(14)$  & $17-19$  & $0.10$ \\
$0.0939(44)$ &  $0.754(11)$ & $17-19$ & $0.66$  &  $0.714(17)$  & $17-19$  & $0.19$ \\
\end{tabular}
\end{ruledtabular}
\vspace{-3pt}
\end{table*}

\begin{table*}[tbp]
\caption{$E0$ form factor of the charged decuplet
  baryons for different $m_\pi^2$
  values. The $E0$ form factor of the $\Delta^-$ at the
  $SU(3)_{\rm flavor}$ limit ($m_\pi^2 = 0.485(3)\ {\rm GeV}^2$)
  provides the $E0$ form factor of 
  $\Omega^-$.}
\label{tab:barE0}
\begin{ruledtabular}
\begin{tabular}{ccccccc}
\noalign{\smallskip}
$\mathit m_\pi^2\ ({\rm GeV}^2)$  & $\Delta^{++}$ & 
$\Delta^+$ & $\Delta^-$ &
$\Sigma^{*+}$ & $\Sigma^{*-}$ & $\Xi^{*-}$ \\
\noalign{\smallskip}
\hline
\noalign{\smallskip}
$0.9972(55)$ & $1.601(10)$ & $0.803(5)$  & $-0.803(5)$  & $0.819(6)$  & $-0.789(7)$  & $-0.780(9)$  \\
$0.8936(56)$ & $1.589(11)$ & $0.794(6)$  & $-0.794(6)$  & $0.810(6)$  & $-0.786(7)$  & $-0.778(10)$ \\
$0.7920(55)$ & $1.577(13)$ & $0.788(7)$  & $-0.788(7)$  & $0.801(7)$  & $-0.782(8)$  & $-0.777(10)$ \\
$0.6920(54)$ & $1.566(16)$ & $0.783(8)$  & $-0.783(8)$  & $0.792(8)$  & $-0.779(9)$  & $-0.775(11)$ \\
$0.6910(35)$ & $1.545(11)$ & $0.773(6)$  & $-0.773(6)$  & $0.781(6)$  & $-0.766(7)$  & $-0.760(8)$  \\
$0.5925(33)$ & $1.527(14)$ & $0.764(7)$  & $-0.764(7)$  & $0.768(7)$  & $-0.761(8)$  & $-0.757(9)$  \\
$0.4854(31)$ & $1.509(19)$ & $0.755(10)$ & $-0.755(10)$ & $0.755(10)$ & $-0.755(10)$ & $-0.755(10)$ \\
$0.3795(31)$ & $1.487(22)$ & $0.744(11)$ & $-0.744(11)$ & $0.740(11)$ & $-0.747(10)$ & $-0.751(10)$ \\
$0.2839(33)$ & $1.465(26)$ & $0.733(13)$ & $-0.733(13)$ & $0.726(13)$ & $-0.741(12)$ & $-0.747(11)$ \\
$0.2153(35)$ & $1.451(31)$ & $0.725(16)$ & $-0.725(16)$ & $0.718(15)$ & $-0.736(13)$ & $-0.745(11)$ \\
$0.1384(43)$ & $1.433(44)$ & $0.717(22)$ & $-0.717(22)$ & $0.708(20)$ & $-0.730(15)$ & $-0.743(12)$ \\
$0.0939(44)$ & $1.386(65)$ & $0.693(33)$ & $-0.693(33)$ & $0.698(27)$ & $-0.722(19)$ & $-0.741(13)$
\end{tabular}
\end{ruledtabular}
\vspace{-3pt}
\end{table*}

\begin{figure}[tbp!]
\begin{center}
  {\includegraphics[height=\hsize,angle=90]{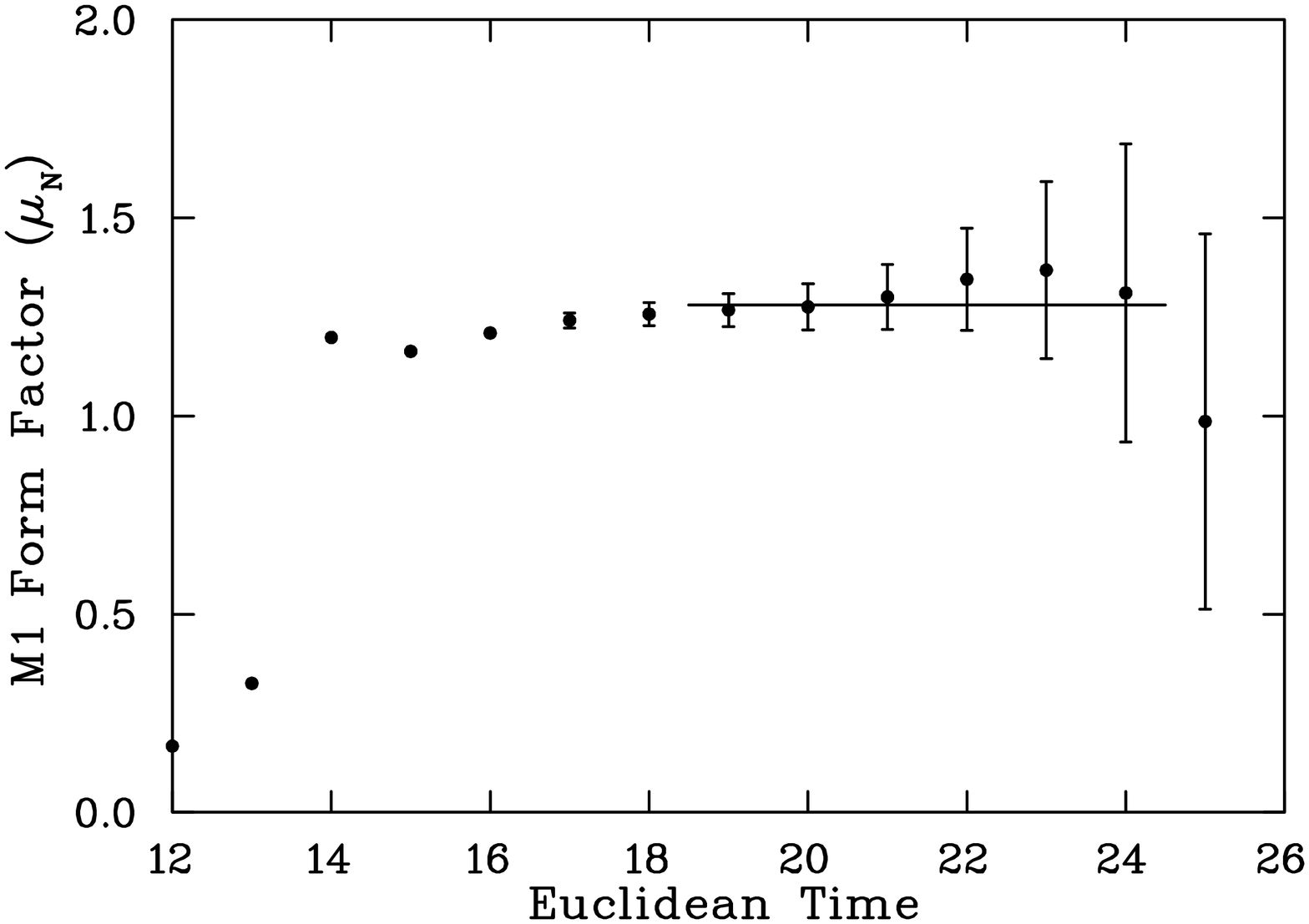}}
\end{center}
\caption{$M1$ magnetic form factor of the $u$ quark in $\Delta$ at
  $Q^2=0.230(1)$ 
  $\mathrm{GeV}^2$ as a function of Euclidean time ($t_2$) for
  $\mathit{m}_{\pi}^{2}=0.485(3)\ \mathrm{GeV}^2$, the $SU(3)$-flavor
  limit. The line indicates the fitting window and the best fit
  value. }
\label{cfm1K3}
\end{figure}

The magnetic form factor $M1$ for the $u$ quark sector in the $\Delta$
at the SU(3) limit is plotted in Fig.~\ref{cfm1K3} as a function of
Euclidean time.
Here the conversion from the natural magneton, $e/(2\, m_B)$, where
the mass of the baryon under investigation appears, to the nuclear
magneton, $e/(2\, m_N)$, where the physical nucleon mass appears, has
been done by multiplying the lattice form factor results by the ratio
$m_N/m_B$.
In this way the form factors are presented in terms of a constant
unit; {\it i.e.} the nuclear magneton.

\begin{figure}[tbp!]
\begin{center}
  {\includegraphics[height=\hsize,angle=90]{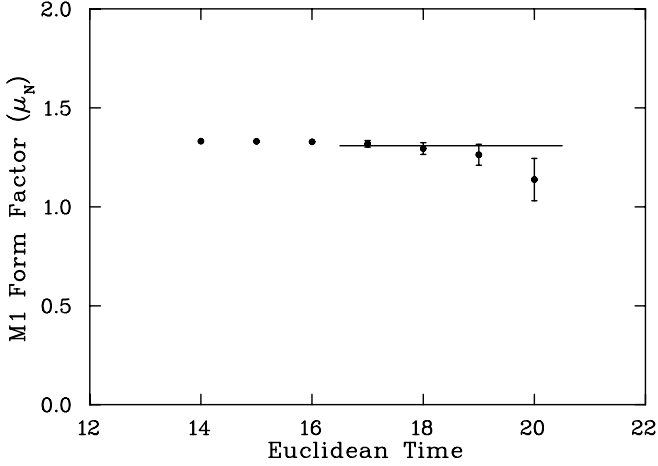}}
\end{center}
\caption{ $M1$ magnetic form factor of the $u$ quark in 
  $\Delta$  at $Q^2 = 0.230(1)\ {\rm GeV}^2$ at the
  ninth quark mass where $m_\pi^2 = 0.215(4)\ {\rm GeV}^2$.  The
  correlator is obtained from the splitting between the ninth and
  eighth quark mass results.  The line indicates the fitting
  window and the best fit value. }
\label{cfm1K6DSp}
\end{figure}

In Fig.~\ref{cfm1K6DSp} we present the Euclidean time dependence of
the the M1 magnetic form factors of $\Delta$ calculated at the ninth
quark mass where $m_\pi^2 = 0.215(4)\ {\rm GeV}^2$ using the
splittings analysis.
The onset of noise at this lighter quark mass is particularly
apparent at time slice 20.

Results for the quark-sector contributions to the M1 magnetic form
factors of decuplet baryons are summarized in Tables~\ref{tab:M1D} to
\ref{tab:M1XS}. While some of the $\chi^2_{\rm dof}$ are some what
large we note that neighboring regimes with acceptable
$\chi^2_{\rm dof}$ have a variation in the central value that is
small with respect to the statistical uncertainty. Results for the
various decuplet baryons are given in Table~\ref{tab:barM1}.

\begin{table*}[tbp]
\caption{Quark sector contributions to the magnetic form factor $M1$ of
  $\Delta$ at $Q^2 = 0.230(1)\ {\rm GeV}^2$.  Sector contributions
  are for a single quark having unit charge.  The fit windows are
  selected using the criteria outlined in Ref.~\cite{Boinepalli:2006xd}. The quark
  contribution at the $SU(3)_{\rm flavor}$ limit when
  $m_\pi^2=0.4854(31)\ {\rm GeV^2}$ provides the $s$ quark contribution
  in $\Omega^-$.} 
\label{tab:M1D}
\begin{ruledtabular}
\begin{tabular}{ccccccc}
\noalign{\smallskip}
$\mathit m_\pi^2\ ({\rm GeV}^2)$  &\multicolumn{3}{c}{$\mathit u_\Delta\
  (\mu_N)$}  & \multicolumn{3}{c}{ $\mathit d_\Delta\ (\mu_N)$ }\\
\noalign{\smallskip}
 & fit value & fit window & $\chi^2_{\rm dof}$ &fit value & fit
 window & $\chi^2_{\rm dof}$ \\
\hline
\noalign{\smallskip}
$0.9960(56)$ & $1.173(25)$ & $19-24$  & $1.45$  &  $1.173(25)$   & $19-24$  & $1.45$ \\
$0.8936(56)$ & $1.201(29)$ & $19-24$  & $1.14$  &  $1.201(29)$   & $19-24$  & $1.14$ \\
$0.7920(55)$ & $1.230(33)$ & $19-24$  & $0.95$  &  $1.230(33)$   & $19-24$  & $0.95$ \\
$0.6920(54)$ & $1.256(40)$ & $19-24$  & $0.75$  &  $1.256(40)$   & $19-24$  & $0.75$ \\
$0.6910(35)$ & $1.248(32)$ & $19-24$  & $1.25$  &  $1.248(32)$   & $19-24$  & $1.25$ \\
$0.5925(33)$ & $1.269(41)$ & $19-24$  & $0.79$  &  $1.269(41)$    & $19-24$  & $0.79$ \\
$0.4854(31)$ & $1.280(56)$ & $19-24$  & $0.31$  &  $1.280(56)$  & $19-24$  & $0.31$  \\
$0.3795(31)$ & $1.301(64)$ & $17-21$ & $1.37$  &  $1.301(64)$  & $17-21$  & $1.37$ \\
$0.2839(33)$ & $1.312(75)$ & $17-19$ & $1.14$  &   $1.312(75)$  & $17-19$  & $1.14$ \\
$0.2153(35)$ & $1.309(89)$ & $17-19$ & $0.91$  &  $1.309(89)$  & $17-19$  & $0.91$ \\
$0.1384(43)$ & $1.28(12)$ & $17-18$ & $1.26$  &  $1.28(12)$  & $17-18$  & $1.26$ \\
$0.0939(44)$ & $1.11(21)$ & $17-18$ & $1.79$  &  $1.11(21)$  & $17-18$  & $1.79$ \\
\end{tabular}
\end{ruledtabular}
\vspace{-3pt}
\end{table*}

\begin{table*}[tbp]
\caption{Quark sector contributions to the magnetic form factor $M1$ of
  $\Sigma^*$ at $Q^2 = 0.230(1)\ {\rm GeV}^2$.  Sector contributions
  are for a single quark having unit charge. The fit windows are
  selected using the criteria outlined in Ref.~\cite{Boinepalli:2006xd}.}
\label{tab:M1SS}
\begin{ruledtabular}
\begin{tabular}{ccccccc}
\noalign{\smallskip}
$\mathit m_\pi^2\ ({\rm GeV}^2)$  
&\multicolumn{3}{c}{$u_{\Sigma^*}\ {\rm or}\ d_{\Sigma^*}\ (\mu_N)$} 
&\multicolumn{3}{c}{$\mathit s_{\Sigma^*}\ (\mu_N)$}
\\
\noalign{\smallskip}
 & fit value & fit window & $\chi^2_{\rm dof}$ &fit value & fit
 window & $\chi^2_{\rm dof}$   \\
\hline
\noalign{\smallskip}
$0.9960(56)$ & $1.191(31)$ & $19-24$ & $1.07$  &  $1.268(40)$ & $19-24$  & $0.51$ \\
$0.8936(56)$ & $1.216(35)$ & $19-24$ & $0.90$  &  $1.275(42)$ & $19-24$  & $0.49$ \\
$0.7920(55)$ & $1.242(39)$ & $19-24$ & $0.82$  &  $1.282(45)$ & $19-24$  & $0.47$ \\
$0.6920(54)$ & $1.264(45)$ & $19-24$ & $0.69$ & $1.289(50)$ & $19-24$ & $0.47$ \\
$0.6910(35)$ & $1.254(37)$ & $19-24$ & $0.94$  &  $1.275(41)$ & $19-24$  & $0.69$ \\
$0.5925(33)$ & $1.272(45)$ & $19-24$ & $0.64$  &  $1.278(47)$ & $19-24$  & $0.51$ \\
$0.4854(31)$ & $1.280(56)$ & $19-24$ & $0.31$  &  $1.280(56)$ & $19-24$  & $0.31$ \\
$0.3795(31)$ & $1.297(62)$ & $17-21$ & $1.72$  &  $1.289(60)$ & $17-21$  & $0.70$ \\
$0.2839(33)$ & $1.306(68)$ & $17-19$ & $1.88$  &  $1.299(65)$ & $17-19$  & $0.04$ \\
$0.2153(35)$ & $1.305(76)$ & $17-19$ & $1.84$  &  $1.309(69)$ & $17-19$  & $0.08$ \\
$0.1384(43)$ & $1.299(89)$ & $17-19$ & $3.44$  &  $1.330(74)$ & $17-19$  & $0.44$ \\
$0.0939(44)$ & $1.25(12)$  & $17-18$ & $2.55$  &  $1.303(88)$ & $17-18$  & $0.16$
\end{tabular}
\end{ruledtabular}
\vspace{-3pt}
\end{table*}

\begin{table*}[tbp]
\caption{Quark sector contributions to the magnetic form factor $M1$ of
  $\Xi^*$ at $Q^2 = 0.230(1)\ {\rm GeV}^2$.  Sector contributions
  are for a single quark having unit charge. The fit windows are
  selected using the criteria outlined in Ref.~\cite{Boinepalli:2006xd}.}
\label{tab:M1XS}
\begin{ruledtabular}
\begin{tabular}{ccccccc}
\noalign{\smallskip}
$\mathit m_\pi^2\ ({\rm GeV}^2)$  &\multicolumn{3}{c}{$\mathit s_{\Xi^*}\
  (\mu_N)$}  & \multicolumn{3}{c}{ $u_{\Xi^*}\ {\rm or}\ d_{\Xi^*}\ (\mu_N)$ }\\
\noalign{\smallskip}
 & fit value & fit window & $\chi^2_{\rm dof}$ &fit value & fit
 window & $\chi^2_{\rm dof}$   \\
\hline
\noalign{\smallskip}
$0.9960(56)$ & $1.286(50)$ & $19-24$  & $0.45$&  $1.208(39)$ & $19-24$ & $0.84$  \\
$0.8936(56)$ & $1.289(52)$ & $19-24$  & $0.46$ & $1.231(42)$ & $19-24$ & $0.79$  \\
$0.7920(55)$ & $1.293(54)$ & $19-24$  & $0.46$ & $1.254(46)$ & $19-24$ & $0.76$  \\
$0.6920(54)$ & $1.297(56)$ & $19-24$ & $0.45$ & $1.273(51)$ & $19-24$ & $0.63$ \\
$0.6910(35)$ & $1.278(48)$ & $19-24$  & $0.48$ & $1.260(44)$ & $19-24$ & $0.68$  \\
$0.5925(33)$ & $1.280(51)$ & $19-24$  & $0.40$ & $1.274(49)$ & $19-24$ & $0.52$  \\
$0.4854(31)$ & $1.280(56)$ & $19-24$  & $0.31$ & $1.280(56)$ & $19-24$ & $0.31$  \\
$0.3795(31)$ & $1.285(58)$ & $17-21$  & $0.70$ & $1.293(60)$ & $17-21$ & $2.56$  \\
$0.2839(33)$ & $1.289(60)$ & $17-19$  & $0.04$ & $1.300(63)$ & $17-19$ & $2.66$  \\
$0.2153(35)$ & $1.293(62)$ & $17-19$  & $0.02$ & $1.303(66)$ & $17-19$ & $1.44$  \\
$0.1384(43)$ & $1.302(64)$ & $17-18$  & $0.73$ & $1.303(72)$ & $17-18$ & $2.36$  \\
$0.0939(44)$ & $1.301(67)$ & $17-18$  & $0.23$ & $1.313(81)$ & $17-18$ & $0.56$
\end{tabular}
\end{ruledtabular}
\vspace{-3pt}
\end{table*}

\begin{table*}[tbp]
\caption{$M1$ form factor of the charged decuplet
  baryons for different $m_\pi^2$
  values. The $M1$ form factor of the $\Delta^-$ at the
  $SU(3)_{\rm flavor}$ limit ($m_\pi^2 = 0.485(3)\ \rm{GeV}^2$)
  provides the $M1$ form factor of $\Omega^-$.}
\label{tab:barM1}
\begin{ruledtabular}
\begin{tabular}{ccccccc}
\noalign{\smallskip}
$\mathit m_\pi^2\ ({\rm GeV}^2)$  & $\Delta^{++}$ & $\Delta^+$ &
$\Delta^-$ & $\Sigma^{*+}$ & $\Sigma^{*-}$ & $\Xi^{*-}$ \\
\noalign{\smallskip}
\hline
\noalign{\smallskip}
$0.9972(55)$ & $2.35(5)$  & $1.17(2)$  & $-1.17(2)$  & $1.16(3)$  & $-1.22(3)$  & $-1.26(5)$ \\
$0.8936(56)$ & $2.40(6)$  & $1.20(3)$  & $-1.20(3)$  & $1.20(3)$  & $-1.24(4)$  & $-1.27(5)$ \\
$0.7920(55)$ & $2.46(7)$  & $1.23(3)$  & $-1.23(3)$  & $1.23(4)$  & $-1.26(4)$  & $-1.28(5)$ \\
$0.6920(54)$ & $2.51(8)$  & $1.26(4)$  & $-1.26(4)$  & $1.26(4)$  & $-1.27(4)$  & $-1.29(5)$ \\
$0.6910(35)$ & $2.50(6)$  & $1.25(3)$  & $-1.25(3)$  & $1.25(4)$  & $-1.26(4)$  & $-1.27(5)$ \\
$0.5925(33)$ & $2.54(8)$  & $1.27(4)$  & $-1.27(4)$  & $1.27(4)$  & $-1.27(5)$  & $-1.28(5)$ \\
$0.4854(31)$ & $2.56(11)$ & $1.28(6)$  & $-1.28(6)$  & $1.28(6)$  & $-1.28(6)$  & $-1.28(6)$ \\
$0.3795(31)$ & $2.60(13)$ & $1.30(6)$  & $-1.30(6)$  & $1.30(6)$  & $-1.29(6)$  & $-1.29(6)$ \\
$0.2839(33)$ & $2.62(15)$ & $1.31(7)$  & $-1.31(7)$  & $1.31(7)$  & $-1.30(7)$  & $-1.29(6)$ \\
$0.2153(35)$ & $2.62(18)$ & $1.31(9)$  & $-1.31(9)$  & $1.30(8)$  & $-1.31(7)$  & $-1.30(6)$ \\
$0.1384(43)$ & $2.56(24)$ & $1.28(12)$ & $-1.28(12)$ & $1.29(10)$ & $-1.31(8)$  & $-1.30(6)$ \\
$0.0939(44)$ & $2.22(43)$ & $1.11(22)$ & $-1.11(22)$ & $1.23(14)$ & $-1.27(10)$ & $-1.30(7)$
\end{tabular}
\end{ruledtabular}
\vspace{-3pt}
\end{table*}
%

%
\section{Discussion of Results}
\label{sec:Results}
\subsection{Charge radii}

It is well known that the experimental electric (and magnetic) form
factor of the proton is described well by a dipole ansatz at small
$Q^2$
\begin{equation}
\quad {\cal G}_E(Q^2) = { {\cal G}_E(0) \over 
             \left (1 + Q^2 / m^2 \right )^2 } \, ; \quad
      Q^2 \ge 0 .
 \quad
\label{dipole}
\end{equation}
This behavior has also been observed in recent lattice calculations
\cite{Gockeler:2003ay}.
Using this observation, together with the standard small $Q^2$
expansion of the Fourier transform of a spherical charge distribution
\begin{equation}
 \left\langle r_E^2 \right\rangle = -6 \frac{d}{{d}Q^2} {\cal
   G}_E(Q^2) \bigm |_{Q^2=0} \ ,
\label{charad}
\end{equation}
we arrive at an expression which allows us to calculate the electric
charge radius of a baryon using our two available values of the Sachs
electric form factor (${\cal G}_E(Q_{\rm min}^2),\ {\cal G}_E(0)$),
namely
\begin{equation}
\frac{\left\langle r_E^2 \right\rangle}{ {\cal G}_E(0)} = \frac{12}{Q^2}
\left ( \sqrt{ \frac{{\cal G}_E(0)}{{\cal G}_E(Q^2) }} - 1 \right )\ .
\label{charad_dipole}
\end{equation}
However to calculate the charge radii of the neutral baryons, the
above equation cannot be used, due to the fact that in those cases
${\cal G}_E = 0$.
For the neutral baryons it becomes a simple matter to construct the
charge radii by first calculating the charge radii for each quark
sector.
These quark sectors are then combined using the appropriate charge and
quark number factors as described in Sec.~\ref{ffCorr} to obtain the
total baryon charge radii.
Indeed, all baryon charge radii including the charged states are
calculated in this manner.

\begin{figure}[tbp!]
\begin{center}
  {\includegraphics[height=\hsize,angle=90]{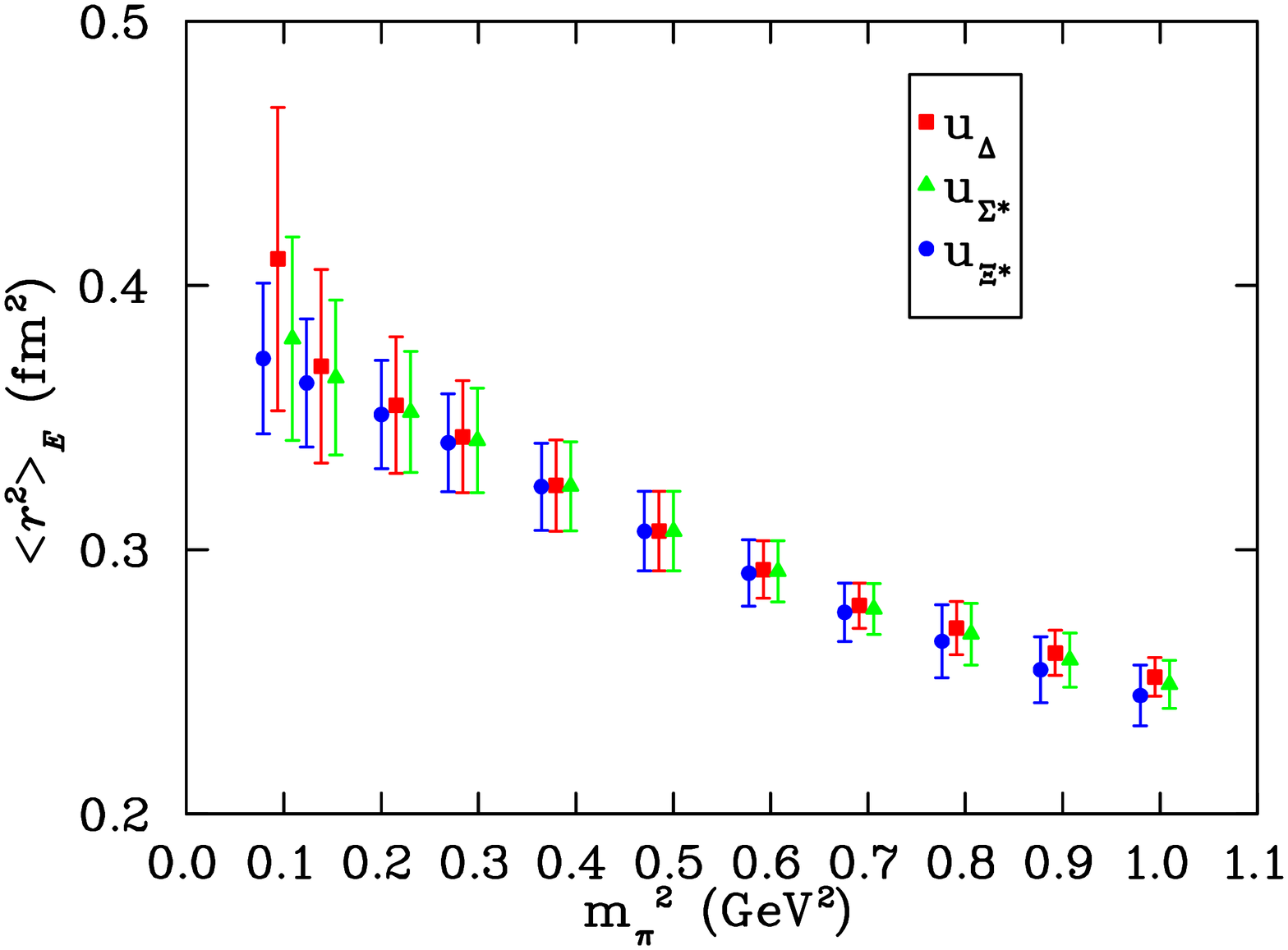}}
\end{center}
\caption{Charge radii of the $u$ quark in $\Delta$, $\Sigma^*$ and
  $\Xi^*$ at different quark masses. The values for $\Delta$ are
  plotted at $m_\pi^2$ while that of the $\Sigma^*$ and $\Xi^*$ are
  plotted at shifted $m_\pi^2$ for clarity.}
\label{dEradu}
\end{figure}

\begin{figure}[tbp!]
\begin{center}
  {\includegraphics[height=\hsize,angle=90]{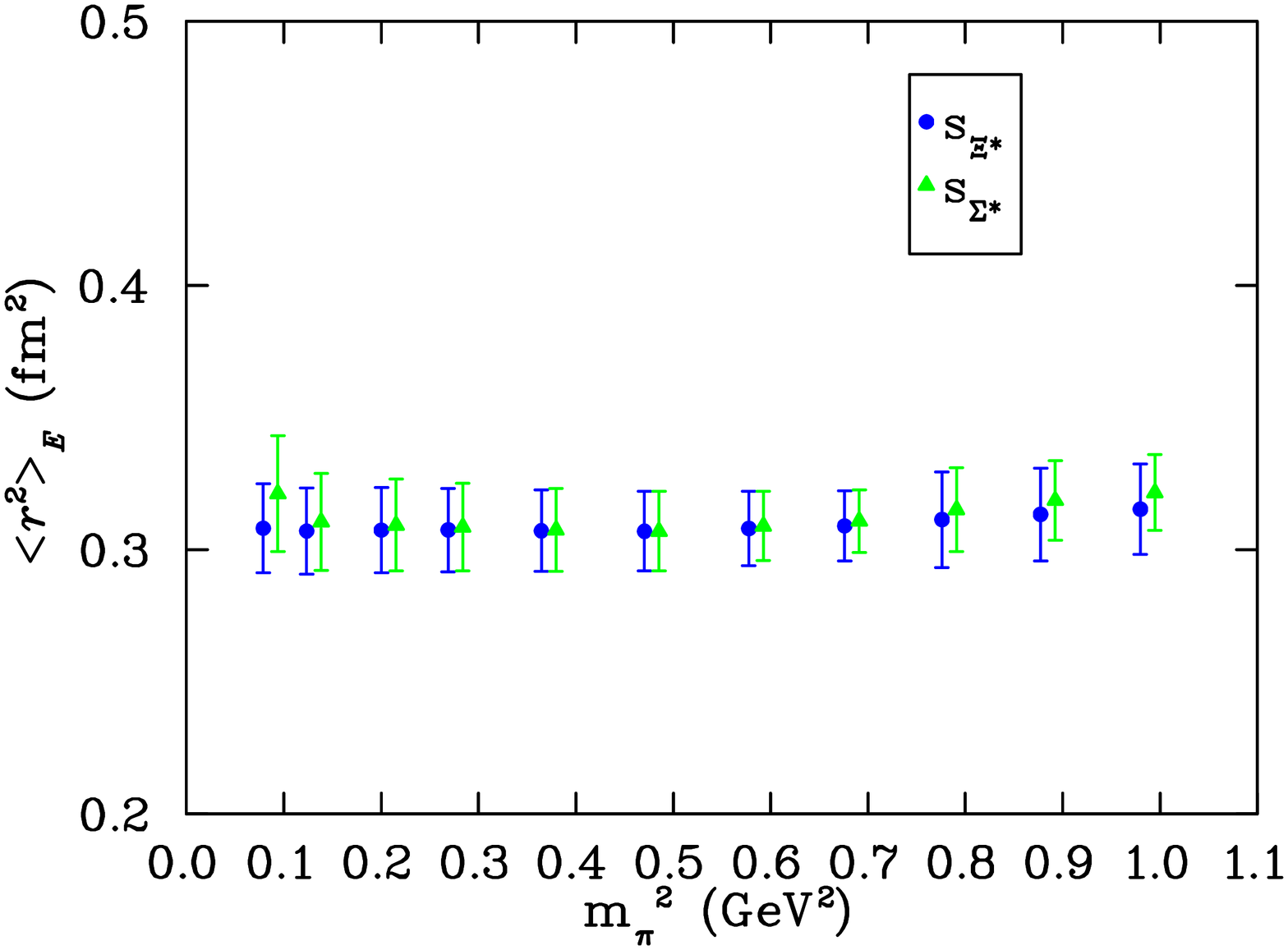}}
\end{center}
\caption{Charge radii of the $s$ quark in $\Sigma^*$ and
  $\Xi^*$ at different quark masses. The values for $\Sigma^*$ are
  plotted for shifted $m_\pi^2$ values for clarity. }
\label{dErads}
\end{figure}

Tables~\ref{tab:eradD} to \ref{tab:eradXS} provide the electric charge
radii of the decuplet baryons and the quark sector contributions.
Figures~\ref{dEradu} and ~\ref{dErads} depict plots of the quark
contributions to the decuplet charge radii. 
At the SU(3) limit (sixth quark mass) the quark contributions are
identical in all cases as expected.

\begin{figure}[tbp!]
\begin{center}
  {\includegraphics[height=6cm,angle=0]{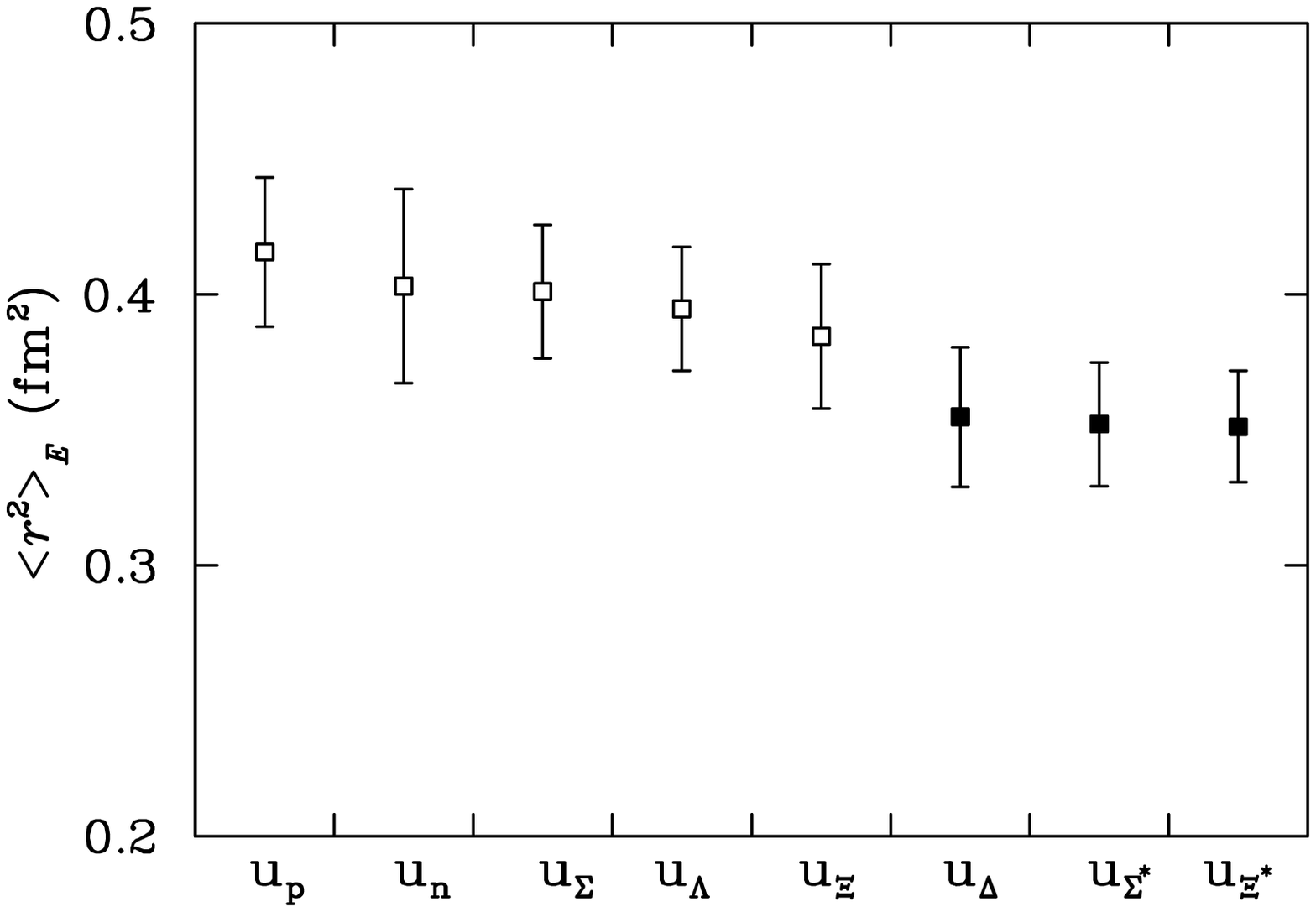}}
\end{center}
\caption{Charge radii of the $u$ quark sector in the octet (open
  squares) and the decuplet (filled squares) baryons at the ninth
  quark mass where $m_\pi^2 = 0.215(4)\ {\rm GeV}^2$.}
\label{CR_quark_Dble}
\end{figure}

\begin{figure}[tbp!]
\begin{center}
  {\includegraphics[height=6cm,angle=0]{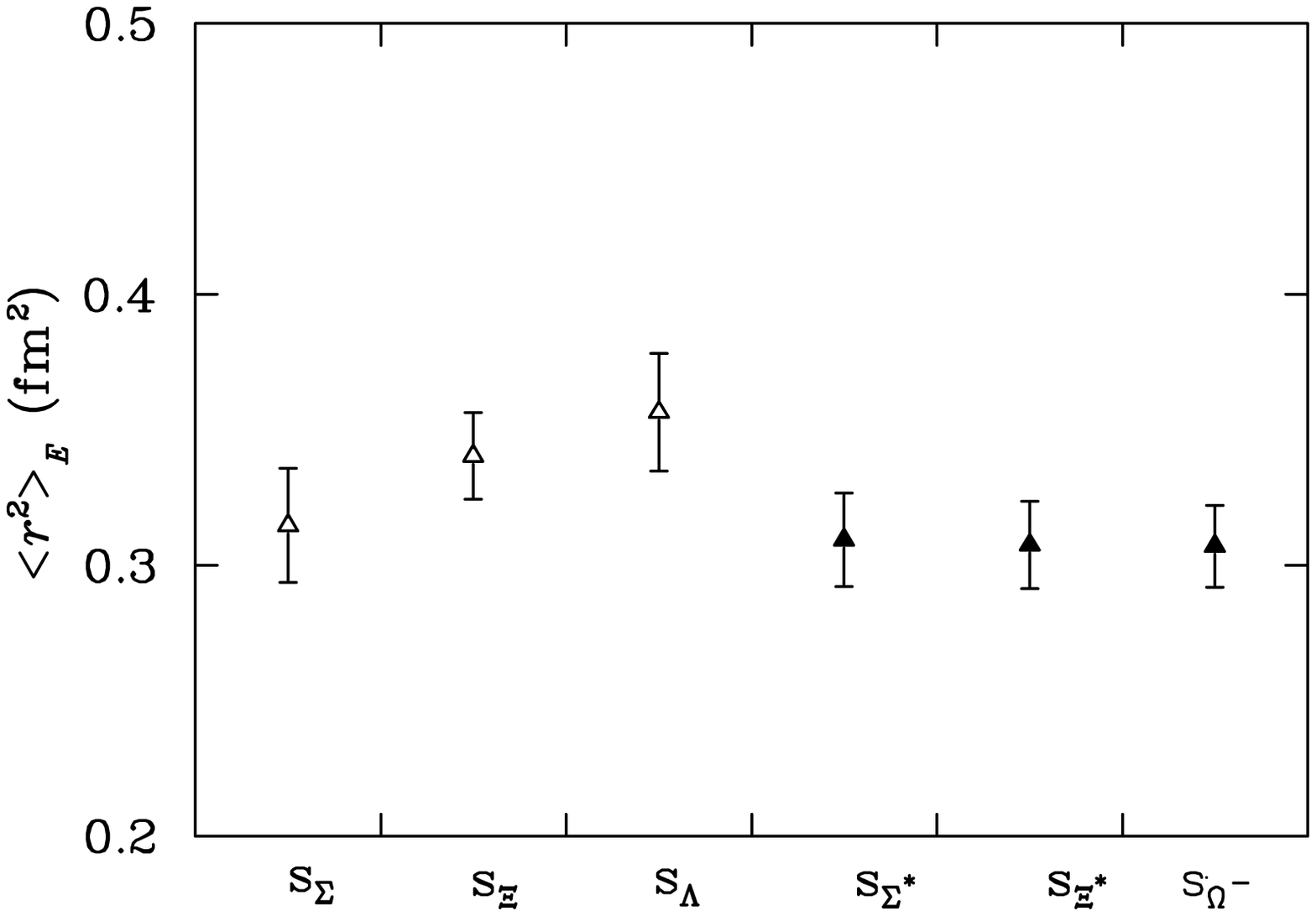}}
\end{center}
\caption{Charge radii of the $s$ quark sector in the octet (open
  squares) and the decuplet (filled squares) baryons at the ninth
  quark mass where $m_\pi^2 = 0.215(4)\ {\rm GeV}^2$. }
\label{CR_quark_Sing}
\end{figure}

In Figs.~\ref{CR_quark_Dble} and \ref{CR_quark_Sing} we compare the
charge radii of the quark sectors in the decuplet baryons to those in
the octet baryons at the ninth quark mass.
From the figures it is evident that the contribution of the quarks is
very much baryon dependent in the octet case, while for the decuplet
baryons there is much less environmental sensitivity to the individual
quark contributions.
More specifically, in the case of the $u$ quark in the octet baryons,
the charge radius decreases with the inclusion of the $s$ quark, while
such an influence of the $s$ quark on $u$ quark charge radius is
absent in the decuplet behavior.
Furthermore, we note that the charge radius of the $u$ quark
distribution in the decuplet baryons is {\it smaller} than that in the octet
baryons.

\begin{figure}[tbp!]
\begin{center}
  {\includegraphics[height=6cm,angle=0]{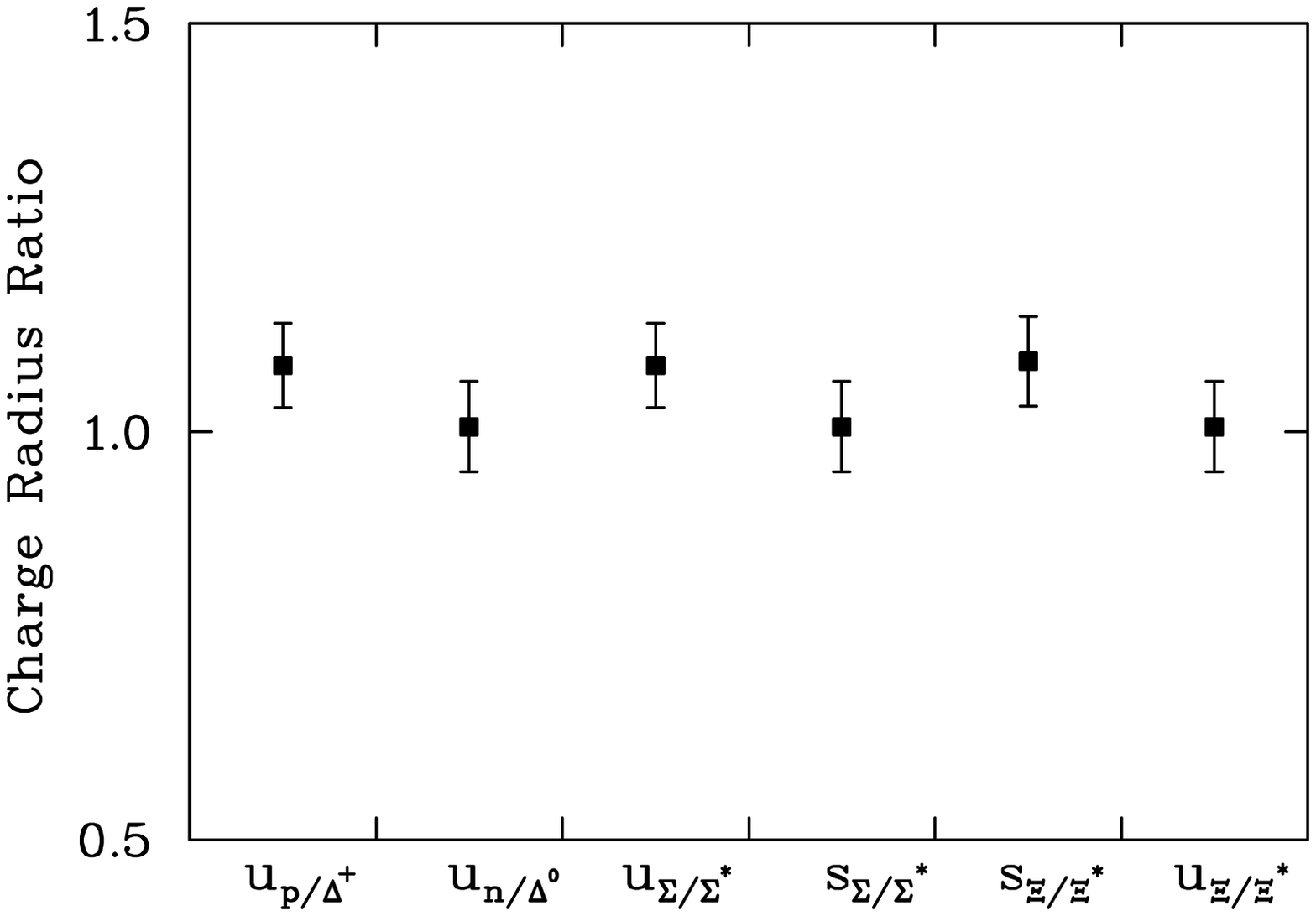}}
\end{center}
\caption{Ratio of charge radii of the quark sector contributions in
  the octet/decuplet baryons at the $SU(3)_{\rm flavor}$ limit where $m_\pi^2
  = 0.485(3)\ {\rm GeV}^2$.}
\label{ratio_CR_quark_su3}
\end{figure}

\begin{figure}[tbp!]
\begin{center}
  {\includegraphics[height=6cm,angle=0]{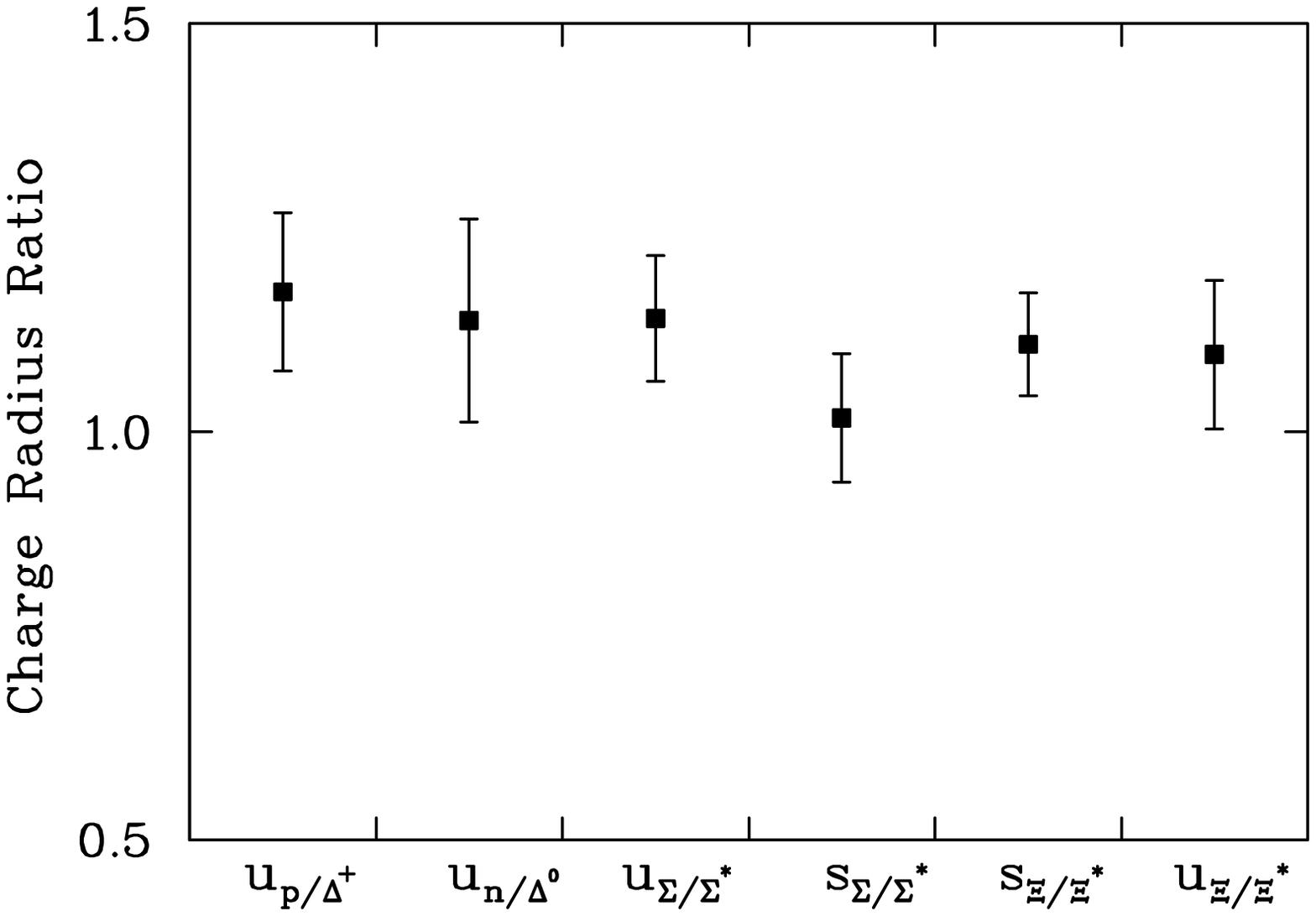}}
\end{center}
\caption{Ratio of charge radii of the quark sector contributions in
  the octet/decuplet baryons at the ninth quark mass where $m_\pi^2 =
  0.215(4)\ {\rm GeV}^2$. }
\label{ratio_CR_quark_9qm}
\end{figure}

From Fig.~\ref{CR_quark_Dble} it is evident that the charge radius of
the $u$ quark in the proton ($u_p$) is larger than that of the $u$
quark in the $\Delta^+$ ($u_{\Delta^+}$).
In order to investigate this difference more accurately, we compute
the ratios of charge distributions of similar quarks in the octet to
that in the decuplet.
The uncertainty in the ratio ${\langle r^2 \rangle} (u_p)/{\langle
  r^2\rangle} (u_{\Delta^+})$ is calculated using the jack-knife
method.
Figures~\ref{ratio_CR_quark_su3} and~\ref{ratio_CR_quark_9qm} depict
the ratio of the quark contributions in the octet baryons to those in
the decuplet baryons at the $SU(3)_{\rm flavor}$ limit and the ninth
quark mass, respectively.
In both cases, the doubly represented $u$ quark contribution to the
charge radius of 
octet baryons is larger than that in case of the singly represented
octet quarks and decuplet quarks.
At the SU(3) limit, all quarks take the strange quark mass, and hence
one would expect the quark model picture to dominate.
This suggests that the $u_\Delta$ should have a broader distribution
than the distribution than that of $u_p$ due to hyperfine interactions. 
Our results contrast this prediction.
The smaller charge radius of $u_\Delta$ compared to that of $u_p$ also
rules out any suggestion of a hyperfine attraction leading to $ud$
diquark clustering in the nucleon or hyperon
states~\cite{Leinweber:1993nr}.

\begin{figure}[tbp!]
\begin{center}
  {\includegraphics[height=\hsize,angle=90]{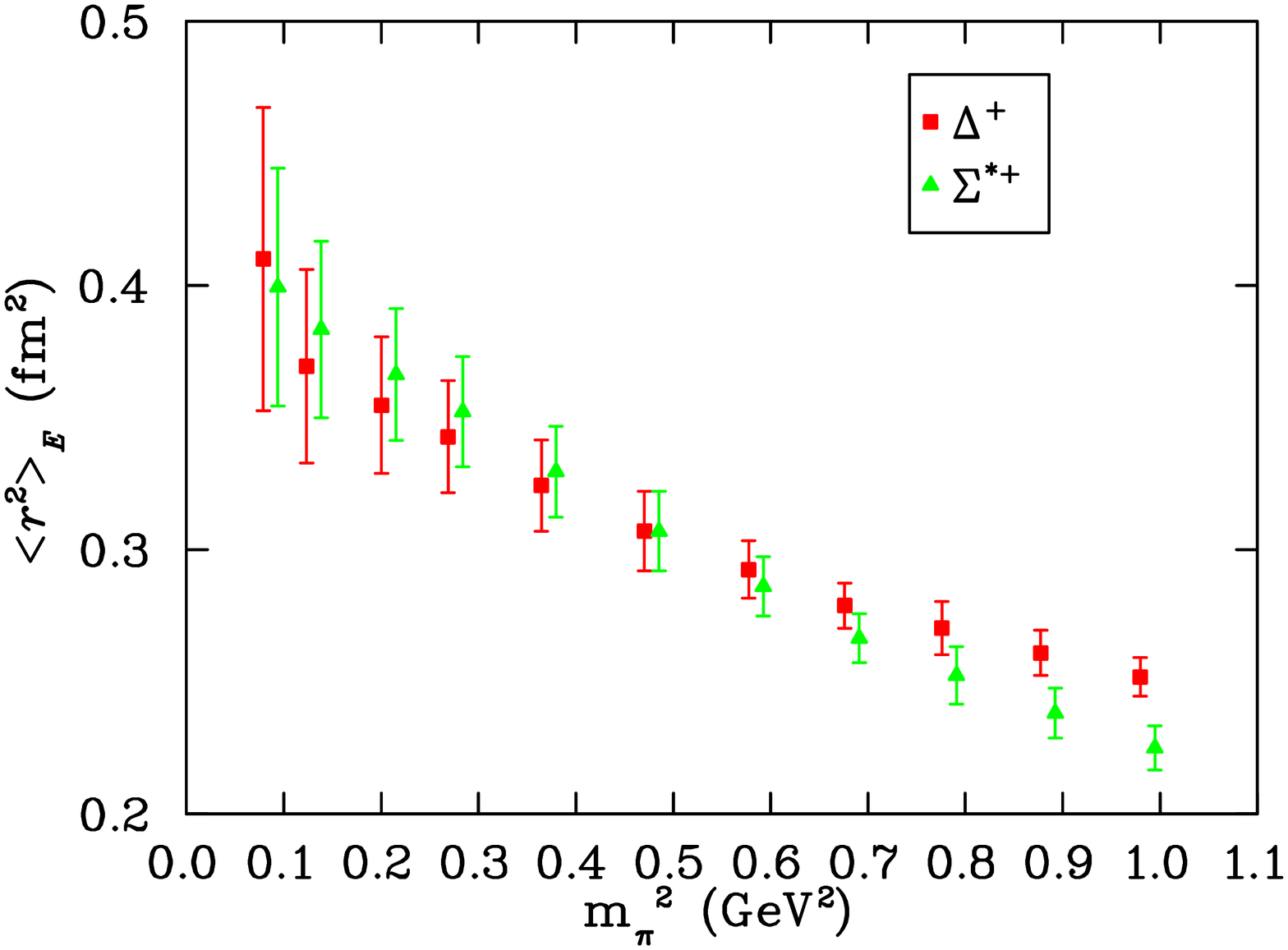}}
\end{center}
\caption{Charge radii of  $\Delta^+$ and  $\Sigma^{*+}$ at different
  quark masses. The values for $\Sigma^*$ are plotted at shifted
  $m_\pi^2$ for clarity.}
\label{dEradp}
\end{figure}

\begin{figure}[tbp!]
\begin{center}
  {\includegraphics[height=\hsize,angle=90]{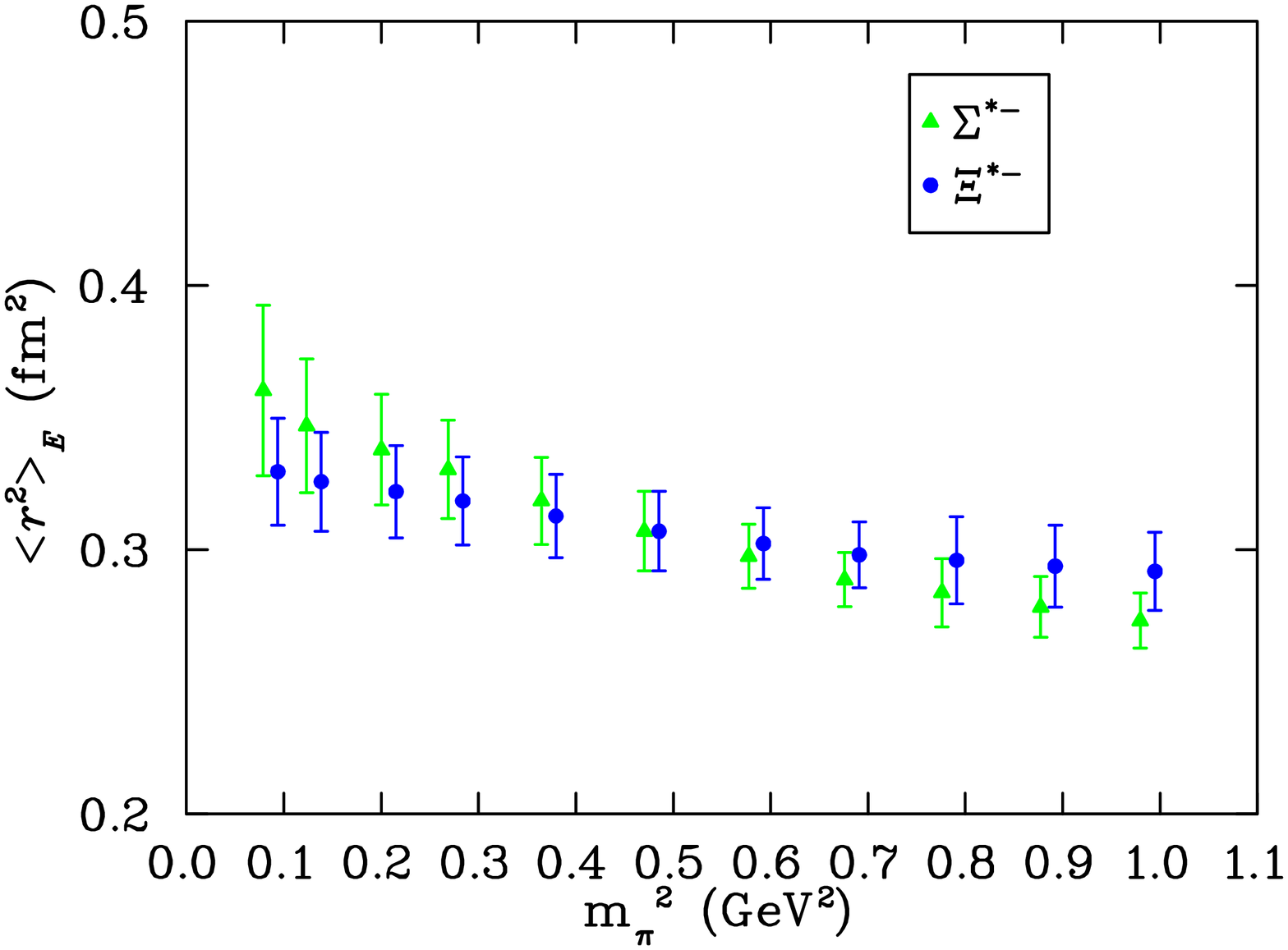}}
\end{center}
\caption{Charge radii (magnitude) of  $\Sigma^{*-}$ and $\Xi^{*-}$ at
  different quark masses. The values for $\Xi^{*-}$ are plotted at
  shifted $m_\pi^2$ for clarity.}
\label{dEradm}
\end{figure}

\begin{figure}[tbp!]
\begin{center}
  {\includegraphics[height=\hsize,angle=90]{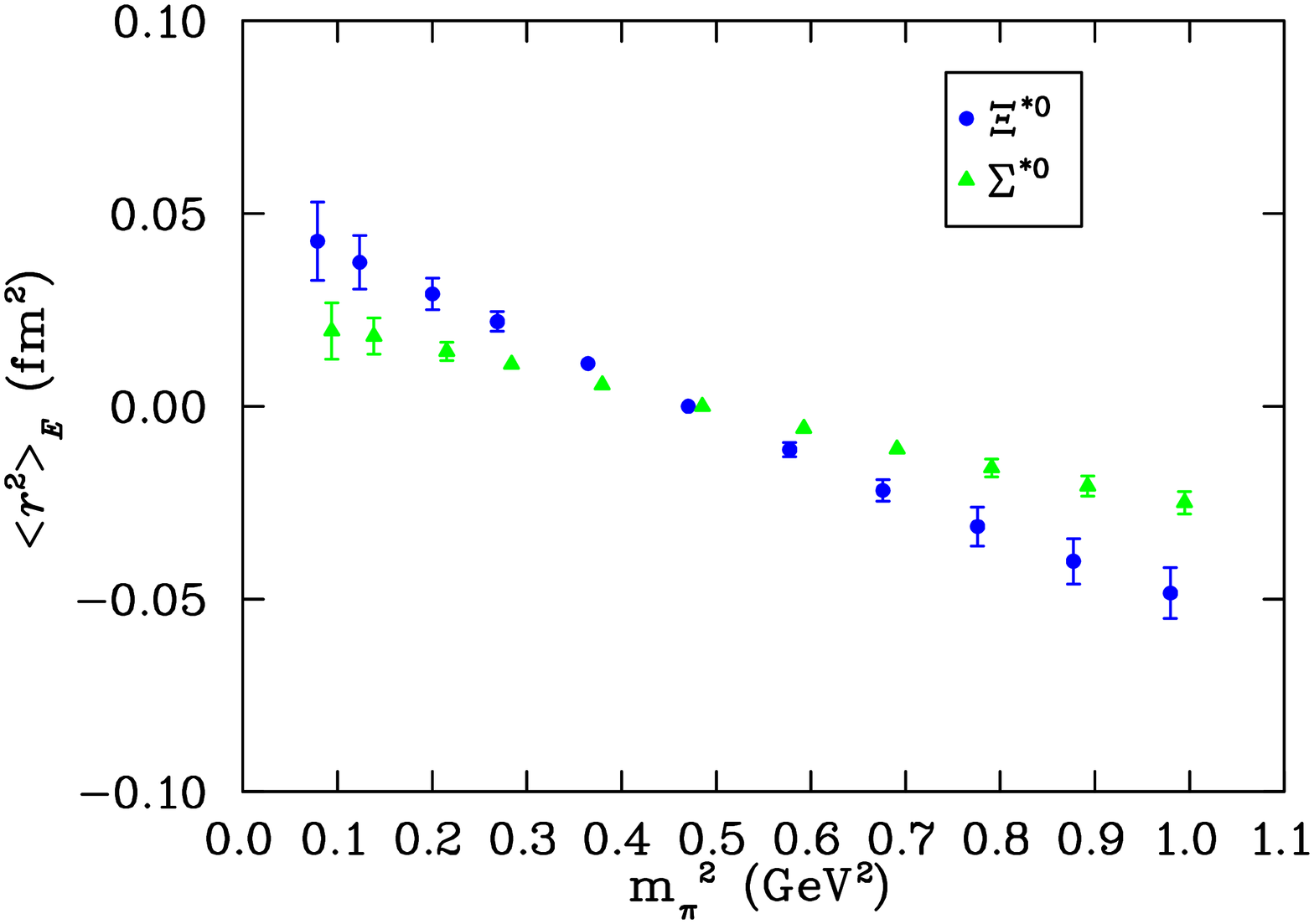}}
\end{center}
\caption{Charge radii of  $\Sigma^{*0}$ and $\Xi^{*0}$ at different
  quark masses.  The values for $\Sigma^{*0}$ are plotted at shifted
  $m_\pi^2$ for clarity.}
\label{dEradn}
\end{figure}

The charge radii of the various decuplet baryon states are shown in
Figs.~\ref{dEradp}, \ref{dEradm} and \ref{dEradn} as a function of
$m_\pi^2$.
The charge radius of $\Delta^-$ is numerically equal to that of the
$\Delta^+$ with a negative sign. 
The charge radius of $\Omega^-$ is taken as that of the $\Delta^{-}$
in the $SU(3)_{\rm flavor}$ limit, and is numerically equal to $-0.307
\pm 0.015~\rm{fm}^2$.
As our calculations neglect the $\Omega^- \rightarrow \Xi^0\,\pi^-$
dressing, we anticipate our result to underestimate the magnitude.

The decuplet baryon form factors are dominated by the net charge of
the light quarks.
For the $\Delta^0$ the symmetry of the $u$ and $d$ quarks makes the
form factors vanish. 
This is in contrast to the neutron where the three quarks are in
mixed-symmetric states, giving rise to a non-zero form factor and
charge radius.
Charge radii of the neutral $\Sigma^*$ and $\Xi^*$ are also close to
zero and are dominated by the light quark sectors.

\begin{figure}[tbp!]
\begin{center}
  {\includegraphics[height=6cm,angle=0]{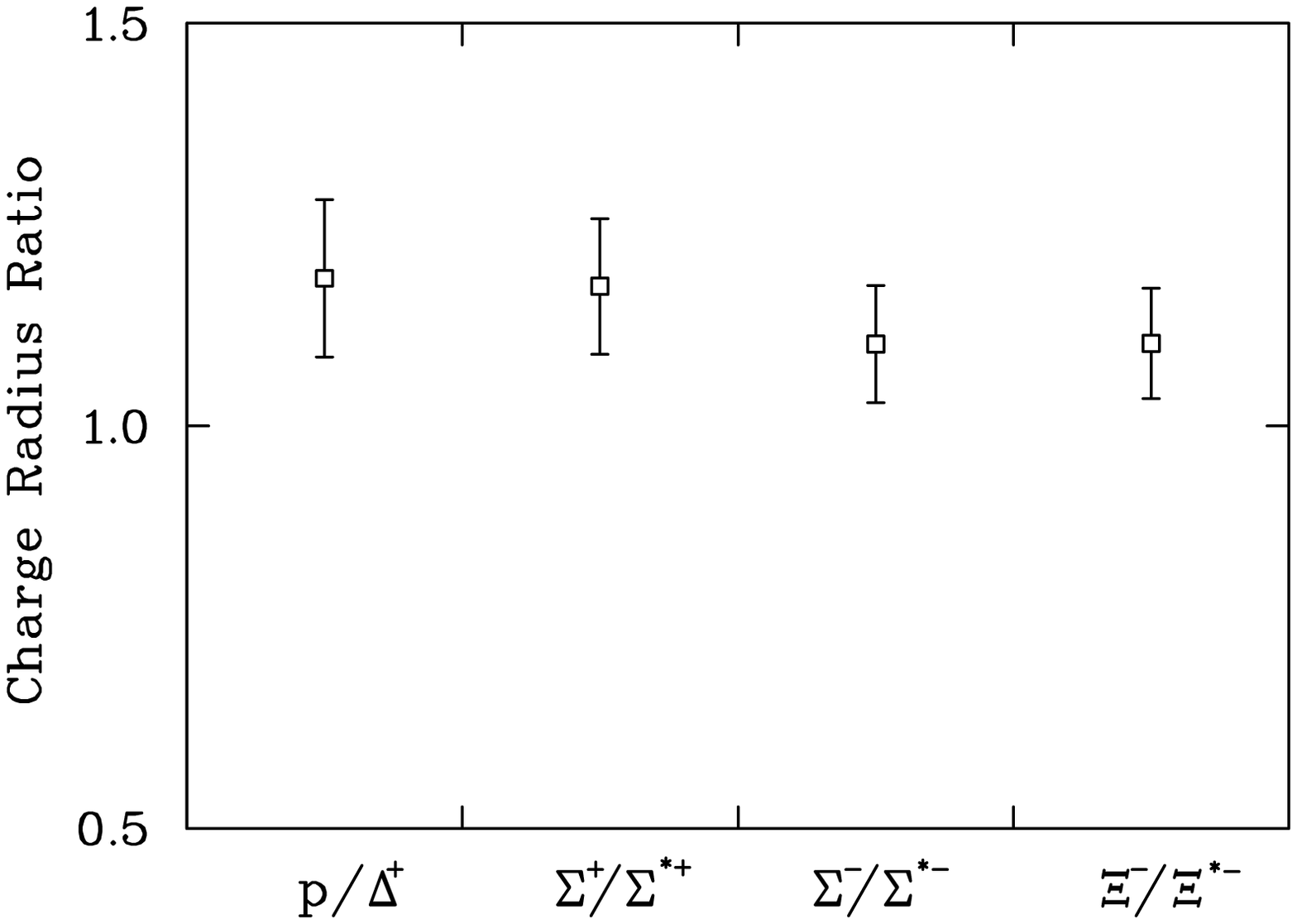}}
\end{center}
\caption{Ratio of charge radii of the octet/decuplet baryons at the
  ninth quark mass where $m_\pi^2 = 0.215(4)\ {\rm GeV}^2$. }
\label{ratio_CR_Bar}
\end{figure}

The presence of the $s$ quark as one moves from $\Delta$ to $\Sigma^*$
and $\Xi^*$ reduces the magnitude of the charge radius as indicated in
Figs.~\ref{dEradp}, \ref{dEradm} and \ref{dEradn}.
By examining the ratio of the charge radii of the octet to decuplet
baryons in Fig.~\ref{ratio_CR_Bar}, we observe that the octet baryons
have a slightly larger charge radius than their decuplet counterparts.

\begin{table*}[tbp]
\caption {Charge radii of the
 $\Delta$ baryons for different $m_\pi^2$. Quark sector contributions for
 a single quark of unit charge are included. The charge radii of the
 $\Delta^{++}$ are twice that of the $\Delta^+$ and the
 results for the $\Delta^0$ are $0$ in QQCD. At the
 $SU(3)_{\rm flavor}$ limit we find the charge radius of the
 $\Omega^-$ to be equal to $-0.307(15)~\rm{fm}^2$.}
\label{tab:eradD}
\begin{ruledtabular}
\begin{tabular}{ccccc}
$\mathit m_\pi^2 (\rm GeV^2)$ &  $\mathit u_{\Delta} (\rm fm^2)$ &
  $\mathit d_{\Delta} (\rm fm^2)$ & $\Delta^+ (\rm fm^2)$ & $\Delta^-
  (\rm fm^2)$ \\
\hline
$0.9960(56)$ & $0.238(7)$   &  $0.238(7)$   & $0.238(7)$   & $-0.238(7)$\\
$0.8936(56)$ & $0.247(8)$   &  $0.247(8)$   & $0.247(8)$   & $-0.247(8)$\\
$0.7920(55)$ & $0.256(10)$  &  $0.256(10)$  & $0.256(10)$  & $-0.256(10)$ \\ 
$0.6920(54)$ & $0.264(12)$ & $0.264(12)$ & $0.264(12)$ & $-0.264(12)$ \\
$0.6910(35)$ & $0.279(9)$   &  $0.279(9)$   & $0.279(9)$   & $-0.279(9)$ \\
$0.5925(33)$ & $0.293(11)$  &  $0.293(11)$  & $0.293(11)$  & $-0.293(11)$ \\ 
$0.4854(31)$ & $0.307(15)$  &  $0.307(15)$  & $0.307(15)$  & $-0.307(15)$ \\
$0.3795(31)$ & $0.324(17)$  &  $0.324(17)$  & $0.324(17)$  & $-0.324(17)$ \\
$0.2839(33)$ & $0.343(21)$  &  $0.343(21)$  & $0.343(21)$  & $-0.343(21)$ \\
$0.2153(35)$ & $0.355(26)$  &  $0.355(26)$  & $0.355(26)$  & $-0.355(26)$\\
$0.1384(43)$ & $0.370(37)$  &  $0.370(37)$  & $0.370(37)$  & $-0.370(37)$\\
$0.0939(44)$ & $0.410(57)$  &  $0.410(57)$  & $0.410(57)$  & $-0.410(57)$
\end{tabular}
\end{ruledtabular}
\vspace{-3pt}
\end{table*}

\begin{table*}[tbp]
\caption{Charge radii of the
 $\Sigma^*$ baryons for different $m_\pi^2$. Quark sector contributions for
 a single quark of unit charge are included. }
\label{tab:eradSS}
\begin{ruledtabular}
\begin{tabular}{cccccc}
$\mathit m_\pi^2 (\rm GeV^2)$ &  $\mathit u_{\Sigma^*} (\rm fm^2) $ &
$\mathit s_{\Sigma^*} (\rm fm^2)$ & $\Sigma^{*+} (\rm fm^2)$ &
$\Sigma^{*0}(\rm fm^2)$ & $\Sigma^{*-} (\rm fm^2)$ \\
\hline
$0.9960(56)$ & $0.233(8)$ & $0.299(13)$ & $0.212(8)$  & $-0.022(3)$ & $-0.255(10)$ \\
$0.8936(56)$ & $0.242(10)$ & $0.296(14)$ & $0.224(9)$  & $-0.018(2)$ & $-0.260(11)$\\ 
$0.7920(55)$ & $0.251(11)$ & $0.293(15)$ & $0.237(10)$ & $-0.014(2)$ & $-0.265(12)$\\
$0.6920(54)$ & $0.260(13)$ & $0.289(16)$ & $0.250(12)$ & $-0.010(2)$ & $-0.270(14)$\\
$0.6910(35)$ & $0.278(10)$ & $0.311(12)$ & $0.267(9)$  & $-0.011(1)$ & $-0.289(10)$ \\
$0.5925(33)$ & $0.292(12)$ & $0.309(13)$ & $0.286(11)$ & $-0.006(1)$ & $-0.298(12)$\\ 
$0.4854(31)$ & $0.307(15)$ & $0.307(15)$ & $0.307(15)$ & $ 0.000(1)$ & $-0.307(15)$\\
$0.3795(31)$ & $0.324(17)$ & $0.308(16)$ & $0.330(17)$ & $0.006(1)$ & $-0.319(16)$ \\
$0.2839(33)$ & $0.341(20)$ & $0.309(17)$ & $0.352(21)$ & $0.011(1)$  & $-0.330(19)$ \\
$0.2153(35)$ & $0.352(23)$ & $0.309(17)$ & $0.366(25)$ & $0.014(2)$  & $-0.338(21)$ \\
$0.1384(43)$ & $0.365(29)$ & $0.311(18)$ & $0.383(33)$ & $ 0.018(5)$ & $-0.347(25)$\\
$0.0939(44)$ & $0.380(39)$ & $0.321(22)$ & $0.399(45)$ & $ 0.020(7)$ & $-0.360(32)$
\end{tabular}
\end{ruledtabular}
\vspace{-3pt}
\end{table*}

\begin{table*}[tbp]
\caption{Charge radii of the
 $\Xi^*$ baryons for different $m_\pi^2$ values. Quark sector contributions
 for a single quark of unit charge are included.}
\label{tab:eradXS}
\begin{ruledtabular}
\begin{tabular}{ccccc}
$\mathit m_\pi^2 (\rm GeV^2)$ &  $\mathit s_{\Xi^*} (\rm fm^2)$ &
$\mathit u_{\Xi^*} (\rm fm^2)$ & $\Xi^{*0} (\rm fm^2)$ & $\Xi^{*-}
(\rm fm^2)$  \\
\hline
$0.9960(56)$ & $0.291(16)$   &  $0.227(11)$ & $-0.042(5)$  & $-0.269(14)$ \\
$0.8936(56)$ & $0.289(16)$   &  $0.236(12)$ & $-0.035(5)$  & $-0.271(14)$ \\
$0.7920(55)$ & $0.287(17)$   &  $0.246(13)$ & $-0.027(4)$  & $-0.273(15)$ \\
$0.6920(54)$ & $0.285(17)$ & $0.256(14)$ & $-0.019(3)$ & $-0.275(16)$ \\
$0.6910(35)$ & $0.309(13)$   &  $0.276(11)$ & $-0.022(3)$  & $-0.298(12)$ \\
$0.5925(33)$ & $0.308(14)$   &  $0.291(13)$ & $-0.011(2)$  & $-0.302(14)$ \\ 
$0.4854(31)$ & $0.307(15)$   &  $0.307(15)$ & $ 0.000(0)$  & $-0.307(15)$ \\
$0.3795(31)$ & $0.307(15)$   &  $0.324(16)$ & $0.011(1)$   & $-0.313(16)$ \\
$0.2839(33)$ & $0.308(16)$   &  $0.341(19)$ & $0.022(3)$   & $-0.319(17)$ \\
$0.2153(35)$ & $0.307(16)$   &  $0.351(21)$ & $0.029(4)$   & $-0.322(17)$ \\
$0.1384(43)$ & $0.307(16)$   &  $0.363(24)$ & $0.037(7)$   & $-0.326(19)$ \\
$0.0939(44)$ & $0.308(17)$   &  $0.372(29)$ & $0.043(10)$  & $-0.330(20)$
\end{tabular}
\end{ruledtabular}
\vspace{-3pt}
\end{table*}

\subsection{Magnetic moments}

The magnetic moment is provided by the value of the magnetic form
factor at zero momentum transfer, $Q^2 = 0$,
\begin{equation}
\mu = {\cal{G}}_M(0)\frac{e}{2M_B}\, ,
\label{magmom}
\end{equation} 
in units of the natural magneton, where $M_B$ is the mass of the
baryon.
Since the magnetic form factors must be calculated at a finite value
of momentum transfer, $Q^2$, the magnetic moment must be inferred from
our results, ${\cal{G}}_M(Q^2)$, obtained at the minimum non-vanishing
momentum transfer available on our periodic lattice. 
We choose to scale our results from ${\cal{G}}_M(Q^2)$ to
${\cal{G}}_M(0)$.
We do this by assuming that the $Q^2$ dependence of the electric and
magnetic form factors are similar at the quark masses simulated herein.
This is supported by experiment where the proton ratio
$\frac{{\cal{G}}_M(Q^2)}{\mu \, {\cal{G}}_E(Q^2)} \simeq 1$ for values
of $Q^2$ similar to that probed here.
In this case
\begin{equation}
{\cal{G}}_M(0) =
\frac{{\cal{G}}_M(Q^2)}{{\cal{G}}_E(Q^2)} \, {\cal{G}}_E(0) \, .
\label{pffm}
\end{equation}
The strange and light sectors of hyperons will scale differently, and
therefore we apply Eq.~(\ref{pffm}) to the individual quark sectors
for all the decuplet baryons.
Decuplet baryon properties are then reconstructed via
\begin{equation}
{\cal{G}}^B_M(0) = {\cal{G}}^l_M(0) + {\cal{G}}^s_M(0),
\label{magmombar}
\end{equation}
where $l$ labels the light quarks and $s$ labels the strange
quark, and quark numbers and charges are included.
Similar calculations are performed for the $u$ and $d$ sectors of the
$\Delta$.

\begin{figure}[tbp!]
\begin{center}
  {\includegraphics[height=\hsize,angle=90]{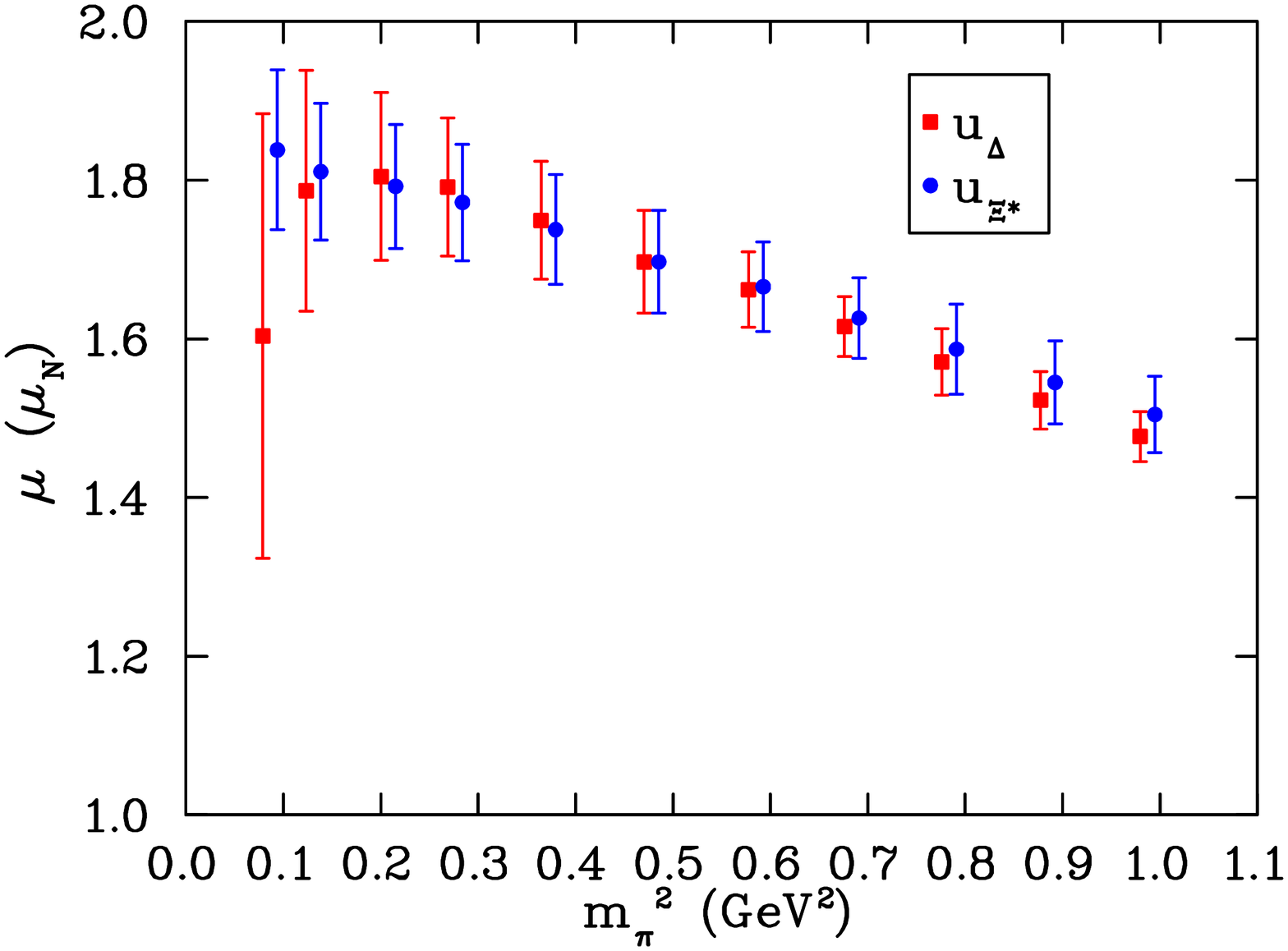}}
\end{center}
\caption{Magnetic moments of a $u$ quark in $\Delta$ and
  $\Xi^*$ as a function of quark mass.  The values for $\Xi^*$ are
  plotted at shifted $m_\pi^2$ for clarity.}
\label{magmomu1}
\end{figure}

\begin{figure}[tbp!]
\begin{center}
  {\includegraphics[height=\hsize,angle=90]{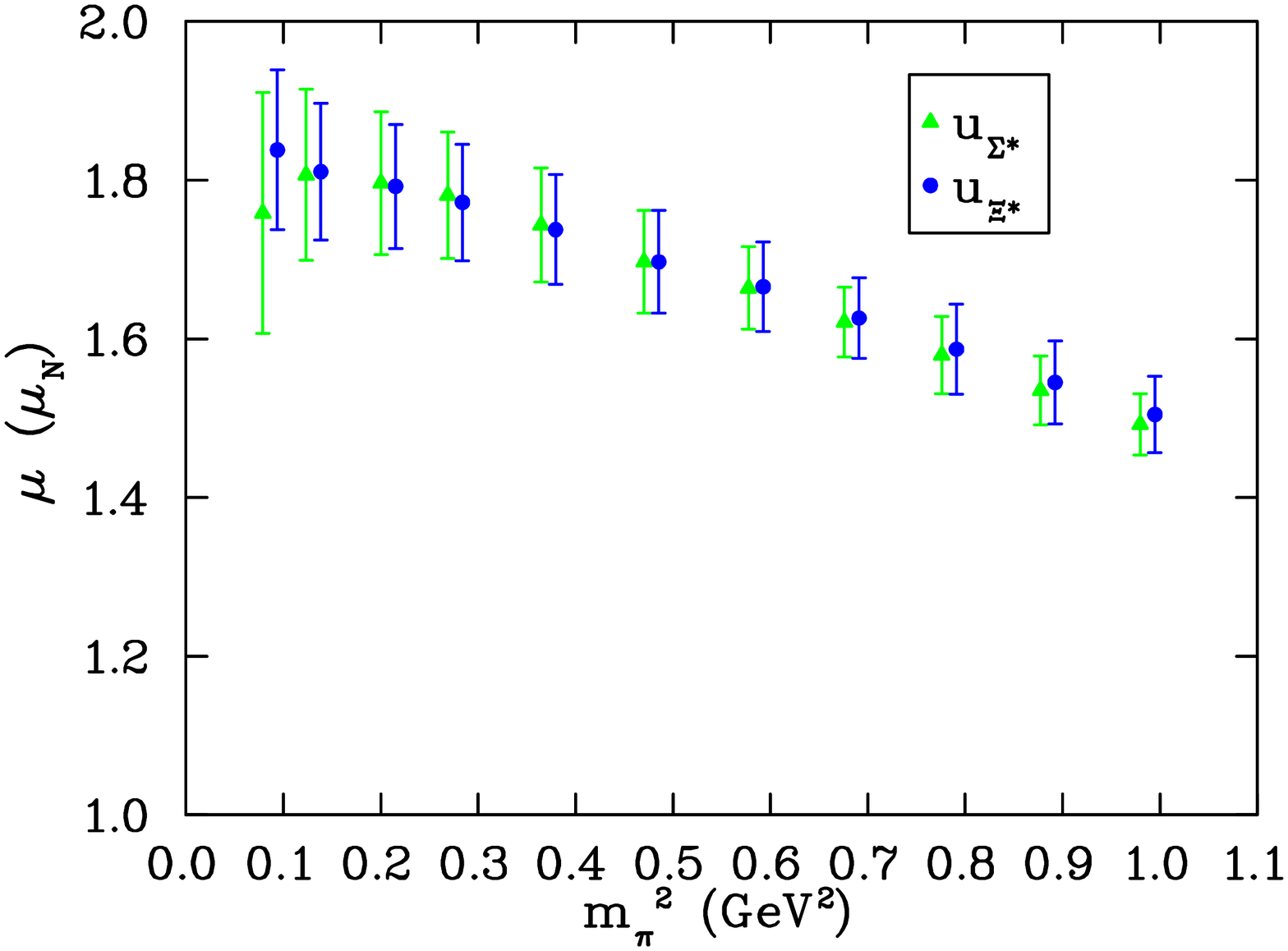}}
\end{center}
\caption{Magnetic moments of a $u$ quark in $\Sigma^*$ and
  $\Xi^*$ as a function of quark mass.  The values for $\Xi^*$ are
  plotted at shifted $m_\pi^2$ for clarity.}
\label{magmomu2}
\end{figure}

\begin{figure}[h]
\begin{center}
  {\includegraphics[height=\hsize,angle=90]{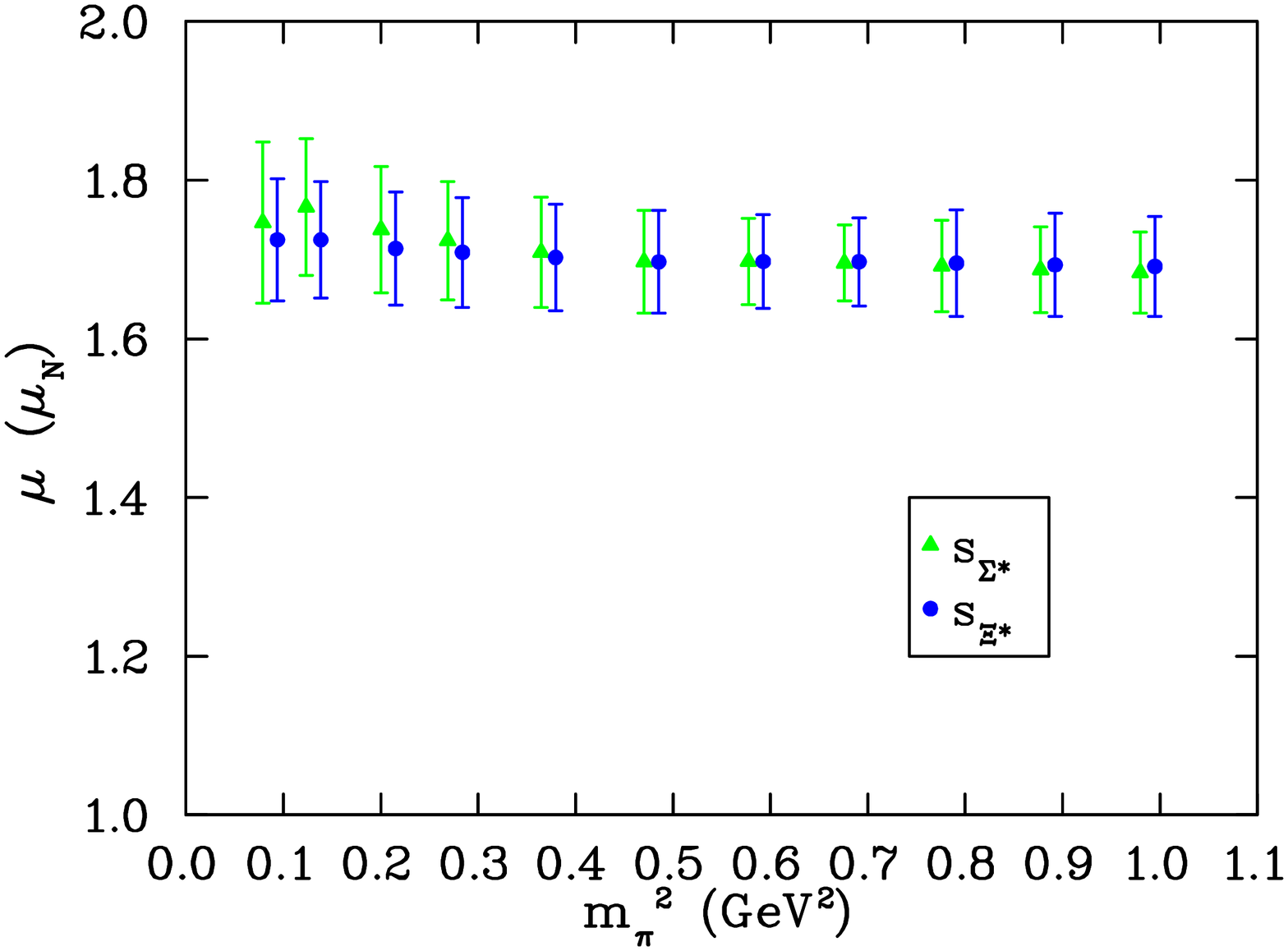}}
\end{center}
\caption{Magnetic moments of an $s$ quark in $\Sigma^*$ and
  $\Xi^*$ for different quark masses.  The values for $\Xi^*$ are
  plotted at shifted $m_\pi^2$ for clarity.}
\label{magmoms}
\end{figure}

In Figs.~\ref{magmomu1}, \ref{magmomu2} and \ref{magmoms}, we display
the quark sector contributions to the decuplet magnetic moments, which
are equal in the $SU(3)_{\rm flavor}$ limit (sixth quark mass) for
single quarks of unit charge.
Here we observe that the quark contributions in the $\Sigma^{*}$ and
$\Xi^{*}$ are very similar, which provides evidence that there is
little environmental sensitivity. The chiral behavior of the
$u_{\Delta}$ is also very interesting.

\subsection{Effective Moments}

In order to compare the decuplet moments with the octet moments from
our previous lattice calculation~\cite{Boinepalli:2006xd}, we
construct so-called effective moments by equating the lattice quark
sector contributions to the same sector of the SU(6)-magnetic moment
formula derived from SU(6)-spin-flavor symmetry wave functions.

The simple quark model formula from SU(6)-spin-flavor symmetry gives
the magnetic moment of the proton as 
\begin{equation}
\mu_p = \frac{4}{3}\mu_u - \frac{1}{3}\mu_d\, ,
\label{eq:mup}
\end{equation}
where $\mu_u$ and $\mu_d$ are the intrinsic magnetic moments of the
doubly
represented $u$ and singly represented $d$ constituent quarks,
respectively, per single quark. 
This can be generalized for any baryon with two doubly represented
quarks $D$ and one singly represented quark $S$.
Focusing on the proton and using the charge factors of the doubly
represented and singly represented quarks as 2/3 and (-1/3)
respectively, we define effective moments for single quarks of unit
charge, $\mu^{\rm Eff}$, by
\begin{equation}
  \mu_u = \frac{2}{3} \mu_D^{\rm Eff} \,,\, \mu_d = -\frac{1}{3}
  \mu_S^{\rm Eff} \,,
\end{equation}
such that Eq.~(\ref{eq:mup}) becomes
\begin{equation}
\mu_p = \left(\frac{4}{3}\right)\left(\frac{2}{3}\right)\mu^{\rm
  Eff}_D -  \left(\frac{1}{3}\right) \left(-\frac{1}{3}\right)\mu^{\rm
  Eff}_S \, ,
\label{Magmom_SQM}
\end{equation}
where the charge factors within $\mu_u$ and $\mu_d$ are now explicit.

On the lattice we calculate the baryon magnetic moments from the
individual quark sector contributions using 
\begin{equation}
\mu_p = 2 \left(\frac{2}{3}\right)\mu_D^{\rm Latt} + 1
\left(-\frac{1}{3}\right)\mu_S^{\rm Latt} \,.
\label{Magmom_Latt}
\end{equation}

In the above equation the factors 2 and 1 in the first and second term
account for the number of doubly and singly represented quarks, 
while the charges are indicated in parentheses.
Equating quark sectors in Eqs.~(\ref{Magmom_SQM}) and
~(\ref{Magmom_Latt}) yields for the effective moments,
\begin{eqnarray}
\mu_S^{\rm Eff}& = & -3 \mu_S^{\rm Latt} \, , \nonumber \\
\mu_D^{\rm Eff}& = & \frac{3}{2} \mu_D^{\rm Latt}\, \,.
\end{eqnarray}
One could also define a constituent quark mass via
\begin{eqnarray}
\mu_S^{\rm Eff}& = &\frac{e}{2 m_S^{\rm Eff}} \, , \nonumber \\
\mu_D^{\rm Eff}& = &\frac{e}{2 m_D^{\rm Eff}}  \, \,,
\end{eqnarray}
revealing that $\mu_S^{\rm Eff} \simeq \mu_D^{\rm Eff}$ in most
constituent quark models.

For the decuplet baryons, the magnetic moment is the sum of the
individual constituent-quark contributions.
Hence Eq.~(\ref{Magmom_SQM}) for the $\Delta$ baryons becomes
\begin{equation}
\mu_{\Delta^+} = 2 \left(\frac{2}{3}\right)\mu_D^{\rm Eff} + 1
\left(-\frac{1}{3}\right)\mu_S^{\rm Eff} \, .
\end{equation} 
On the lattice this is exactly the equation we use to build the
decuplet baryon moments from the quark sector contributions.
Therefore, the quark level magnetic moments that we calculate are the
effective moments of the quarks for both the doubly and singly
represented quarks, {\it i.e.,}
\begin{equation}
\mu^{\rm Eff} =  \mu^{\rm Latt} \, .\\
\end{equation}

\begin{figure}[tbp!]
\begin{center}
  {\includegraphics[height=6cm,angle=0]{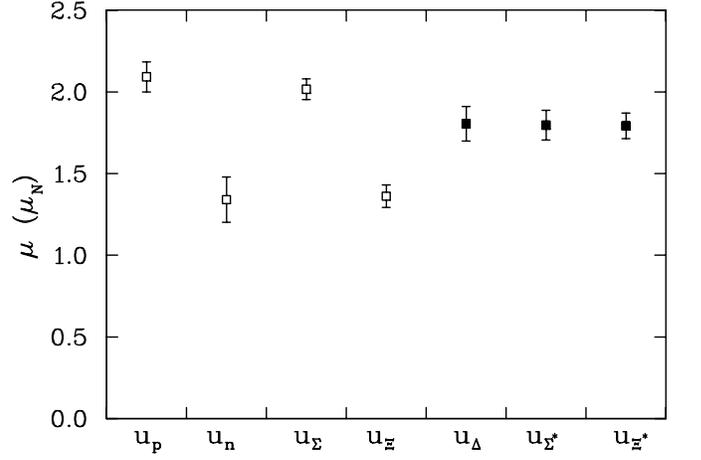}}
\end{center}
\caption{Effective moments of the $u$ quark sector in the octet and
  the decuplet baryons at the ninth quark mass where $m_\pi^2 =
  0.215(4)\ {\rm GeV}^2$.}
\label{Mom_quark_Dble}
\end{figure}

\begin{figure}[tbp!]
\begin{center}
  {\includegraphics[height=6cm,angle=0]{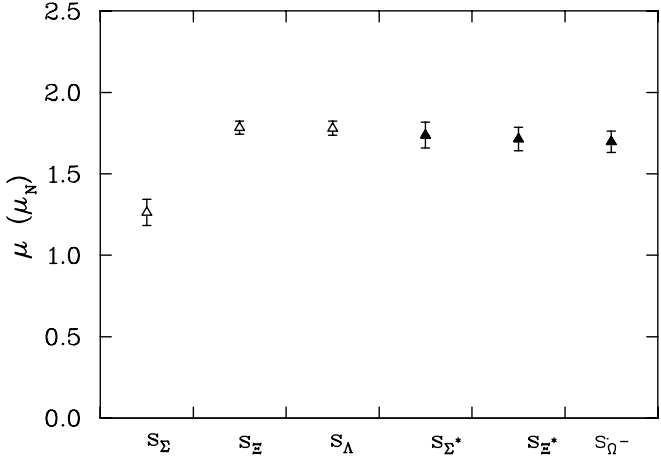}}
\end{center}
\caption{Effective moments of the $s$ quark sector in the octet and
  the decuplet baryons at the ninth quark mass where $m_\pi^2 =
  0.215(4)\ {\rm GeV}^2$.}
\label{Mom_quark_Sing}
\end{figure}

In Figs.~\ref{Mom_quark_Dble} and ~\ref{Mom_quark_Sing} we plot the
effective moments of the $u$ and $s$ quark sectors of the octet and
decuplet baryons at the ninth quark mass.
Here we observe that the quarks in the octet baryons show far more
environmental sensitivity than their counterparts in the decuplet
baryons.

\begin{figure}[tbp!]
\begin{center}
  {\includegraphics[height=\hsize,angle=90]{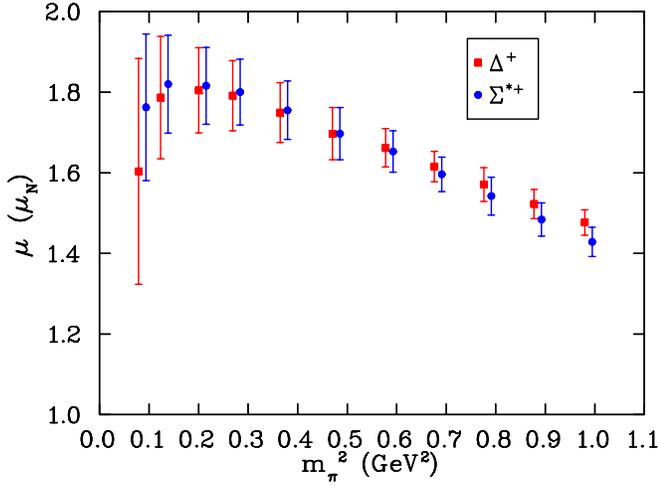}}
\end{center}
\caption{Magnetic moments of  $\Delta^+$ and
  $\Sigma^{*+}$ at different quark masses. The values for
  $\Sigma^{*+}$ are plotted at shifted $m_\pi^2$ for clarity.}
\label{magmomp}
\end{figure}

\begin{figure}[tbp!]
\begin{center}
  {\includegraphics[height=\hsize,angle=90]{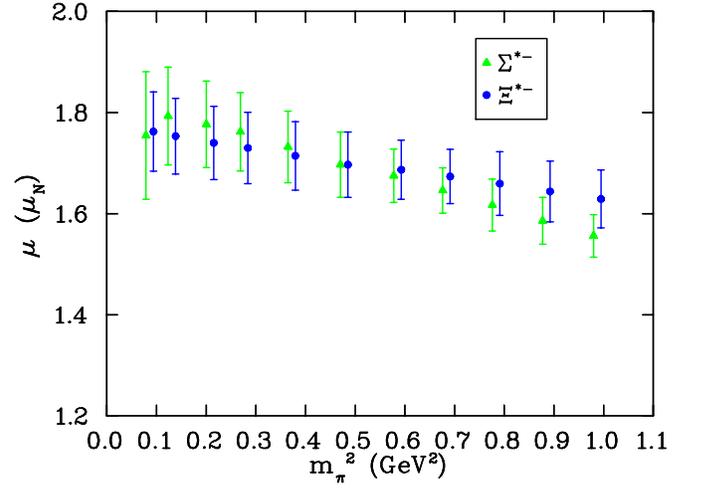}}
\end{center}
\caption{Magnetic moments (magnitude) of  $\Sigma^{*-}$
  and $\Xi^{*-}$ at different quark masses. The values for $\Xi^{*-}$
  are plotted at shifted $m_\pi^2$ for clarity. }
\label{magmomm}
\end{figure}

\begin{figure}[tbp!]
\begin{center}
  {\includegraphics[height=\hsize,angle=90]{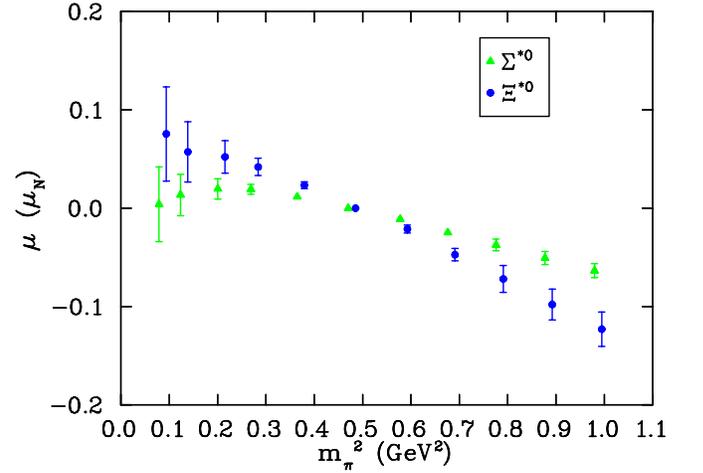}}
\end{center}
\caption{Magnetic moments of  $\Sigma^{*0}$
  and $\Xi^{*0}$ at different quark masses.  The values for $\Xi^{*0}$
  are plotted at shifted $m_\pi^2$ for clarity. }
\label{magmomn}
\end{figure}

The baryon magnetic moments are plotted in Figs.~\ref{magmomp},
\ref{magmomm} and \ref{magmomn}.
For the magnetic moment of $\Omega^-$ we take the value of
$\mu_{\Delta^-}$ at the SU(3) limit viz., $-1.697 \pm 0.065~\mu_N$, which
is smaller than the value given by the Particle Data Group ($-2.02 \pm
0.05~\mu_N$). 

This discrepancy could be partly due to the fact that the mass of
$\Omega^-$ from our lattice calculation
($1.73 \pm 0.012~{\rm GeV}$) is slightly larger than the
experimentally measured value ($1.67~{\rm GeV}$).

Another reason for this discrepancy is likely to reside in the absence
of $K\Xi$ loops in the virtual decay of $\Omega^-$.
The virtual transition $\Omega \rightarrow \Xi K$
requires the presence of a light sea-quark flavor, while in QQCD,
there is only a heavy valence strange quark. 
In reality this would provide an important contribution, since $\Xi$
is a lower mass state than $\Omega^-$.
The predominant contribution is $\Omega^0 \rightarrow \Xi^0 K^-$ with
the $z$-component of angular momentum in the positive direction. This
process will act to enhance the magnitude of the negative moment.
The absence of such loops in QQCD represents missing physics and
causes the discrepancy from the values of full QCD.
This is certainly a good place to search for dynamical sea quark
effects.

\begin{figure}[tbp!]
\begin{center}
  {\includegraphics[height=\hsize,angle=90]{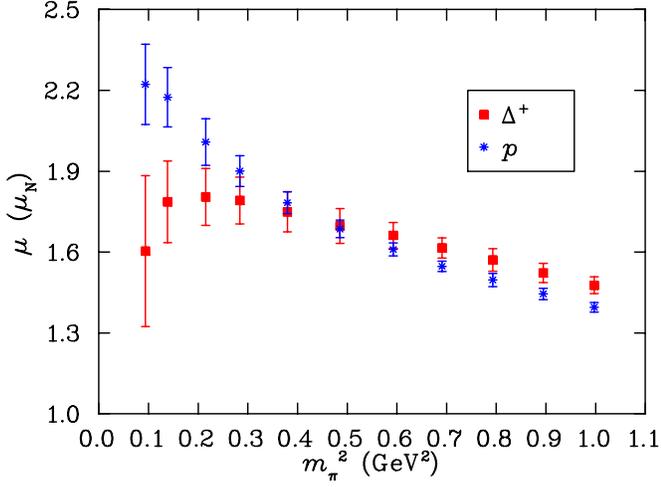}}
\end{center}
\caption{Magnetic moments of  $\Delta^+$ and the proton
  at different quark masses. }
\label{magmomPD}
\end{figure}

Figure~\ref{magmomPD} compares the magnetic moment of the $\Delta^+$
with our earlier result for the proton magnetic moment on the same set
of gauge field configurations~\cite{Boinepalli:2006xd}.
A simple quark model predicts that the proton and the $\Delta^+$ have
equal magnetic moments. 
However the interplay between the different pion-loop contributions to
the $\Delta^+$ magnetic moments indicate that the proton magnetic
moment should be greater than that of the $\Delta^+$ in Full
QCD~\cite{Cloet:2003jm}.

The presence of the $\Delta \to N \pi$ decay channel is particularly
important for the quark mass dependence of $\Delta$
properties~\cite{Leinweber:2003ux}.  
Rapid curvature associated with non-analytic behavior is shifted to
larger pion masses near the $N$-$\Delta$ mass splitting, $m_\pi \sim
M_\Delta - M_N$.  
As described below, quenched-QCD decay-channel contributions come with
a sign opposite to that of full QCD.  
This artifact holds tremendous promise for revealing unmistakable
signatures of the quenched meson cloud.

\begin{figure}[tb]
\begin{center}
{\includegraphics[height=\hsize,angle=90]{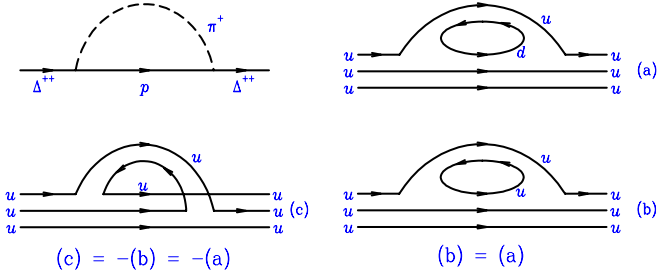}}
\caption{Quark-flow diagrams for meson-cloud contributions to 
  $\Delta^{++}$ in full QCD.} 
\label{deltaFlow}
\end{center}
\end{figure}

The change in sign for the decay-channel contributions is easily
understood through the consideration of the quark flow diagrams in
Fig.~\ref{deltaFlow}, illustrating the meson-cloud contributions to
the $\Delta^{++}$ resonance in full QCD.
Quark flow diagram (a) corresponds to the hadronic process described
in the top left diagram of Fig.~\ref{deltaFlow}.
Since QCD is flavor-blind, the process illustrated in diagram (b) is
equivalent to diagram (a) provided the masses of the $u$ and $d$
quarks are taken to be equal.
On its own, diagram (b) describes the decay of the $\Delta^{++}$ to a
doubly-charged $uuu$ ``proton,'' which we denote $p^{++}$.  
Of course, such states do not exist in full QCD and diagram (c)
provides a contribution which is exactly equal but opposite in sign to
diagram (b) when the intermediate state is a $uuu$ proton.
Upon quenching the theory, both diagrams (a) and (b) are eliminated,
leaving only diagram (c).  
Hence the physics of the $\Delta \to N \pi$ decay is present in the
quenched approximation \cite{Labrenz:1996jy} but its contribution has
the wrong sign. This signature of quenched chiral physics is manifest
in our results.

All the decuplet baryons show, to some extent, a turn over of the
magnetic moment in the low quark mass region just above the opening of
the $N \pi$ decay channel.
The magnitude of the turn over is dampened by the presence of a strange
quark, which is seen by the fact that $\Sigma^*$ has a smaller turn
over than $\Delta$. The $\Xi$ baryons with two $s$ quarks only admit
Kaon loops and do not display a turn over, further clarifying a link
to chiral physics.

Magnetic moments of the decuplet baryons are listed in
Tables~\ref{tab:magmomD} to~\ref{tab:magmomXS}.

\begin{table*}[tbp]
\caption{ Magnetic moments of the $\Delta^+$ in nuclear magnetons for
  different $m_\pi^2$ in $\rm{GeV}^2$. Quark sector contributions for
  single quarks of unit charge are also provided.
  The magnetic moment of $\Delta^-$ is
  equal in magnitude 
  to that of $\Delta^+$ with a negative sign and that of $\Delta^0$ is
  0. Charge symmetry also requires that the $\Delta^{++}$ has a magnetic
  moment twice that of 
  $\Delta^+$. The magnetic moment of $\Omega^-$ is that of $\Delta^-$
  at the $SU(3)_{\rm flavor}$ limit where $m_\pi^2 =
  0.4854(31)~\rm{GeV}^2$, and takes the value $-1.697(65)\ \mu_N$.} 
\label{tab:magmomD}
\begin{ruledtabular}
\begin{tabular}{cccc}
$\mathit m_\pi^2$ &  $\mathit u_{\Delta}~(\mu_N)$ & $\mathit
d_{\Delta}~(\mu_N)$ & $\Delta^+~(\mu_N)$ \\
\hline
$0.9960(56)$ & $1.466(31)$  & $1.466(31)$  & $1.466(31)$ \\
$0.8936(56)$ & $1.512(36)$  & $1.512(36)$  &$1.512(36)$  \\
$0.7920(55)$ & $1.559(42)$  & $1.559(42)$  & $1.559(42)$ \\ 
$0.6920(54)$ & $1.604(50)$ & $1.604(50)$ & $1.604(50)$ \\
$0.6910(35)$ & $1.615(38)$  & $1.615(38)$  & $1.615(38)$ \\
$0.5925(33)$ & $1.662(48)$  & $1.662(48)$  & $1.662(48)$ \\ 
$0.4854(31)$ & $1.697(65)$  & $1.697(65)$  & $1.697(65)$ \\
$0.3795(31)$ & $1.749(74)$  & $1.749(74)$  & $1.749(74)$ \\
$0.2839(33)$ & $1.792(87)$  & $1.792(87)$  & $1.792(87)$ \\
$0.2153(35)$ & $1.80(11)$   & $1.80(11)$   & $1.80(11)$  \\
$0.1384(43)$ & $1.79(15)$   & $1.79(15)$   & $1.79(15)$  \\
$0.0939(44)$ & $1.60(28)$   & $1.60(28)$   & $1.60(28)$
\end{tabular}
\end{ruledtabular}
\end{table*}

\begin{table*}[tbp]
\caption{ Magnetic moments of $\Sigma^*$, $\Sigma^{*0}$ and
  $\Sigma^{*-}$ in nuclear magnetons with quark sector
  contributions for a single quark of unit charge at different
  $m_\pi^2$ values.} 
\label{tab:magmomSS}
\begin{ruledtabular}
\begin{tabular}{cccccc}
$\mathit m_\pi^2$ &  $\mathit u_{\Sigma^*}~(\mu_N)$ & $\mathit
s_{\Sigma^*}~(\mu_N)$ & $\Sigma^{*+}~(\mu_N)$ & $\Sigma^{*0}~(\mu_N)$
& $\Sigma^{*-}~(\mu_N)$ \\
\hline
$0.9960(56)$ & $1.482(38)$ & $1.671(51)$ & $1.418(36)$ & $-0.063(7)$  & $-1.545(42)$ \\
$0.8936(56)$ & $1.524(43)$ & $1.675(54)$ & $1.474(41)$ & $-0.050(7)$  & $-1.574(46)$\\ 
$0.7920(55)$ & $1.568(48)$ & $1.680(57)$ & $1.531(47)$ & $-0.037(6)$  & $-1.605(51)$\\
$0.6920(54)$ & $1.609(55)$ & $1.683(62)$ & $1.585(54)$ & $-0.025(5)$ & $-1.634(57)$ \\
$0.6910(35)$ & $1.621(44)$ & $1.696(48)$ & $1.596(43)$ & $-0.025(3)$  & $-1.646(45)$ \\
$0.5925(33)$ & $1.664(52)$ & $1.698(54)$ & $1.653(51)$ & $-0.011(2)$  & $-1.675(53)$\\ 
$0.4854(31)$ & $1.697(65)$ & $1.697(65)$ & $1.697(65)$ & $ 0.000(0)$  & $-1.697(65)$\\
$0.3795(31)$ & $1.744(72)$ & $1.709(69)$ & $1.755(72)$ & $0.012(2)$   & $-1.732(71)$ \\
$0.2839(33)$ & $1.781(80)$ & $1.724(75)$ & $1.800(82)$ & $0.019(5)$   & $-1.762(78)$ \\
$0.2153(35)$ & $1.796(90)$ & $1.738(80)$ & $1.816(95)$ & $0.020(10)$  & $-1.777(85)$ \\
$0.1384(43)$ & $1.81(11)$ & $1.766(86)$ & $1.82(12)$  & $ 0.013(21)$ & $-1.793(97)$\\
$0.0939(44)$ & $1.76(15)$ & $1.75(10)$  & $1.76(18)$  & $ 0.004(38)$ & $-1.75(13)$
\end{tabular}
\end{ruledtabular}
\end{table*}

\begin{table*}[tbp]
\caption{ Magnetic moments of $\Xi^{*0}$ and $\Xi^{*-}$ in nuclear
  magnetons with quark sector contributions for a single quark of unit
  charge at different
  $m_\pi^2$ values.}
\label{tab:magmomXS}
\begin{ruledtabular}
\begin{tabular}{ccccc}
$\mathit m_\pi^2$ &  $\mathit s_{\Xi^*}~(\mu_N)$ & $\mathit
u_{\Xi^*}~(\mu_N)$ & $\Xi^{*0}~(\mu_N)$ & $\Xi^{*-}~(\mu_N)$ \\
\hline
$0.9960(56)$ & $1.681(63)$ & $1.494(48)$  & $-0.124(18)$  & $-1.619(57)$ \\
$0.8936(56)$ & $1.683(65)$ & $1.534(52)$  & $-0.099(16)$  & $-1.633(60)$ \\
$0.7920(55)$ & $1.685(67)$ & $1.576(56)$  & $-0.073(14)$  & $-1.649(63)$ \\
$0.6920(54)$ & $1.687(69)$ & $1.615(62)$  & $-0.048(11)$  & $-1.663(67)$ \\
$0.6910(35)$ & $1.697(56)$ & $1.626(51)$  & $-0.047(6)$   & $-1.674(54)$\\
$0.5925(33)$ & $1.698(59)$ & $1.666(57)$  & $-0.021(4)$   & $-1.687(58)$\\ 
$0.4854(31)$ & $1.697(65)$ & $1.697(65)$  & $ 0.000(0)$   & $-1.697(65)$\\
$0.3795(31)$ & $1.703(67)$ & $1.738(69)$  & $0.023(4)$    & $-1.714(68)$ \\
$0.2839(33)$ & $1.709(69)$ & $1.772(73)$  & $0.042(9)$    & $-1.730(70)$\\
$0.2153(35)$ & $1.714(71)$ & $1.792(78)$  & $0.052(16)$   & $-1.740(73)$ \\
$0.1384(43)$ & $1.725(73)$ & $1.811(86)$  & $0.057(31)$   & $-1.753(75)$ \\
$0.0939(44)$ & $1.725(77)$ & $1.84(10)$   & $0.076(48)$  & $-1.763(79)$
\end{tabular}
\end{ruledtabular}
\end{table*}

\clearpage

\subsection{Electric Quadrupole Form Factors}

\begin{figure}[tbp!]
\begin{center}
  {\includegraphics[height=\hsize,angle=90]{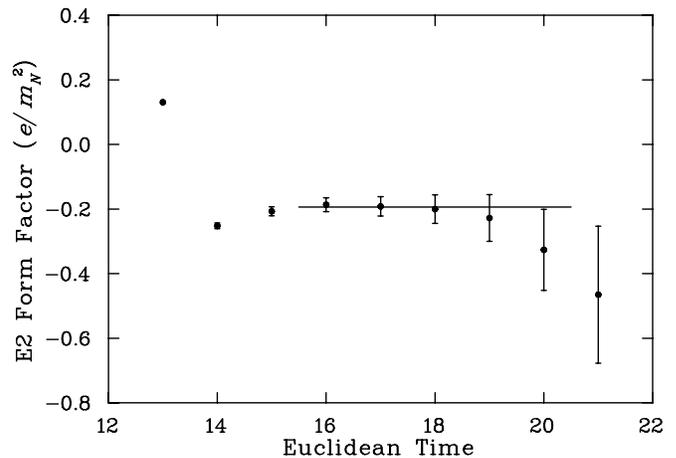}}
\end{center}
\caption{$E2$ electric form factor of the $u$ or $d$ quark sector of the
  $\Delta $ at the $SU(3)_{\rm flavor}$ limit as a function of time. }
\label{cfe2K3}
\end{figure}

\begin{figure}[tbp!]
\begin{center}
  {\includegraphics[height=\hsize,angle=90]{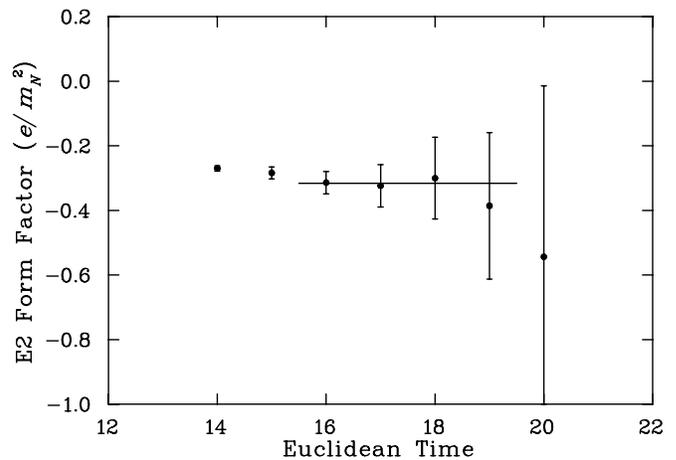}}
\end{center}
\caption{$E2$ electric form factor of the $u$ or $d$ quark sector 
 of the $\Delta $ as a function of time at the ninth quark mass where
 $m_\pi^2 = 0.215(4) {\rm GeV^2}$. }
\label{cfe2K6}
\end{figure}

The $E2$ form factors of the spin-3/2 decuplet baryons provide
interesting information about the distribution of charge and its
deviation from spherical symmetry.
In Fig.~\ref{cfe2K3} we show the correlation function proportional to
the $E2$ quadrupole form factor of the 
$u$ quark in $\Delta$ at the $SU(3)_{\rm flavor}$ limit, in units
of $e/M_N^2$, as a function of Euclidean time.
Figure~\ref{cfe2K6} indicates the quadrupole form factor of the $u$
quark in $\Delta$ at the ninth quark mass.  
Here the employment of the splittings technique facilitates the
extraction of the signal.
In both cases a nontrivial result is obtained.
As mentioned in Sec.~\ref{subsec:3ptFn}, we consider the symmetry of
the last two terms in Eq.~(\ref{ge2_mod}) as the deciding factor in
selecting the upper limit of the fit-window.

The quark sector contributions to the form factors in units of
$e/M_N^2$ of all the decuplet members are indicated in
Tables~\ref{tab:E2D} to \ref{tab:E2XS}.
For an axially deformed object the quadrupole form factor is related
to the charge distribution in the Breit frame
through~\cite{Leinweber:1992hy} 
\begin{equation}
{\mathcal{G}}_{E2}(0) =M_B^2 \int d^3
  r\overline{\psi}(r)(3z^2-r^2)\psi(r)\, ,
\end{equation}
where $3z^2 - r^2$ is the standard operator used for quadrupole
moments. 
A positive quadrupole form factor for a positively charged baryon
indicates a prolate charge distribution, while a negative quadrupole
form factor indicates an oblate charge distribution.
In non-relativistic models, the $E2$ form factor vanishes unless some
configuration mixing of higher orbital-angular momentum states is
included in the baryon ground state.

The $E2$ form factors of the charged decuplet baryons in units of
${\rm fm^2}$ for different values of $m_\pi^2$ are listed in
table~\ref{tab:barE2}.
The $E2$ form factor of the $\Delta^0$ is identically equal to zero
and for the other neutral baryons, it is close to zero.
The results for the charged decuplet baryons are non-zero, indicating
that they have a deformed shape.

\begin{figure}
\begin{center}
  {\includegraphics[height=\hsize,angle=90]{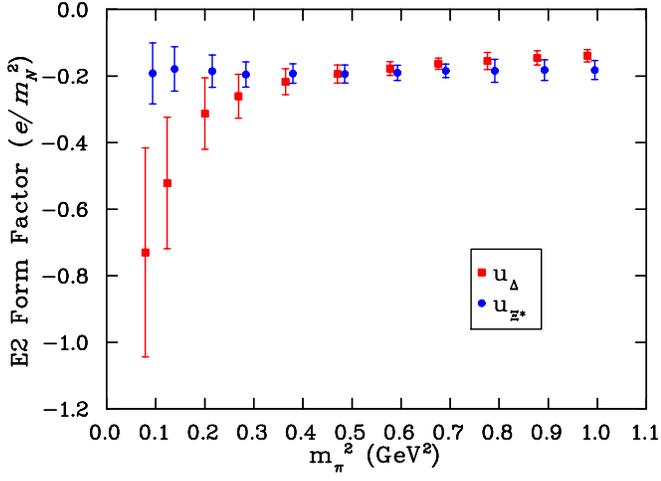}}
\end{center}
\caption{$E2$ form factor contributions from the $u$ quark sectors of
  the $\Delta$ and $\Xi^*$.}
\label{ge2u.dx}
\end{figure}

\begin{figure}
\begin{center}
  {\includegraphics[height=\hsize,angle=90]{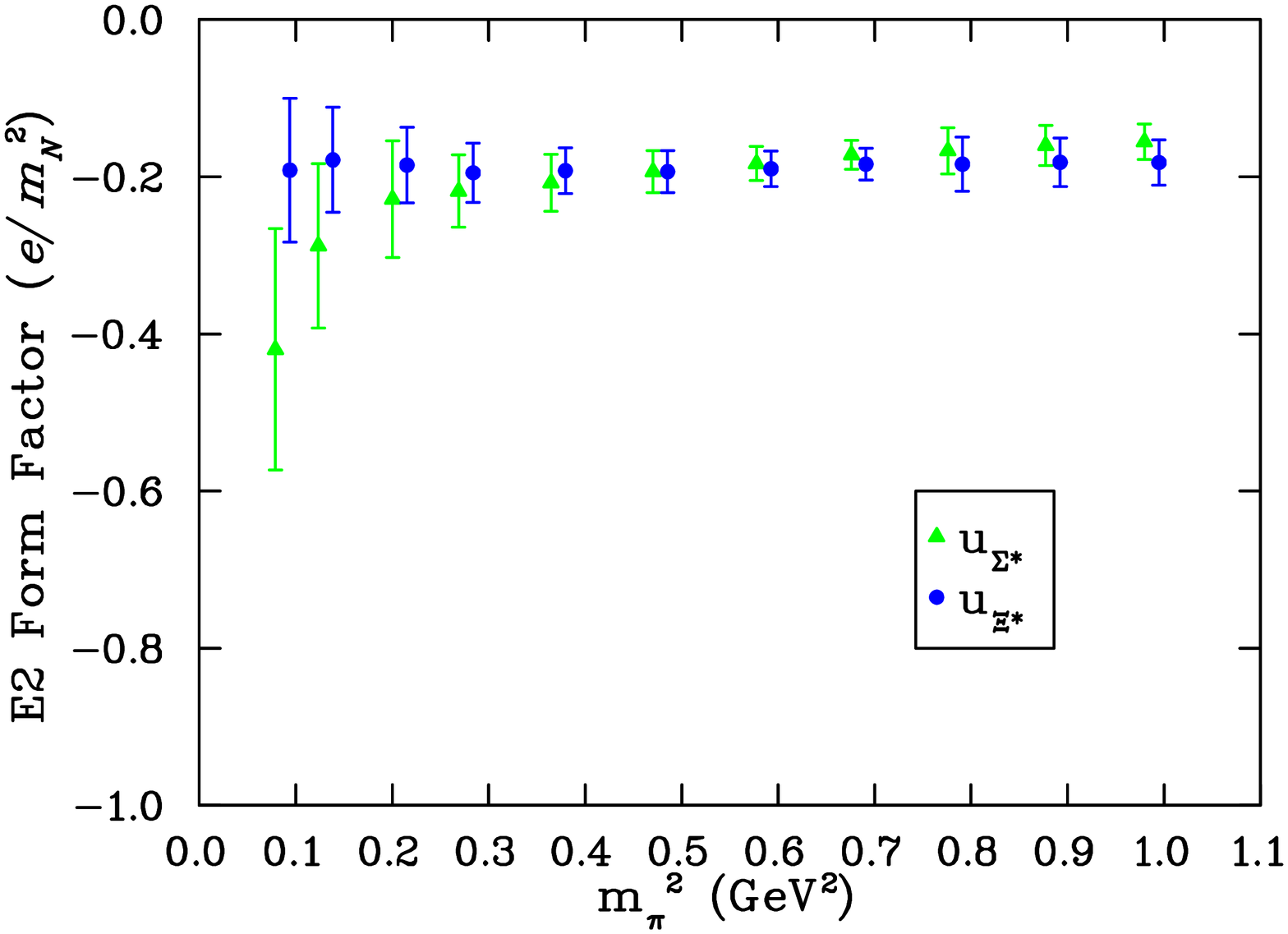}}
\end{center}
\caption{$E2$ form factor contributions from the $u$ quark sectors of
  the $\Sigma^*$ and $\Xi^*$.}
\label{ge2u.sx}
\end{figure}

\begin{figure}
\begin{center}
  {\includegraphics[height=\hsize,angle=90]{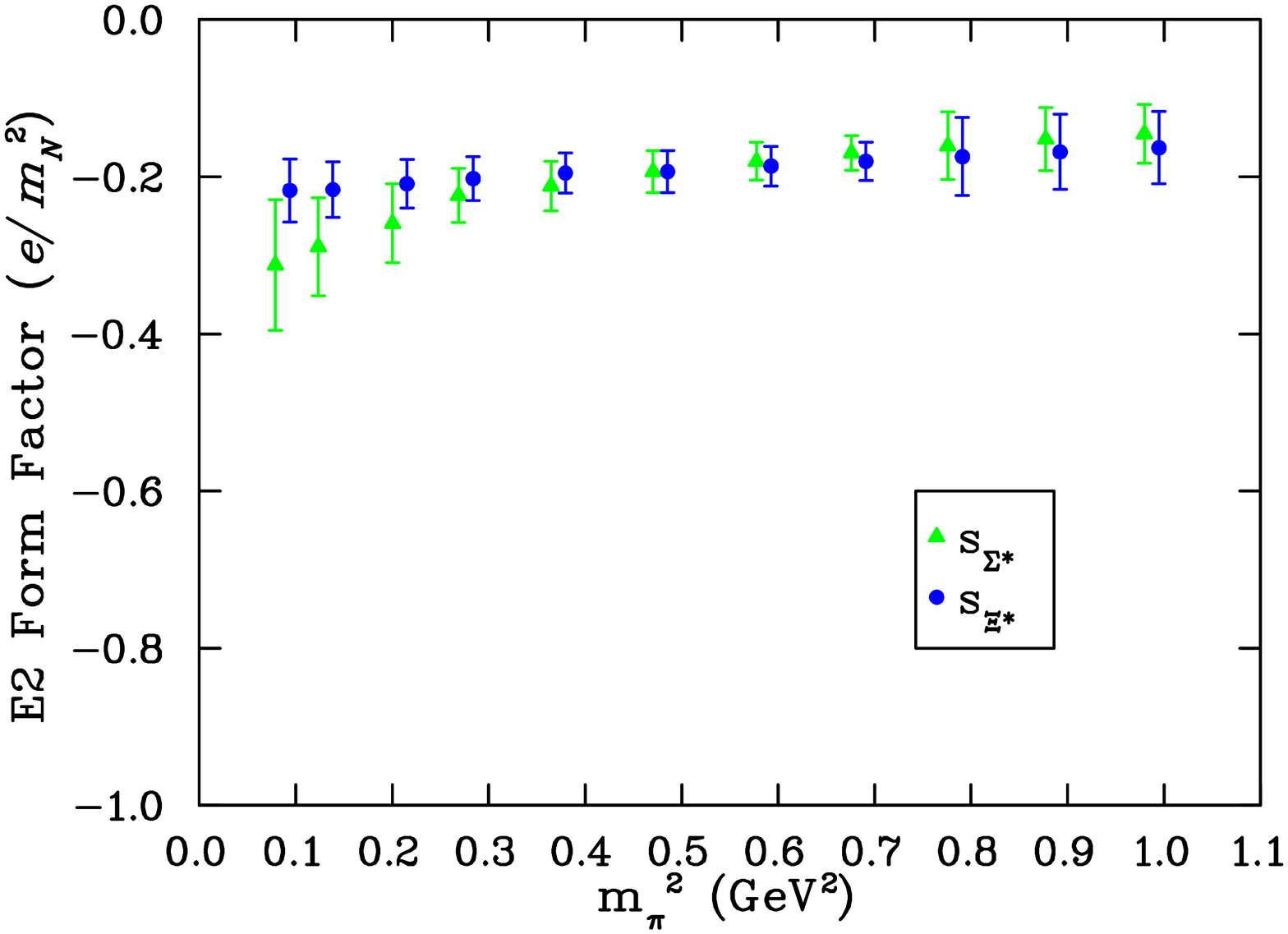}}
\end{center}
\caption{$E2$ form factor contributions from the $s$ quark sectors of
  the $\Sigma^*$ and $\Xi^*$.}
\label{ge2s.sx}
\end{figure}

\begin{figure}
\begin{center}
  {\includegraphics[height=\hsize,angle=90]{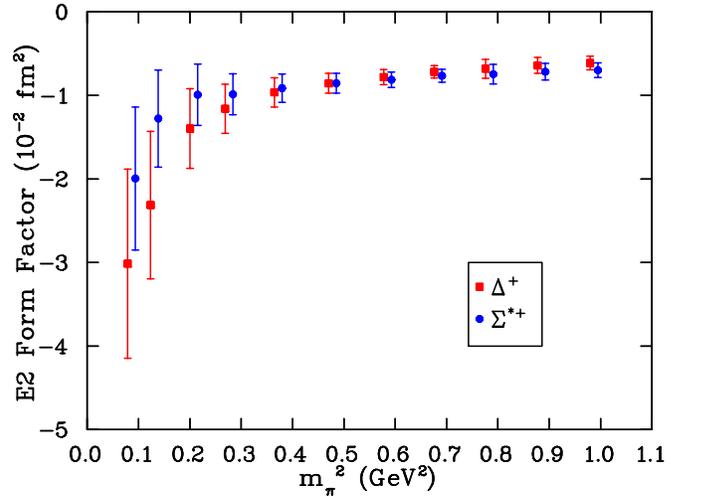}}
\end{center}
\caption{Values of the $E2$ form factors in units of
  $10^{-2}\mathrm{fm}^2$ for the 
  $\Delta^+$ and 
  $\Sigma^{*+}$ at different quark masses. The values for
  $\Delta^{+}$ are plotted at shifted $m_\pi^2$ for clarity.}
\label{ge2p}
\end{figure}

\begin{figure}
\begin{center}
  {\includegraphics[height=\hsize,angle=90]{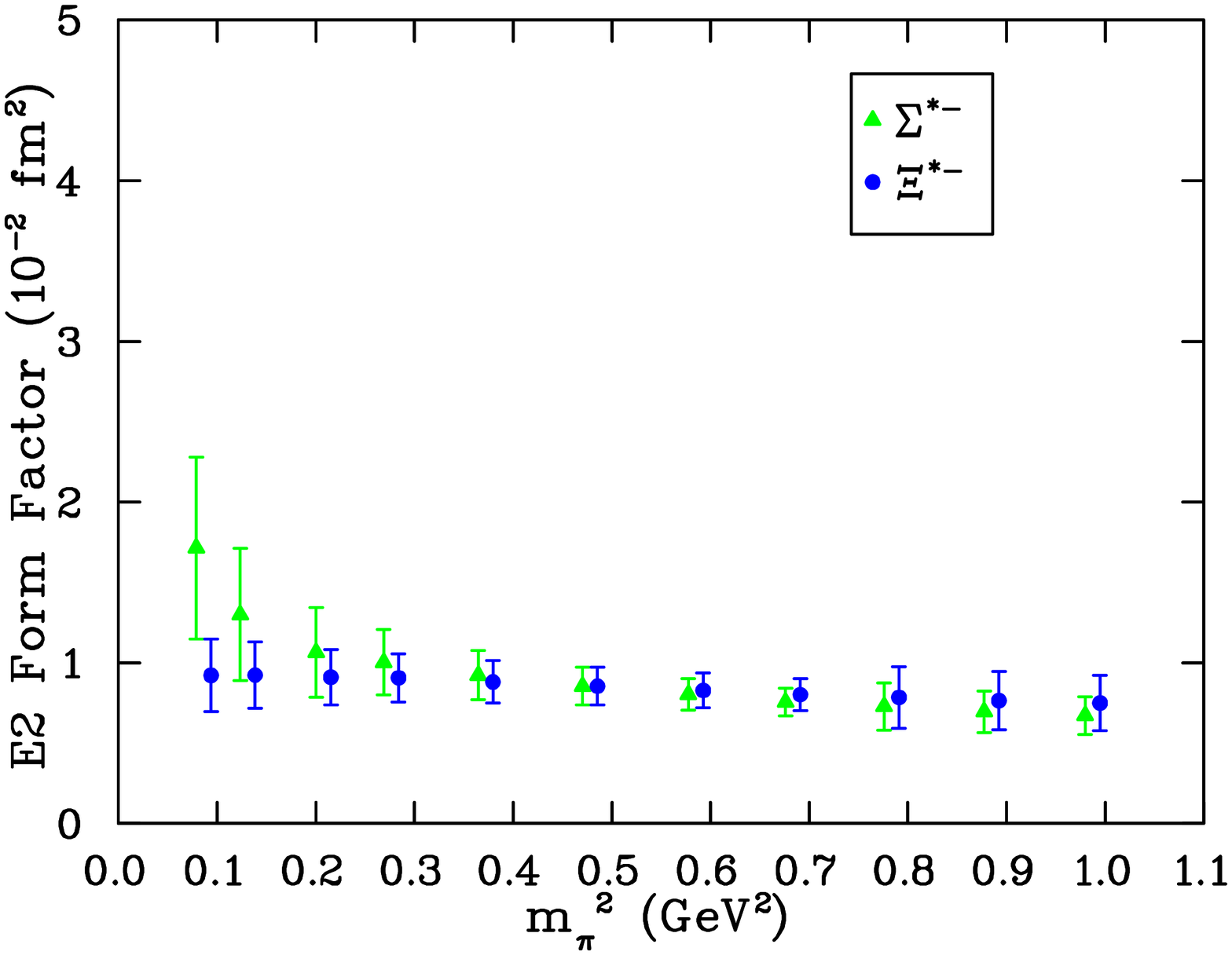}}
\end{center}
\caption{$E2$ form factors in units of $10^{-2}\mathrm{fm}^2$ for the
  $\Sigma^{*-}$ 
  and $\Xi^{*-}$ at different quark masses. The values for $\Sigma^{*-}$
  are plotted at shifted $m_\pi^2$ for clarity. }
\label{ge2m}
\end{figure}

\begin{figure}
\begin{center}
  {\includegraphics[height=\hsize,angle=90]{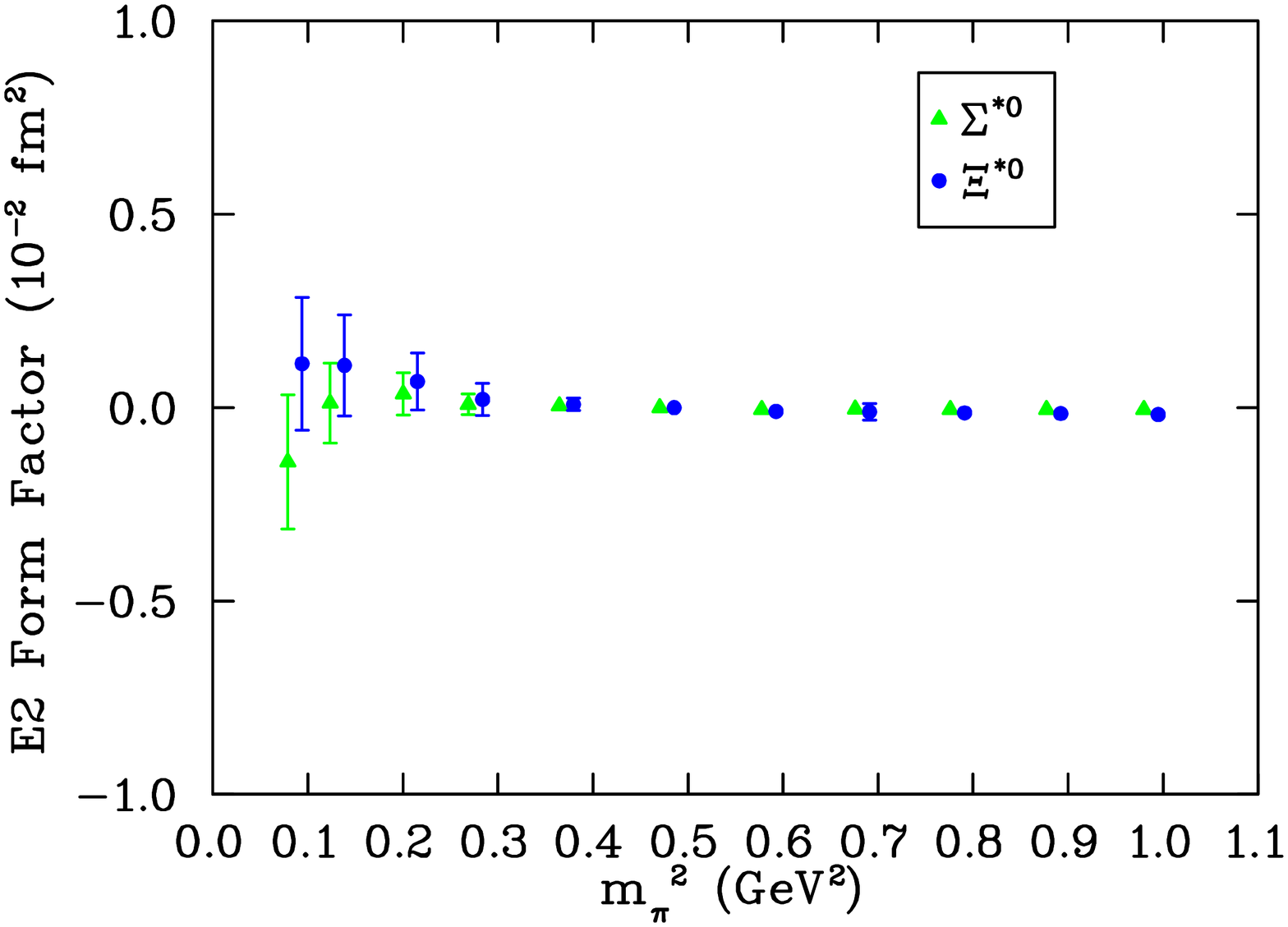}}
\end{center}
\caption{$E2$ form factors in units of $10^{-2}\mathrm{fm}^2$ for the
  $\Sigma^{*0}$ 
  and $\Xi^{*0}$ at different quark masses.  The values for $\Sigma^{*0}$
  are plotted at shifted $m_\pi^2$ for clarity. }
\label{ge2n}
\end{figure}

The quark sector contributions to the $E2$ form factors are shown in
Figs.~\ref{ge2u.dx}~to~\ref{ge2s.sx}.
Once again, the importance of chiral physics is manifest in these
results. A significant enhancement of the magnitude of the light-quark
sector contribution to the $E2$ form factor is observed in the
$\Delta$ as the opening of the $N \pi$ decay channel is approached. A
similar effect is seen, to a lesser extent, in the $\Sigma^*$, while
for the $\Xi^-$ no chiral curvature is observed due to the two $s$
quarks admitting only Kaon loops, as discussed in the previous
section.
The $E2$ form factors for the various decuplet baryons are shown in
Figs.~\ref{ge2p}~to~\ref{ge2n}.

From our simulation we conclude that the $E2$ form factor of the
$\Omega^-$ baryon (the value of the $\Delta^-$ form factor at
$SU(3)_{\rm flavor}$ limit) is $(0.86 \pm 0.12) \times 10^{-2}~\rm fm^2$. 
The accuracy of our result indicates a definite non-zero value of the $E2$ form
factor of $\Omega^-$, and we favor a positive value.
Since $\Omega^-$ is a negatively charged baryon, this result implies
that it has an oblate shape, with the equatorial axis being larger
than the polar axis.

Similarly the $E2$ form factor for the $\Delta^+$ is $(-0.86 \pm 0.12)
\times 10^{-2}~\rm fm^2$ at the $SU(3)$ flavor symmetry point.  Our
results for the $\Delta^+$ compare favorably with the results of
Ref.~\cite{Alexandrou:2007we}. Using the closest available pion masses
of $533(3)$~MeV, and $563(4)$~MeV for this study,
and~\cite{Alexandrou:2007we} respectively, we find the $E2$ form
factor to be $(-1.16 \pm 29) \times 10^{-2}~\rm fm^2$, to be compared
with $(-1.08 \pm 40) \times 10^{-2}~\rm fm^2$. We note however, that
this study is performed at a finite $Q^2=0.230(1)\ {\rm GeV}^2$,
whilst~\cite{Alexandrou:2007we} reports results at $Q^2 = 0$.

The negative $E2$ form factor of a positive $\Delta^+$ baryon implies an
oblate shape in the Breit frame.
As illustrated in Fig.~\ref{ge2p}, the $E2$ form factor grows
substantially in magnitude as the chiral limit is approached, taking
the value $-0.030(11)~\rm fm^2$ at our lightest quark mass.
We note that the $E2$ form factor of  $\Delta^{++}$ is twice that of
the $\Delta^+$ $E2$ form factor and hence takes the value
$-0.060(23)~\rm fm^2$ at our lightest quark mass.

\begin{table*}[tbp]
\caption{Quark sector contributions to the $E2$ form factor of
  $\Delta$  at $Q^2 = 0.230(1)\ {\rm GeV}^2$ in fixed units of
  $e/m_N^2$.  Sector contributions 
  are for a single quark having unit charge. The fit windows are
  selected using the criteria outlined in Ref.~\cite{Boinepalli:2006xd}.}
\label{tab:E2D}
\begin{ruledtabular}
\begin{tabular}{ccccccc}
\noalign{\smallskip}
$\mathit m_\pi^2\ ({\rm GeV}^2)$  &\multicolumn{3}{c}{$\mathit u_\Delta$}  & \multicolumn{3}{c}{ $\mathit d_\Delta$ }\\
\noalign{\smallskip}
 & fit value & fit window & $\chi^2_{\rm dof}$ &fit value & fit
 window & $\chi^2_{\rm dof}$  \\
\hline
\noalign{\smallskip}
$0.9960(56)$ & $-0.117(16)$ & $16-20$ & $1.63$  & $-0.117(16)$ & $16-20$ & $1.63$ \\
$0.8936(56)$ & $-0.123(18)$ & $16-20$ & $1.39$  & $-0.123(18)$ & $16-20$ & $1.39$ \\
$0.7920(55)$ & $-0.130(22)$ & $16-20$ & $1.16$  & $-0.130(22)$ & $16-20$ & $1.16$ \\
$0.6920(54)$ & $-0.137(26)$ & $16-20$ & $1.01$  & $-0.137(26)$ & $16-20$ & $1.01$ \\
$0.6910(35)$ & $-0.163(17)$ & $16-20$ & $0.76$  & $-0.163(17)$ & $16-20$ & $0.76$ \\
$0.5925(33)$ & $-0.177(21)$ & $16-20$ & $0.75$  & $-0.177(21)$ & $16-20$ & $0.75$ \\
$0.4854(31)$ & $-0.194(27)$ & $16-20$ & $0.86$  & $-0.194(27)$ & $16-20$ & $0.86$ \\
$0.3795(31)$ & $-0.218(40)$ & $16-19$ & $1.03$  & $-0.218(40)$ & $16-19$ & $1.03$ \\
$0.2839(33)$ & $-0.263(67)$ & $16-19$ & $1.57$  & $-0.263(67)$ & $16-19$ & $1.57$ \\
$0.2153(35)$ & $-0.32(11)$  & $16-19$ & $1.20$  & $-0.32(11)$  & $16-19$ & $1.20$ \\
$0.1384(43)$ & $-0.52(20)$  & $16-18$ & $0.72$  & $-0.52(20)$  & $16-18$ & $0.72$ \\
$0.0939(44)$ & $-0.68(26)$  & $15-16$ & $1.06$  & $-0.68(26)$  & $15-16$ & $1.06$
\end{tabular}
\end{ruledtabular}
\vspace{-3pt}
\end{table*}

\begin{table*}[tbp]
\caption{Quark sector contributions to the $E2$ form factor of
  $\Sigma^*$ at $Q^2 = 0.230(1)\ {\rm GeV}^2$ in units of $e/m_N^2$.
  Sector contributions 
  are for a single quark having unit charge. The fit windows are
  selected using the criteria outlined in Ref.~\cite{Boinepalli:2006xd}.}
\label{tab:E2SS}
\begin{ruledtabular}
\begin{tabular}{ccccccc}
\noalign{\smallskip}
$\mathit m_\pi^2\ ({\rm GeV}^2)$  
&\multicolumn{3}{c}{$u_{\Sigma^*}\ {\rm or}\ d_{\Sigma^*}$} 
&\multicolumn{3}{c}{$\mathit s_{\Sigma^*}$}
\\
\noalign{\smallskip}
 & fit value & fit window & $\chi^2_{\rm dof}$ &fit value & fit
 window & $\chi^2_{\rm dof}$ \\
\hline
\noalign{\smallskip}
$0.9960(56)$ & $-0.132(19)$ & $16-20$ & $1.52$ & $-0.113(29)$ & $16-20$ & $0.75$ \\
$0.8936(56)$ & $-0.136(22)$ & $16-20$ & $1.36$ & $-0.118(31)$ & $16-20$ & $0.71$ \\
$0.7920(55)$ & $-0.145(25)$ & $16-20$ & $1.20$ & $-0.124(33)$ & $16-20$ & $0.66$ \\
$0.6920(54)$ & $-0.147(29)$ & $16-20$ & $1.07$ & $-0.131(36)$ & $16-20$ & $0.65$ \\
$0.6910(35)$ & $-0.172(18)$ & $16-20$ & $0.91$ & $-0.170(22)$ & $16-20$ & $0.48$ \\
$0.5925(33)$ & $-0.183(22)$ & $16-20$ & $0.89$ & $-0.180(24)$ & $16-20$ & $0.61$ \\
$0.4854(31)$ & $-0.194(27)$ & $16-20$ & $0.86$ & $-0.194(27)$ & $16-20$ & $0.86$ \\
$0.3795(31)$ & $-0.208(36)$ & $16-20$ & $0.48$ & $-0.211(32)$ & $16-20$ & $1.89$ \\
$0.2839(33)$ & $-0.225(51)$ & $16-17$ & $0.41$ & $-0.231(38)$ & $16-17$ & $0.61$ \\
$0.2153(35)$ & $-0.233(73)$ & $16-19$ & $1.08$ & $-0.257(48)$ & $16-19$ & $1.84$ \\
$0.1384(43)$ & $-0.29(11)$  & $16-17$ & $1.71$ & $-0.300(67)$ & $16-17$ & $1.07$ \\
$0.0939(44)$ & $-0.42(16)$  & $16-17$ & $0.94$ & $-0.325(88)$ & $16-17$ & $0.31$
\end{tabular}
\end{ruledtabular}
\vspace{-3pt}
\end{table*}

\begin{table*}[tbp]
\caption{Quark sector contributions to the $E2$ form factor of
  $\Xi^*$ at $Q^2 = 0.230(1)\ {\rm GeV}^2$ in units of $e/m_N^2$.
  Sector contributions 
  are for a single quark having unit charge. The fit windows are
  selected using the criteria outlined in Ref.~\cite{Boinepalli:2006xd}.}
\label{tab:E2XS}
\begin{ruledtabular}
\begin{tabular}{ccccccc}
\noalign{\smallskip}
$\mathit m_\pi^2\ ({\rm GeV}^2)$  &\multicolumn{3}{c}{$\mathit s_{\Xi^*}$}  & \multicolumn{3}{c}{ $u_{\Xi^*}\ {\rm or}\ d_{\Xi^*}$ }\\
\noalign{\smallskip}
 & fit value & fit window & $\chi^2_{\rm dof}$ &fit value & fit
 window & $\chi^2_{\rm dof}$  \\
\hline
\noalign{\smallskip}
$0.9960(56)$ & $-0.127(36)$ & $16-20$  & $0.64$ & $-0.157(25)$ & $16-20$ & $1.75$ \\
$0.8936(56)$ & $-0.131(37)$ & $16-20$  & $0.64$ & $-0.157(27)$ & $16-20$ & $1.52$ \\
$0.7920(55)$ & $-0.136(39)$ & $16-20$  & $0.64$ & $-0.159(30)$ & $16-20$ & $1.35$ \\
$0.6920(54)$ & $-0.141(40)$ & $16-20$  & $0.66$ & $-0.159(33)$ & $16-20$ & $1.18$ \\
$0.6910(35)$ & $-0.180(24)$ & $16-20$  & $0.65$ & $-0.184(20)$ & $16-20$ & $1.17$ \\
$0.5925(33)$ & $-0.186(25)$ & $16-20$  & $0.73$ & $-0.190(23)$ & $16-20$ & $1.04$ \\
$0.4854(31)$ & $-0.194(27)$ & $16-20$  & $0.86$ & $-0.194(27)$ & $16-20$ & $0.86$ \\
$0.3795(31)$ & $-0.201(29)$ & $16-17$  & $0.82$ & $-0.198(33)$ & $16-21$ & $0.62$ \\
$0.2839(33)$ & $-0.208(31)$ & $16-17$  & $0.59$ & $-0.200(41)$ & $16-17$ & $0.23$ \\
$0.2153(35)$ & $-0.214(34)$ & $16-17$  & $0.80$ & $-0.191(52)$ & $16-17$ & $0.06$ \\
$0.1384(43)$ & $-0.222(38)$ & $16-17$  & $0.92$ & $-0.184(71)$ & $16-18$ & $0.34$ \\
$0.0939(44)$ & $-0.222(41)$ & $15-16$  & $0.17$ & $-0.183(82)$ & $15-16$ & $0.70$
\end{tabular}
\end{ruledtabular}
\vspace{-3pt}
\end{table*}

\begin{table*}[tbp]
\caption{$E2$ form factors at $Q^2 = 0.230(1)\ {\rm GeV}^2$ of the charged
  decuplet 
  baryons in units of $10^{-2}\rm fm^2$ for different $m_\pi^2$
  values. The $E2$ form factor of the $\Delta^-$ at the
  $SU(3)_{\rm flavor}$ limit where $m_\pi^2 = 0.485(3)$ provides the $E2$
  form factor of $\Omega^-$.}
\label{tab:barE2}
\begin{ruledtabular}
\begin{tabular}{ccccccc}
\noalign{\smallskip}
$\mathit m_\pi^2\ ({\rm GeV}^2)$  & $\Delta^{++}$ & 
$\Delta^+$ & $\Delta^-$ &
$\Sigma^{*+}$ & $\Sigma^{*-}$ & $\Xi^{*-}$ \\
\noalign{\smallskip}
\hline
\noalign{\smallskip}
$0.9972(55)$ & $-1.03(14)$ & $-0.517(69)$ & $0.517(69)$ & $-0.613(77)$ & $0.555(97)$& $0.60(14)$   \\
$0.8936(56)$ & $-1.08(16)$ & $-0.541(80)$ & $0.541(80)$ & $-0.629(88)$ & $0.57(11)$  & $0.62(15)$   \\
$0.7920(55)$ & $-1.15(19)$ & $-0.575(96)$ & $0.575(96)$ & $-0.65(10)$  & $0.60(12)$ & $0.63(16)$  \\
$0.6920(54)$ & $-1.21(23)$ & $-0.61(12)$  & $0.61(12)$  & $-0.67(12)$  & $0.63(14)$ & $0.65(17)$ \\
$0.6910(35)$ & $-1.44(15)$ & $-0.718(75)$ & $0.718(75)$ & $-0.765(77)$ & $0.757(86)$  & $0.80(10)$    \\
$0.5925(33)$ & $-1.56(18)$ & $-0.782(91)$ & $0.782(91)$ & $-0.813(92)$ &  $0.804(99)$ & $0.83(11)$  \\
$0.4854(31)$ & $-1.71(24)$ & $-0.86(12)$  & $0.86(12)$  & $-0.86(12)$  & $0.86(12)$ & $0.86(12)$  \\
$0.3795(31)$ & $-1.93(35)$ & $-0.96(18)$  & $0.96(18)$  & $-0.91(17)$  & $0.92(15)$& $0.88(13)$   \\
$0.2839(33)$ & $-2.32(59)$ & $-1.16(29)$  & $1.16(29)$  & $-0.99(25)$ & $1.00(20)$ & $0.91(15)$   \\
$0.2153(35)$ & $-2.79(95)$ & $-1.40(48)$  & $1.40(48)$  & $-0.99(37)$ & $1.07(28)$ & $0.91(17)$   \\
$0.1384(43)$ & $-4.6(1.8)$ & $-2.31(88)$  & $2.31(88)$  & $-1.28(58)$ & $1.30(41)$ & $0.92(21)$   \\
$0.0939(44)$ & $-6.0(2.3)$ & $-3.0(1.1)$ & $3.0(1.1)$   & $-1.99(86)$ & $1.71(57)$ & $0.92(23)$
\end{tabular}
\end{ruledtabular}
\vspace{-3pt}
\end{table*}

\subsection{Magnetic Octupole Moments}

\begin{figure}[tbp!]
\begin{center}
  {\includegraphics[height=\hsize,angle=90]{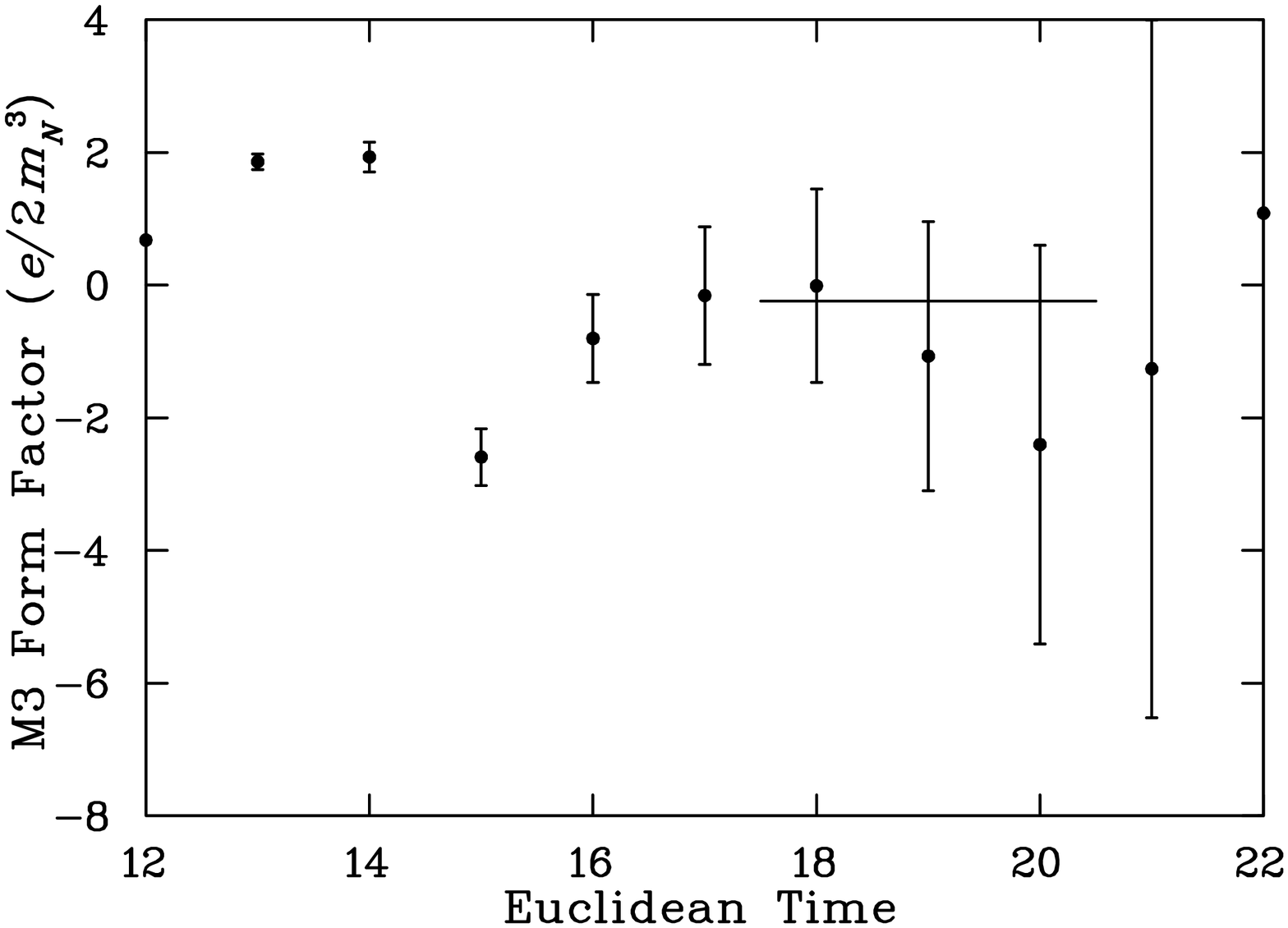}}
\end{center}
\caption{$M3$ form factor of the $\Delta$ at the 
  $SU(3)_{\rm flavor}$ limit as a function of time. }
\label{cfm3K3}
\end{figure}

The magnetic octupole form factors are calculated on the
lattice by considering a combination of ratios of three and two-point
functions as given in Eq.~(\ref{gm3}).

\begin{figure}[tbp!]
\begin{center}
  {\includegraphics[height=\hsize,angle=90]{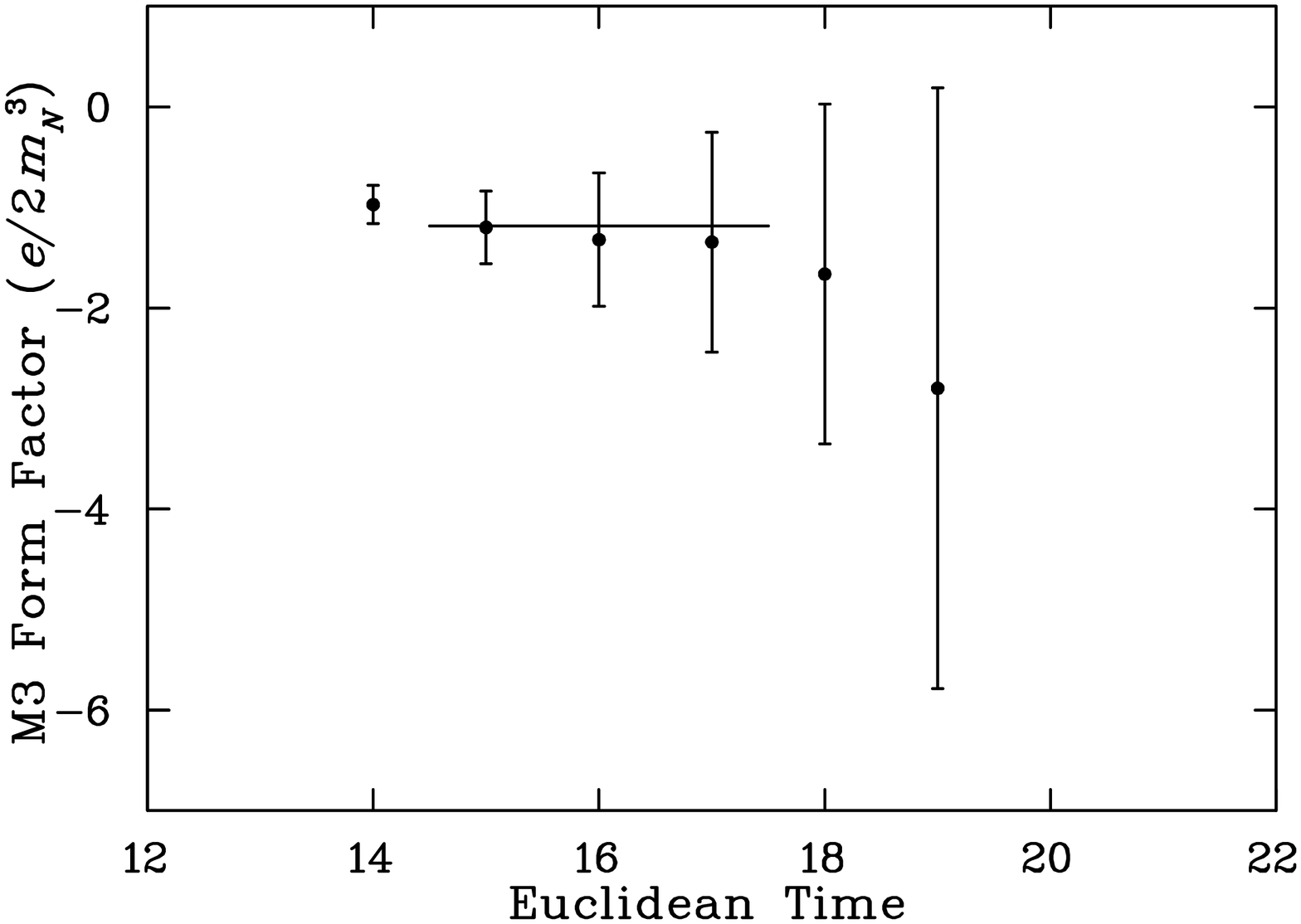}}
\end{center}
\caption{$M3$ form factor (splitting) of the $u$ quark sector of the
  $\Delta$ at the ninth quark mass where
 $m_\pi^2 = 0.215(4) {\rm GeV^2}$ as a function of time. }
\label{cfm3K6}
\end{figure}

Figure~\ref{cfm3K3} provides a plot of the correlator proportional to
the $M3$ form factor of a $u$ quark in the $\Delta$ as a function of
Euclidean time at the $SU(3)$ flavor limit.  Figure~\ref{cfm3K6}
provides the $M3$ form factor at the ninth quark mass, where a plateau
is realized using the splittings method.  Tables~\ref{tab:M3D}
to~\ref{tab:M3XS} list the quark sector $M3$ form factors. The
magnetic octupole form factors ($M3$) of the decuplet baryons are
listed in Table~\ref{tab:barM3}.

\begin{figure}
\begin{center}
  {\includegraphics[height=\hsize,angle=90]{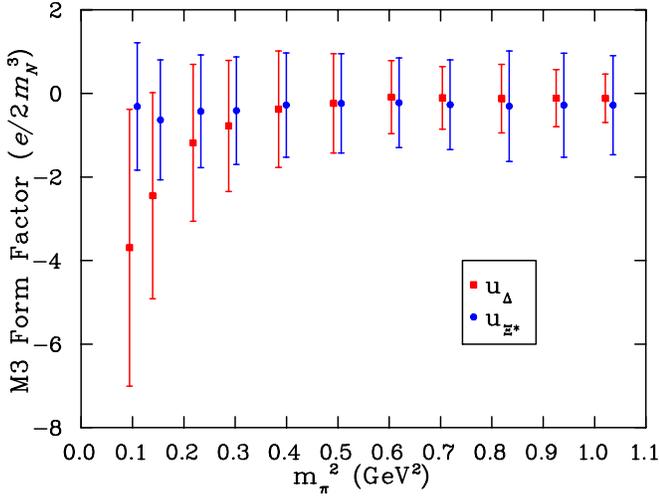}}
\end{center}
\caption{$M3$ form factor contributions from the $u$ quark sectors of
  the $\Delta$ and $\Sigma^*$. The results for the $\Sigma^*$ are
  offset for clarity.}
\label{gm3u.dx}
\end{figure}

\begin{figure}
\begin{center}
  {\includegraphics[height=\hsize,angle=90]{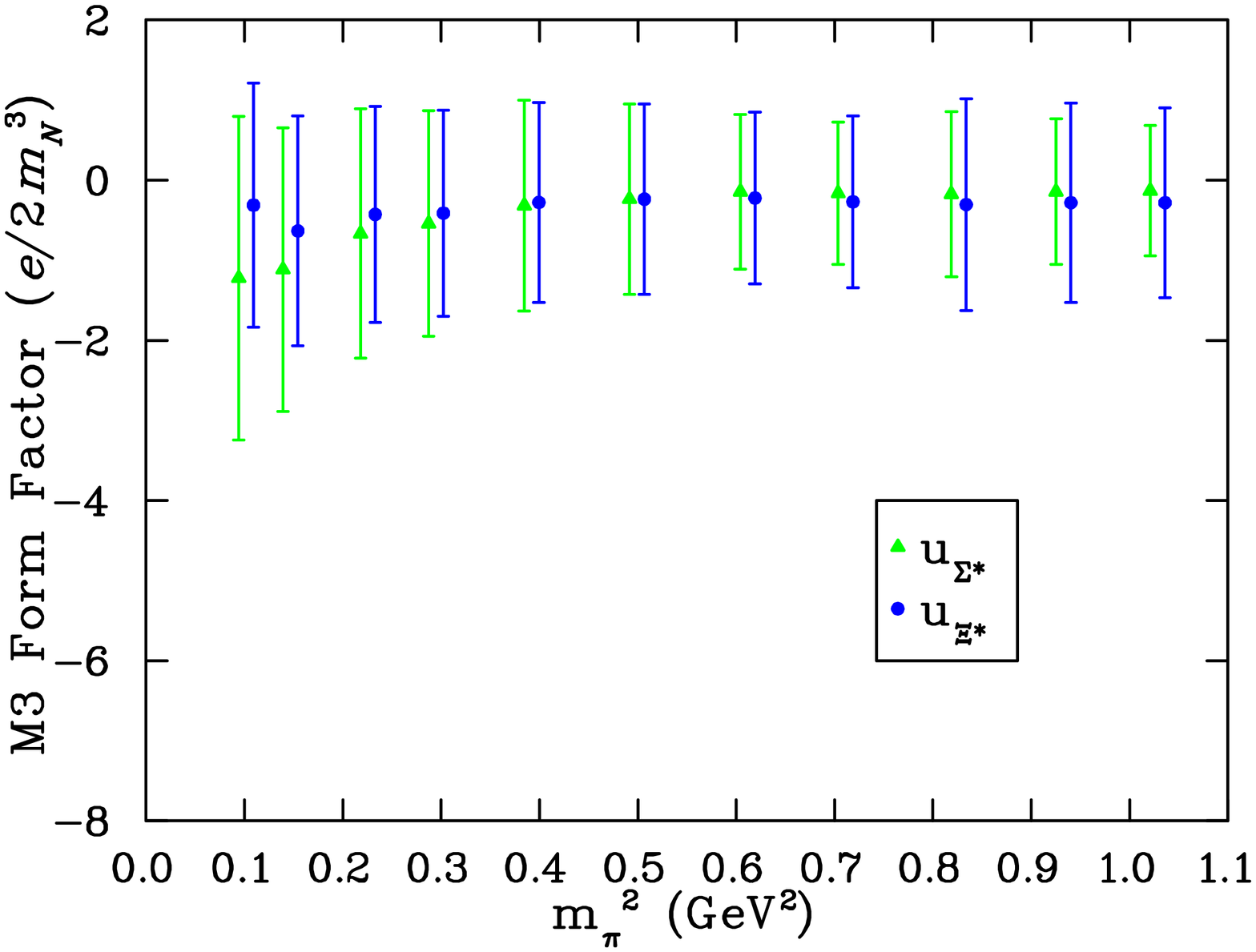}}
\end{center}
\caption{$M3$ form factor contributions from the $u$ quark sectors of
  the $\Sigma^*$ and $\Xi^*$. The results for the $\Xi^*$ have been
  offset for clarity.}
\label{gm3u.sx}
\end{figure}

\begin{figure}
\begin{center}
  {\includegraphics[height=\hsize,angle=90]{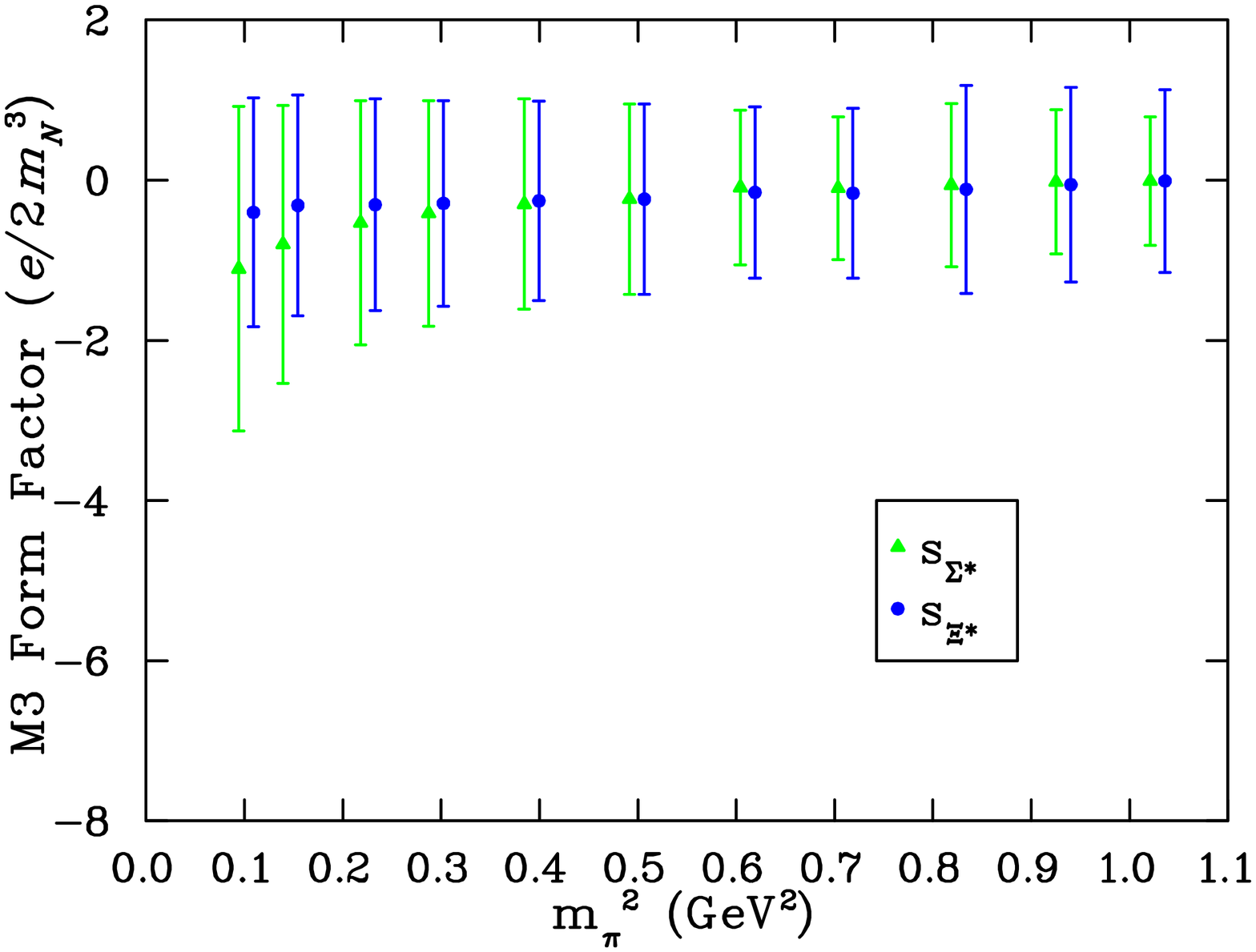}}
\end{center}
\caption{$M3$ form factor contributions from the $s$ quark sectors of
  the $\Sigma^*$ and $\Xi^*$. The results for the $\Xi^*$ have been
  offset for clarity.}
\label{gm3s.sx}
\end{figure}

\begin{figure}
\begin{center}
  {\includegraphics[height=\hsize,angle=90]{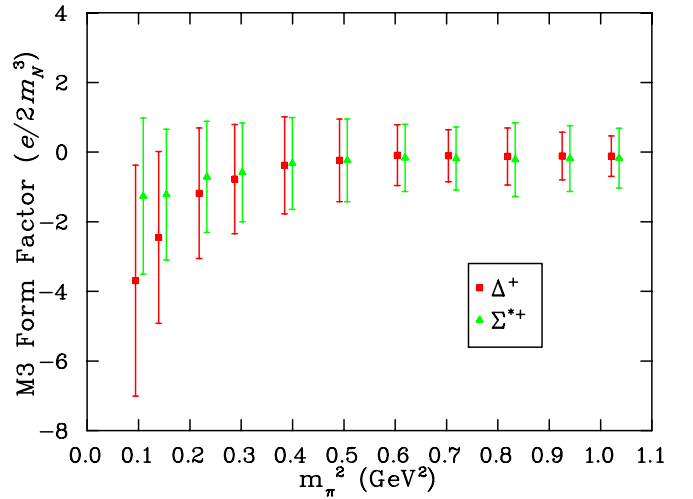}}
\end{center}
\caption{$M3$ form factors for the
  $\Delta^+$ and 
  $\Sigma^{*+}$ at different quark masses. The values for
  $\Sigma^{*+}$ are plotted at shifted $m_\pi^2$ for clarity.}
\label{gm3p}
\end{figure}

\begin{figure}
\begin{center}
  {\includegraphics[height=\hsize,angle=90]{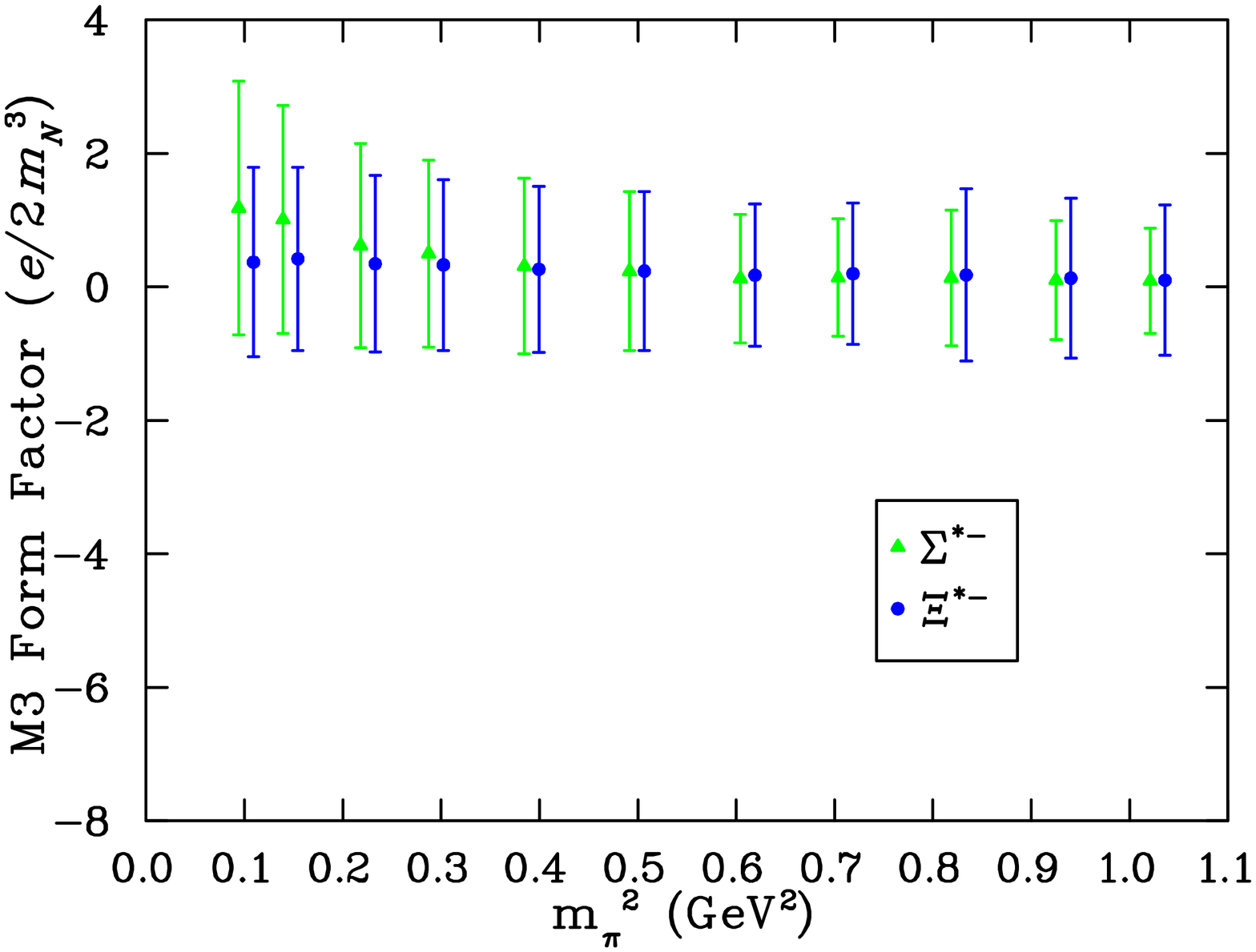}}
\end{center}
\caption{$M3$ form factors  for the $\Sigma^{*-}$ 
  and $\Xi^{*-}$ at different quark masses. The values for $\Xi^{*-}$
  are plotted at shifted $m_\pi^2$ for clarity. }
\label{gm3m}
\end{figure}

\begin{figure}
\begin{center}
  {\includegraphics[height=\hsize,angle=90]{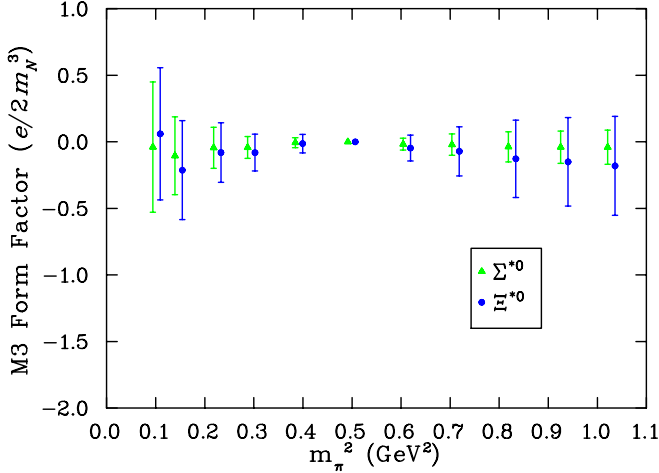}}
\end{center}
\caption{$M3$ form factors for the $\Sigma^{*0}$ 
  and $\Xi^{*0}$ at different quark masses.  The values for $\Xi^{*0}$
  are plotted at shifted $m_\pi^2$ for clarity. }
\label{gm3n}
\end{figure}

Plots of the quark sector contributions to the $M3$ form factors are
provided in
Figs.~\ref{gm3u.dx}~to~\ref{gm3s.sx}. Figs.~\ref{gm3p}~to~\ref{gm3n}
show the $M3$ form factors for the decuplet baryons.

Like $E2$, the $M3$ form factors require nonzero orbital angular
momentum admixtures in the ground state wave
function~\cite{Leinweber:1992hy}.  Our statistics are sufficient to
reveal a non-trivial result for the $M3$ form factor of the $\Delta$
for the first time. We find a result of $-3.7(3.3)~e/2 m_N^3$ for the
$\Delta^+$ at a squared pion mass of $0.094(4)~\mathrm{GeV}^2$, close
to the physical limit. At larger masses the results are consistent
with zero, but systematically negative, with enhancement at the
lightest masses for the $\Delta$, and to a lesser extent the
$\Sigma^*$.

\begin{table*}[tbp]
\caption{Quark sector contributions to the $M3$ form factor of
  $\Delta$  at $Q^2 = 0.230(1)\ {\rm GeV}^2$ in units of $e/2 m_N^3$.
  Sector contributions 
  are for a single quark having unit charge. The fit windows are
  selected using the criteria outlined in Ref.~\cite{Boinepalli:2006xd}.}
\label{tab:M3D}
\begin{ruledtabular}
\begin{tabular}{ccccccc}
\noalign{\smallskip}
$\mathit m_\pi^2\ ({\rm GeV}^2)$  &\multicolumn{3}{c}{$\mathit u_\Delta$}  & \multicolumn{3}{c}{ $\mathit d_\Delta$ }\\
\noalign{\smallskip}
 & fit value & fit window & $\chi^2_{\rm dof}$ &fit value & fit
 window & $\chi^2_{\rm dof}$  \\
\hline
\noalign{\smallskip}
$0.9972(55)$ &  $0.19(58)$ & $18-22$ & $0.83$ & $0.19(58)$ & $18-22$ & $0.83$ \\
$0.8936(56)$ &  $0.19(68)$ & $18-22$ & $0.63$ & $0.19(68)$ & $18-22$ & $0.63$ \\
$0.7920(55)$ &  $0.18(82)$ & $18-22$ & $0.50$ & $0.18(82)$ & $18-22$ & $0.50$ \\
$0.6920(54)$ &  $0.2(1.0)$ & $18-22$ & $0.44$ & $0.2(1.0)$ & $18-22$ & $0.44$ \\
$0.6910(35)$ & $-0.10(75)$ & $18-22$ & $0.74$  & $-0.10(75)$ & $18-22$ & $0.74$ \\
$0.5925(33)$ & $-0.09(87)$ & $18-20$ & $1.01$  & $-0.09(87)$ & $18-20$ & $1.01$ \\
$0.4854(31)$ & $-0.2(1.2)$ & $18-20$ & $0.92$  & $-0.2(1.2)$ & $18-20$ & $0.92$ \\
$0.3795(31)$ & $-0.4(1.4)$ & $16-18$ & $0.44$  & $-0.4(1.4)$ & $16-18$ & $0.44$ \\
$0.2839(33)$ & $-0.8(1.6)$ & $15-17$ & $0.01$  & $-0.8(1.6)$ & $15-17$ & $0.01$ \\
$0.2153(35)$ & $-1.2(1.9)$ & $15-17$ & $0.12$  & $-1.2(1.9)$ & $15-17$ & $0.12$ \\
$0.1384(43)$ & $-2.4(2.5)$ & $15-17$ & $0.37$  & $-2.4(2.5)$ & $15-17$ & $0.37$ \\
$0.0939(44)$ & $-3.7(3.3)$ & $15-17$ & $0.32$  & $-3.7(3.3)$ & $15-17$ & $0.32$
\end{tabular}
\end{ruledtabular}
\vspace{-3pt}
\end{table*}

\begin{table*}[tbp]
\caption{Quark sector contributions to the $M3$ form factor of
  $\Sigma^*$ baryons at $Q^2 = 0.230(1)~{\rm GeV}^2$ in units of
  $e/2 m_N^3$.  Sector contributions 
  are for a single quark having unit charge. The fit windows are
  selected using the criteria outlined in Ref.~\cite{Boinepalli:2006xd}.}
\label{tab:M3SS}
\begin{ruledtabular}
\begin{tabular}{ccccccc}
\noalign{\smallskip}
$\mathit m_\pi^2\ ({\rm GeV}^2)$  
&\multicolumn{3}{c}{$u_{\Sigma^*}\ {\rm or}\ d_{\Sigma^*}$} 
&\multicolumn{3}{c}{$\mathit s_{\Sigma^*}$}
\\
\noalign{\smallskip}
 & fit value & fit window & $\chi^2_{\rm dof}$ &fit value & fit
 window & $\chi^2_{\rm dof}$   \\
\hline
\noalign{\smallskip}
$0.9972(55)$ & $0.13(81)$ & $18-22$ & $0.56$ & $0.64(80)$ & $18-22$ & $0.50$ \\
$0.8936(56)$ & $0.12(91)$ & $18-22$ & $0.50$ & $0.63(90)$ & $18-22$ & $0.44$ \\
$0.7920(55)$ & $0.1(1.0)$ & $18-22$ & $0.48$ & $0.6(1.0)$ & $18-22$ & $0.39$ \\
$0.6920(54)$ & $0.1(1.2)$ & $18-22$ & $0.47$ & $0.6(1.2)$ & $18-22$ & $0.38$ \\
$0.6910(35)$ & $-0.16(89)$ & $18-22$ & $0.96$  & $0.10(89)$  & $18-22$ & $0.64$ \\
$0.5925(33)$ & $-0.1(1.0)$ & $18-20$ & $1.03$  & $-0.1(1.0)$ & $18-20$ & $0.90$ \\
$0.4854(31)$ & $-0.2(1.2)$ & $18-20$ & $0.92$  & $-0.2(1.2)$ & $18-20$ & $0.92$ \\
$0.3795(31)$ & $-0.3(1.3)$ & $16-18$ & $0.43$  & $-0.3(1.3)$ & $16-18$ & $0.61$ \\
$0.2839(33)$ & $-0.5(1.4)$ & $15-17$ & $0.01$  & $-0.4(1.4)$ & $15-17$ & $0.30$ \\
$0.2153(35)$ & $-0.7(1.6)$ & $15-17$ & $0.01$  & $-0.5(1.5)$ & $15-17$ & $0.38$ \\
$0.1384(43)$ & $-1.1(1.8)$ & $15-17$ & $0.21$  & $-0.8(1.7)$ & $15-17$ & $0.07$ \\
$0.0939(44)$ & $-1.2(2.0)$ & $15-17$ & $0.07$  & $-1.1(2.0)$ & $15-17$ & $0.89$
\end{tabular}
\end{ruledtabular}
\vspace{-3pt}
\end{table*}

\begin{table*}[tbp]
\caption{Quark sector contributions to the $M3$ form factor of
  $\Xi^*$ baryons at $Q^2 = 0.230(1)\ {\rm GeV}^2$ in units of
  $e/2 m_N^3$.  Sector contributions 
  are for a single quark having unit charge. The fit windows are
  selected using the criteria outlined in Ref.~\cite{Boinepalli:2006xd}.}
\label{tab:M3XS}
\begin{ruledtabular}
\begin{tabular}{ccccccc}
\noalign{\smallskip}
$\mathit m_\pi^2\ ({\rm GeV}^2)$  &\multicolumn{3}{c}{$\mathit s_{\Xi^*}$}  & \multicolumn{3}{c}{ $u_{\Xi^*}\ {\rm or}\ d_{\Xi^*}$ }\\
\noalign{\smallskip}
 & fit value & fit window & $\chi^2_{\rm dof}$ &fit value & fit
 window & $\chi^2_{\rm dof}$   \\
\hline
\noalign{\smallskip}
$0.9972(55)$ & $0.6(1.1)$ & $18-22$ & $0.39$ & $-0.1(1.2)$ & $18-22$ & $0.45$ \\
$0.8936(56)$ & $0.6(1.2)$ & $18-22$ & $0.39$ & $-0.1(1.2)$ & $18-22$ & $0.48$ \\
$0.7920(55)$ & $0.5(1.3)$ & $18-22$ & $0.40$ & $-0.1(1.3)$ & $18-22$ & $0.52$ \\
$0.6920(54)$ & $0.5(1.4)$ & $18-22$ & $0.41$ & $0.0(1.4)$  & $18-22$ & $0.53$ \\
$0.6910(35)$ & $-0.2(1.1)$ & $18-22$ & $0.83$ & $-0.3(1.1)$ & $18-22$ & $1.27$ \\
$0.5925(33)$ & $-0.2(1.1)$ & $18-20$ & $0.92$ & $-0.2(1.1)$ & $18-20$ & $1.03$ \\
$0.4854(31)$ & $-0.2(1.2)$ & $18-20$ & $0.92$ & $-0.2(1.2)$ & $18-20$ & $0.92$ \\
$0.3795(31)$ & $-0.3(1.2)$ & $16-18$ & $0.62$ & $-0.3(1.2)$ & $16-18$ & $0.54$ \\
$0.2839(33)$ & $-0.3(1.3)$ & $15-17$ & $0.52$ & $-0.4(1.3)$ & $15-17$ & $0.17$ \\
$0.2153(35)$ & $-0.3(1.3)$ & $15-17$ & $0.63$ & $-0.4(1.4)$ & $15-17$ & $0.05$ \\
$0.1384(43)$ & $-0.3(1.4)$ & $15-17$ & $0.23$ & $-0.6(1.4)$ & $15-17$ & $0.47$ \\
$0.0939(44)$ & $-0.4(1.4)$ & $15-17$ & $0.44$ & $-0.3(1.5)$ & $15-17$ & $0.82$
\end{tabular}
\end{ruledtabular}
\vspace{-3pt}
\end{table*}

\begin{table*}[tbp]
\caption{$M3$ form factor results at $Q^2 = 0.230(1)\ {\rm GeV}^2$ of
  the charged decuplet baryons in units of $e/2 m_N^3$ for different
  $m_\pi^2$ values.
  The $M3$ form factor of the $\Delta^-$ at the
  $SU(3)_{\rm flavor}$ limit where $m_\pi^2 = 0.485(3)$ provides the $M3$
  form factor of $\Omega^-$.}
\label{tab:barM3}
\begin{ruledtabular}
\begin{tabular}{ccccccc}
\noalign{\smallskip}
$\mathit m_\pi^2\ ({\rm GeV}^2)$  & $\Delta^{++}$ & 
$\Delta^+$ & $\Delta^-$ &
$\Sigma^{*+}$ & $\Sigma^{*-}$ & $\Xi^{*-}$ \\
\noalign{\smallskip}
\hline
\noalign{\smallskip}
$0.9972(55)$ & $0.4(1.2)$ & $0.19(58)$ & $-0.19(58)$ & $-0.04(86)$ & $-0.30(79)$ & $-0.4(1.1)$ \\
$0.8936(56)$ & $0.4(1.4)$ & $0.19(68)$ & $-0.19(68)$ & $-0.05(94)$ & $-0.29(89)$ & $-0.4(1.2)$ \\
$0.7920(55)$ & $0.4(1.6)$ & $0.18(82)$ & $-0.18(82)$ & $-0.1(1.1)$ & $-0.3(1.0)$ & $-0.3(1.3)$ \\
$0.6920(54)$ & $0.4(2.0)$ & $0.2(1.0)$ & $-0.2(1.0)$ & $-0.1(1.2)$ & $-0.3(1.2)$ & $-0.3(1.4)$ \\
$0.6910(35)$ & $-0.2(1.5)$ & $-0.10(75)$ & $0.10(75)$ & $-0.18(90)$ & $0.14(88)$ & $0.2(1.1)$ \\
$0.5925(33)$ & $-0.2(1.7)$ & $-0.09(87)$ & $0.09(87)$ & $-0.2(1.0)$ & $0.1(1.0)$ & $0.2(1.1)$ \\
$0.4854(31)$ & $-0.5(2.4)$ & $-0.2(1.2)$ & $0.2(1.2)$ & $-0.2(1.2)$ & $0.2(1.2)$ & $0.2(1.2)$ \\
$0.3795(31)$ & $-0.8(2.8)$ & $-0.4(1.4)$ & $0.4(1.4)$ & $-0.3(1.3)$ & $0.3(1.3)$ & $0.3(1.2)$ \\
$0.2839(33)$ & $-1.6(3.1)$ & $-0.8(1.6)$ & $0.8(1.6)$ & $-0.6(1.4)$ & $0.5(1.4)$ & $0.3(1.3)$ \\
$0.2153(35)$ & $-2.4(3.8)$ & $-1.2(1.9)$ & $1.2(1.9)$ & $-0.7(1.6)$ & $0.6(1.5)$ & $0.3(1.3)$ \\
$0.1384(43)$ & $-4.9(4.9)$ & $-2.4(2.5)$ & $2.4(2.5)$ & $-1.2(1.9)$ & $1.0(1.7)$ & $0.4(1.4)$ \\
$0.0939(44)$ & $-7.4(6.6)$ & $-3.7(3.3)$ & $3.7(3.3)$ & $-1.3(2.2)$ & $1.2(1.9)$ & $0.4(1.4)$
\end{tabular}
\end{ruledtabular}
\vspace{-3pt}
\end{table*}

\begin{table*}
\caption{Collected results for the $\Omega^-$. Results are obtained from
  the $\Delta^-$ at the $SU(3)_{\rm flavor}$ limit, where $m_\pi^2
   = 0.4854(31)~{\rm GeV}^2$.}
\label{tab:omegaresults}
\begin{ruledtabular}
\begin{tabular}{lc}
\noalign{\smallskip}
quantity  & fit value \\
\noalign{\smallskip}
\hline
\noalign{\smallskip}
mass $({\rm GeV})$  & $1.732(12)$  \\
charge radius $({\rm fm}^2)$ & $-0.307(15)$  \\
magnetic moment $(\mu_N)$  & $-1.697(65)$  \\
$E2$ form factor $(10^{-2}{\rm fm}^2)$ & $0.86(12)$  \\
$M3$ form factor $(e/2 m_N^3)$ & $0.2(1.2)$
\end{tabular}
\end{ruledtabular}
\vspace{-3pt}
\end{table*}

\clearpage

%
\section{Summary}
\label{sec:Summary}

We have performed an extensive calculation of the electromagnetic
properties of decuplet baryons at both the quark level and the
baryon level, including the quadrupole and octupole form factors of
the spin-3/2 baryons.
For the first time we obtain non-trivial results for both the $E2$ and
$M3$ form factors. In particular, we find decuplet baryons to be
oblate in shape.

We find that the quarks in the decuplet are not as sensitive to their
environment as their octet counterparts.
Of particular note, is the discovery that the decuplet-baryon radii
are \emph{smaller} than that of the octet baryons, contradicting the simple
quark model, but substantiating hints in the early study of
Ref.~\cite{Leinweber:1992hy}.

A particularly interesting finding is that the suppression of
sea-quark loop
contributions in QQCD reduces the decuplet magnetic moment
considerably, resulting in a turnover in the magnetic moment at light
quark masses, as illustrated in Fig.~\ref{magmomPD}.
At large pion masses, the $\Delta^+$ moment is enhanced relative to
the proton moment in accord with earlier quenched lattice QCD
calculations \cite{Leinweber:1992hy,Leinweber:1990dv} and model
expectations.
However, as the chiral regime is approached, the non-analytic behavior
of the quenched meson cloud is revealed, enhancing the proton and
suppressing the $\Delta^+$, in accord with the expectations of
$\rm{Q}\chi{\rm PT}$.
This suppression should be absent in full QCD. Hence this is one
particular case that we have identified as a place to look for effects
of unquenching in future dynamical simulations.
We also predict that unquenching effects should be observed in the
$\Omega^-$ magnetic moment, which we find to be suppressed in quenched
QCD with regard to the experimentally measured value due to the
absence of $K\Xi$ loops in the virtual decay of $\Omega^-$.

Through a calculation of the decuplet E2 form factors and electric
quadrupole moments, we predict oblate shapes for the
decuplet baryons.
It will be interesting to confront this prediction with an
experimental measurement of the $\Omega^-$ quadrupole form factor. We
provide a summary of all the $\Omega^-$ measurements in
Table~\ref{tab:omegaresults}.

Finally we have obtained non-trivial values for the $M3$ form factor
for the first time in lattice QCD studies. These results provide an
interesting and novel forum for the further development of our
understanding of non-perturbative QCD.

\begin{acknowledgments}

We thank the Australian Partnership for Advanced Computing
(APAC) and the South Australian Partnership for Advanced
Computing (SAPAC) for generous grants of supercomputer time which have
enabled this project.  This work was supported by the Australian
Research Council.
J.Z. is supported by STFC grant PP/F009658/1.
JBZ is supported by Chinese NSFC-Grant No. 10675101 and 10835002.

\end{acknowledgments}

%

\end{document}